\documentclass[a4paper,12pt]{article}
\usepackage{graphicx}
\usepackage{bm}
\usepackage{amsmath}
\usepackage{cases}
\begin{document}
\setcounter{secnumdepth}{3}
\setcounter{tocdepth}{2}


\title{Scaling perspective on intramolecular vibrational energy flow: analogies, insights, and challenges\footnote{{\tt Published in Adv. Chem. Phys. {\bf 153}, 43-110 (2013). DOI: 10.1002/9781118571767.ch2}}}
\author{Srihari Keshavamurthy \\
Department of Chemistry, Indian Institute of Technology,\\
Kanpur, Uttar Pradesh 208 016, India}

\maketitle

\tableofcontents

\section{Introduction: motivation and historical overview}
\label{sec:intro}

This is a review, certainly not the first one, about the phenomenon of intramolecular vibrational energy
redistribution (IVR) - its mechanism and its impact on the reaction rate
theories. For a subject that is nearly a century old, several reviews at appropriate times have appeared in the literature\cite{ivr_rev1,ivr_rev2,ivr_rev3,ivr_rev4,ivr_rev5,ivr_rev6,ivr_rev7,ivr_rev8,ivr_rev9,ivr_rev10,ivr_rev11,ivr_rev12,ivr_rev13}.
It is not entirely surprising that experimental and theoretical
attempts to understand IVR started nearly half a century ago. After all, IVR
is at the heart of almost every bond making and bond breaking event. Over the years experiments have become\cite{expt_adv1,expt_adv2,expt_adv3} sophisticated and detailed enough to measure several quantities of interest, providing ample challenges for theory.
Therefore, it was crucial to make some assumption about the rate and extent
of IVR in order to formulate theoretical models. In this context one of the early and very influential approximation is enshrined as the Rice-Ramsperger-Kassels-Marcus (RRKM) theory\cite{rrkm1,rrkm2,rrkm3}. In a remarkable series of papers\cite{rrkm3} in $1952$ Marcus put the earlier RRK theory\cite{rrkm1,rrkm2} on a firm basis by including the other cornerstone of reaction dynamics - transition state theory\cite{tst1,tst2,tst3} (TST) and with one stroke banished dynamics from rate calculations. The resulting microcanonical rate expression for a reaction with threshold energy $E_{0}$
\begin{equation}
k_{\rm RRKM}(E) = \frac{N^{\dagger}(E-E_{0})}{2 \pi \hbar \rho(E)}
\label{rrkmcl}
\end{equation}
 involves the number of states at the transition state $N^{\dagger}(E-E_{0})$ with energy less then or equal to $E-E_{0}$ and the total reactant density of states $\rho(E)$. Dynamics disappears in the assumption of ``instantaneous" IVR on the reaction timescales. In fact Marcus invoked the notion of active, adiabatic and inactive degrees of freedom (DOF) with respect to their roles in IVR. The active DOFs contribute their energy unhindered to the breaking bond whereas the adiabatic DOFs contribute little to none of their energy to the breaking bond. The IVR between inactive DOFs and the breaking bond is assumed to be sufficiently fast at the transition state. I mention this classification scheme for two main reasons. Firstly, it goes beyond the usual bland statement, often made, of instantaneous IVR involving all DOFs. Secondly, such a classification will later on prove to be invaluable in terms of gaining insights into the dynamical deviations from the statistical RRKM theory. The simplicity and elegance of Eq.~\ref{rrkmcl} and its ease of implementation has led to several checks\cite{rrkm_validity} on the validity of the statistical approximation. 
 
Dynamical corrections to Eq.~\ref{rrkmcl} can be succinctly expressed in terms of a correction factor\cite{qns_stilbene,refcor1}
 \begin{equation}
 \kappa(E) = \frac{k_{\rm IVR}(E)}{k_{\rm IVR}(E) + \nu(E)}
 \end{equation}
 which involves the IVR rate $k_{\rm IVR}(E)$ and the barrier crossing, from reactant to product, frequency $\nu(E)$. The above correction, although seemingly simple bur far from trivial, is at the heart of the present review. In fact all of the dynamical complexity is hidden in $\kappa(E)$ and much of the modern development in chemical reaction dynamics is focused on understanding the mechanistic origins of $\kappa(E)$. Fro instance, during conformational changes the barrier crossing frequency is of the order of $10$ ps$^{-1}$ whereas the IVR rate is of the order of about $1$ ps$^{-1}$\cite{qns_stilbene}.
 
Before proceeding further, however, it is relevant to mention another pioneering work that appeared in $1955$, essentially around the time that RRKM was proposed. This is the famous Fermi-Pasta-Ulam-Tsingou (FPUT) numerical experiment\cite{fput1,fput2} wherein the energy spreading in a chain of masses connected by nonlinear springs was investigated. Surprisingly, an initially plucked mode did not ``thermalize" over long timescales. In other words, modes did not share energy actively and the equipartition principle was apparently violated.  Intrigued by this lack of ergodicity, a flurry of activity ensued and the paradox was eventually resolved. The original parameters of FPUT put the system below the threshold of strong chaos and hence the lack of ergodicity. Interestingly, en route to resolving the FPUT paradox, several new ideas like solitons, intrinsic localized modes, nonlinear Schr\"{o}dinger equation and the importance of nonlinear resonances and chaos emerged from various studies\cite{fputinspire}. Indeed it would not be an overstatement to say that the FPUT problem inspired much of the significant advances in the field of nonlinear dynamics and quantum chaos. It is worth noting that the celebrated Kolmogorov-Arnold-Moser (KAM) theorem\cite{kam} on the stability of regular Hamiltonian dynamics under the influence of small perturbations also came about during the same time frame.  
 Notwithstanding the irony of RRKM and FPUT being proposed around the same period, the two fields of research came together in the $1976$ seminar paper\cite{oxrice} by Oxtoby and Rice. Apart from appreciating the importance of KAM and the Chirikov resonance overlap criterion\cite{chirioverlap}, proposed in 1959, to the phenomenon of IVR, Oxtoby and Rice were keen ``to examine the dynamics which underlie a statistical description" using models of coupled Morse oscillators. The conclusions of Oxtoby and Rice are remarkably prescient in the context of the current modern state space viewpoint of IVR - the main topic of this review.

The close parallel between the issue raised by FPUT and the phenomenon of IVR is undeniable. In essence, IVR is equivalent to asking the FPUT question, only now for a three dimensional network of coupled anharmonic vibrational modes.
 Thus, it is not surprising that in our attempts to understand IVR in molecular systems we invariably come face to face with some of the deep fundamental issues like quantum chaos, quantum ergodicity, and origins of irreversibility.  What might come as a surprise, however, is the fact that despite extensive classical, semiclassical, and quantum studies on IVR for nearly half a century, our theoretical understanding is still being severely tested and challenged by recent experiments.  Schofield and Wolynes have summed up the reasons for
this unabated interest in IVR very elegantly\cite{schowoly_quote}:
``{\em Interest in how quantum mechanics affects vibrational energy flow
within a single molecule predates the discovery of the Schr\"{o}dinger equation.
Nevertheless, understanding intramolecular vibrational 
energy redistribution (IVR)
remains an area of investigation in chemical physics today. 
How can a problem have such
longevity? Clearly it must be both important, so that the interest remains
sustained, and either very difficult, so that no approach makes much
headway, or very rich in complexity, so there is much to do. Intramolecular
vibrational energy flow shares all three of these features.}" 

In order to set the tone for this review we start by highlighting some of the recent experiments that clearly point towards the need for understanding the dynamics of IVR in greater detail then ever before. A  comprehensive list of earlier experiments can be found in several excellent reviews, especially the one by Nesbitt and Field\cite{ivr_rev6}. Two recent reviews\cite{haserev1,haserev2} on non-RRKM dynamics in unimolecular and bimolecular reactions give several examples wherein a critical rethinking of the traditional reaction mechanisms are required. An overarching theme of the current review is to emphasize the utility of a classical-quantum correspondence approach to IVR. Classical-quantum correspondence approach to IVR has a long history and several beautiful insights into the IVR mechanism have originated from this viewpoint. For example, an earlier review by Uzer\cite{ivr_rev4} summarizes the key developments from this perspective. Note that exact quantum dynamical approaches have also developed rapidly over the last few decades. Thus, starting with Wyatt's recursive residue generation method (RRGM) approach\cite{rrgm} towards the IVR dynamics in the full $30$-mode benzene\cite{wyatt_benzene} to the recent multi-configuration time dependent Hartree\cite{mctdh} (MCTDH) study of the combined influence of conical intersection and IVR on photoisomerization of fulvene by\cite{mctdh_fulvene} Blancafort {\it et al.} one has now very powerful tools for exact quantum studies. 
However, since the review by Uzer, several significant advances have occurred in the field of nonlinear dynamics and phase space transport in systems with three or more degrees of freedom as well. Fortunately, these advances in classical dynamics have happened in parallel with the equally impressive advances in our ability to do quantum dynamics on large scale systems. Although the advances in computational quantum dynamics will not be reviewed here, one has now an ideal platform to take the classical-quantum correspondence to the next level. The hope is that such studies on higher DoF systems will once again yield valuable mechanistic insights into the IVR dynamics and help interpret the detailed quantum computations. In addition, a clear understanding of quantum effects like tunneling and coherence and their relevance to control requires detailed classical-quantum correspondence studies. Several examples\cite{tuncoh_control} can be found in the literature and efforts along these lines can provide insights into the anatomy of the quantum control fields.

The importance, difficulty, and complexity of IVR are being brought out
more clearly and definitively over the last decade. Ironically, the 
need for detailed dynamical studies is required more in the present times
due to the increasing focus on nano and meso scale systems. 
The comfort of being able to make
statistical assumptions about the dynamics of large molecules is no longer
there in such systems. For example, the issue of heat flow in molecular
electronic devices is intimately connected to the IVR mechanism. In a recent
experiment\cite{dlott_heat}, Dlott and coworkers showed that the heat transport through
monomolecular hydrocarbon layers is ballistic, as opposed to diffusive, in 
nature. In order to understand the origins of the ballistic transport and
the transition to diffusive transport with increasing system size it is
imperative to establish the mechanism of intramolecular energy flow
in the system. Similarly, there is experimental evidence now for the
energy flow in proteins to be anomalous. In particular, the work\cite{leit_ivrproteins}
of Leitner and coworkers, based on recent formulations
of IVR (see below) as diffusion in quantum number space, are yielding
important insights into the origin of the anomalous energy flow behavior. 
Again, the details of the vibrational dynamics are expected to be
key in understanding the heat flow in proteins.

As another example consider the experiments of Zwier and coworkers\cite{zwier_natma} on
the conformational dynamics of a small ($36$ atom)
dipeptide called NATMA (N-acetly-tryptophan methylamide). In their combined 
IR and UV study they found, remarkably, that the population ratios of
the three nearly degenerate
conformers depended on which specific NH-stretching vibration of a conformer was
excited by the IR pulse. In other words, the molecule exhibits
mode-specific effects despite the very high density of states. Given that
the NATMA has $164$ minima connected by $714$ transition states, the
energy landscape even for this small biomolecule is fairly complicated. 
Therefore, the existence of distinguishable, conformer-specific, pathways
on the energy landscape is surprising and indicative of nonstatistical
behavior. Clearly, at the molecular level, IVR out of the two
different NH-stretching modes into the ``bath" of low frequency modes
must be sufficiently different. The experiment is also hinting at
the sensitivity of IVR to the geometry of the molecule. It is useful to
remark at this stage that the mode-specificity of NATMA leads to the
exciting possibility of controlling the conformational dynamics. However,
it is then important to understand the IVR dynamics of the initial
NH-stretching modes in the three conformers and hence the crucial
differences which result in mode and conformer specificity- a daunting task for
this molecule. 

In this context, the recent breakthrough experiments\cite{expt_adv3,pate_cpftmws} by
Brooks Pate's group allows for the possibility to track the molecular
geometry during an isomerization reaction. Their results on the
rate of {\em syn} $\leftrightarrow$ {\em anti} interconversion in
cyclopropane carboxaldehyde establishes the nonstatistical nature
of the reaction. Thanks to the development
of the chirped-pulse Fourier transform microwave spectrometer,
for the first time one hopes to be able
to monitor and unravel the complex interplay between IVR and molecular
geometry changes. 
Needless to say, our present state of theoretical understanding of
the mechanism of IVR is going to be severely tested in the years to come.

The above examples have been chosen intentionally
to illustrate the surprises and
the need for a mechanistic understanding of IVR in large molecules - a regime
once thought to be ideal for statistical assumptions\cite{rrkm_validity}. 
Small molecules, due to the low density of states, have
always been prime candidates for the observation of non-RRKM dynamics.
However, even for 
relatively small molecules recent studies are revealing the
complicated but incredibly rich IVR dynamics which precedes the
actual bond breaking/making event. Undoubtedly, a careful reexamination
of both TST and RRKM is required in order to understand the recent results.
We mention a few examples to emphasize the statements above. First
example pertains to the unimolecular decay of formaldehyde. 
Pioneering studies\cite{moore_hcho} in the early nineties by Brad Moore's group had
already hinted towards the existence of two different mechanisms for the
decay of formaldehyde into H$_{2}$ and CO moieties. However, a clear picture
of the possible mechanisms emerged 
very recently\cite{hcho_science} due to the work by the experimental group of Suits\cite{suits_hcho} in 
collaboration with the theoretical efforts by Bowman\cite{bowman_hcho} and coworkers. 
Interestingly, the resolution of the puzzle led to the formulation
of a so called ``roaming" mechanism which completely avoids the
conventional transition state for the reaction. Subsequent to this work
there are several other studies\cite{others_roaming} which strongly point towards
alternative, transition state skirting, pathways for different reactions.  
Two points are worth noting before proceeding further. The first point
is that the roaming mechanism has initmate connections to IVR. The
key issue has to do with the IVR in the partner
HCO radical which somehow makes it weakly coupled to the roaming H-atom.
Thus, they never separate. Instead, the hydrogen atom remains tethered
to the HCO moiety. The weak tether in this case is reminiscent of
earlier studies on the vibrational predissociation
dynamics\cite{vpvdw_ivr} of van der Waals molecules wherein
IVR dynamics plays a central role.
Second point is that the roaming mechanism was theoretically confirmed
by Bowman and coworkers using classical dynamics on the ab initio
potential energy surface. This latter point is essential since as the
next example shows, classical dynamics is the ``weapon of choice"
in uncovering the detailed dynamical pathways in such systems.

Another class of reactions which have been studied in great detail to highlight and understand the
links between IVR and reactivity are the gas phase S$_{N}$2  nucleophilic reactions. In particular, Bill Hase's group has focused\cite{hase_sn2}
on this class of reactions for more than a decade and have come up with results that challenge some
of the earlier traditional thinking. Again,from the perspective of this review, we note at the outset that
these findings emerge from classical dynamics calculations - one is still very far from an all quantum
dynamical investigation of such systems. The first example\cite{hase_sn2} involves the reaction
\begin{equation}
OH^{-} + CH_{3}F \rightarrow CH_{3}OH + F^{-} \nonumber
\end{equation}
where the potential energy surface has a deep ($\sim$ 30 kcal/mol below the asymptotic product energies)
global minimum in the product exit channel due to the
hydrogen bonded complex CH$_{3}$OH$\ldots$F$^{-}$. Assuming that the reaction would follow the intrinsic
reaction path, the system is expected to enter this deep well and get trapped for a significant amount of time.
If facile IVR occurs in the complex during the trapped time then RRKM theory is expected to yield reasonable
rates for the reaction. However, surprisingly, the dynamical calculations revealed that a great majority of the
trajectories simply avoided this deep minimum. In other words, most of the classical trajectories did not follow
the intrinsic reaction path and react via a direct route. Why do the trajectories avoid the deep minimum? Hase and 
coworkers propose that the inefficient IVR in the $[$HO$\ldots$CH$_{3}\ldots$F$]^{-}$ complex and a flat 
O-C$\ldots$F$^{-}$ bending potential might be responsible for the observations. We note that such flat potential
regions and ``faster than RRKM" cases might be more common than one thinks. For instance, Carpenter has recently
highlighted\cite{ivr_rev13} a large class of thermal reactions involving organic molecules wherein nonstatistical effects are expected to
play a key role in understanding the reaction mechanisms. The examples provided by Carpenter demand a fairly
detailed understanding of the dynamics, and in such large systems, ignoring the complexity of
obtaining reliable potential energy surfaces, classical dynamics is the only feasible approach. Finally, it is
interesting to note that even for the roaming mechanism mentioned earlier the potential energy
surface is rather flat in the roaming regions. A recent review\cite{bowshep} by Bowman and Shepler provides several other
examples of unusual reaction dynamics reported over the last few decades. 

It is becoming increasingly clear that
static features on the potential energy surfaces such as well depths and intrinsic reaction coordinates are
not going to be enough if one wants to understand the mechanisms. Dynamical effects are significant and in
some instances can completely alter mechanisms deduced from traditional approaches based on
`static' energetic criteria alone. An interesting example comes from a very recent work\cite{skodge_hfelim} by Skodje and
coworkers on the possible role of water as catalyst of photochemical reactions, in this instance HF-elimination
of fluromethanol in small water clusters. Traditional viewpoints argue in
favor of the catalytic role of water  based on
the significant lowering of transition state barrier due to hydrogen bonding. However, dynamical calculations
revealed the reaction to be direct, competitive with the evaporative process,
with very little IVR within the cluster species. Not surprisingly, the computed
RRKM rates are four orders of magnitude smaller than the dynamical rates. However, the critical issue of why
the IVR is highly restricted in the cluster species is left unanswered. The combination of the increasing appreciation
for the importance of dynamics and the interest in gaining insights into reaction mechanisms of moderately sized
systems points to a critical direction for further advances - a detailed understanding of molecular dynamics from the phase space
perspective. 

The above, admittedly select, examples on specific molecules clearly highlight the importance of dynamics to IVR - not only from the perspective of deviations from the RRKM theory but also from the more exciting viewpoint of being able to control IVR and hence reaction dynamics. Although studies on individual molecules have been instrumental in enhancing our understanding of the mechanistic aspects of IVR, it is also equally important to identify generic features of energy flow dynamics in a class of molecules. Note that the power and  utility of RRKM and TST comes from identifying generic features such as statisticality and recrossing-free dividing surfaces in various systems. 
 Is it at all possible to identify generic features of IVR in cases where the dynamics cannot be ignored? The answer, perhaps surprisingly, is yes and is inspired by an analogy between seemingly unrelated fields - Anderson localization\cite{andloc} (AL) and IVR. Nearly two decades ago, a key paper\cite{loganwoly} by Logan and Wolynes established the close parallels between the two topics
and paved the way for the so called state space diffusion model of IVR. The key result of the Logan-Wolynes
paper is a criterion for the transition between localized and extended eigenstates, akin to
the mobility edge in AL, with emphasis on the local nature of the energy flow. Thus, the
local density of states coupled to a specific initial zeroth-order state and their
connectivity play a central role in formulating the criterion. Subsequently, based on
the Logan-Wolynes work, Leitner, Schofield,
and Wolynes formulated\cite{lrmt} the local random matrix theory (LRMT) leading to the transition criterion as the ``quantum ergodicity threshold'' which signals the onset of
facile energy flow. Application of LRMT to several systems, especially low barrier
isomerization reactions\cite{gruebleit_conf}, has established the validity of the scaling approach and hence the
appropriateness of the analogy to AL. 

The first part of this review will be concerned with the state space scaling model of IVR. The IVR $\leftrightarrow$ AL analogy, however, involves several ideas and subtle approximations.
Some of the approximations are borrowed from AL theory whereas others are specific to the
IVR process. For the purpose of this review it is important to appreciate some of the key
features and approximations since one expects that a classical-quantum correspondence viewpoint
of the analogy might lead to deeper insights. Note that we will not elaborate upon the
technical details of the AL itself and instead refer the reader to the original literature in this regard.
Our purpose here is to highlight the salient features and, what we consider to be, the critical
approximations that deserve a closer inspection.  

In the second part of the review the focus will be on classical-quantum correspondence between the quantum state space scaling model and the classical phase space transport.  Based on recent advances, both conceptual and computational, in our understanding of phase space transport in systems with three or more DoF we will argue that  there is a very natural connection between the quantum state space and classical phase space perspectives on IVR. From one viewpoint, the analogy between AL and IVR on one hand and between AL and quantum chaos on the other hand closes the chain of analogies. The various examples drawn from our work and those of others will strengthen this viewpoint and show that there are several interesting analogies and challenges that still remain to be explored and understood.

\section{IVR: Analogy to Anderson localization}
\label{sec:ivrscaling}

Anderson localization\cite{andloc} is a phenomenon that arises in the study of electron transport in disordered a medium. In 1958\footnote{Note that it is hard to ignore the timing of this seminal work in comparison to the works of Marcus, FPUT, KAM, and Chirikov. It is also interesting to note that Anderson used a tight-binding model Hamiltonian in his study and such Hamiltonians are very similar to the effective spectroscopic Hamiltonians used in IVR studies.} Anderson showed that, in three dimensions, beyond a certain critical amount of disorder the electron is completely localized and hence the system becomes an insulator - the so called metal-insulator transition. In this article we will make no attempt to provide details of AL theory and instead refer readers to several excellent reviews of AL that already exist\cite{andloc_rev}. However, it is worth noting that the phenomenon of AL is rather general and has been observed in light waves\cite{andloc_light} and matter waves\cite{andloc_bec}.

We start by introducing the state space\cite{ivr_rev8} (also known as the quantum number space or QNS for short)
using a simplified vibration-only Hamiltonian and then briefly outline the IVR $\leftrightarrow$ AL 
mapping. Next we introduce the scaling
ansatz\cite{sw} due to Schofield and Wolynes and end by highlighting some of the key predictions and successes of the state space approach. Note that recently Leitner {\it et al.} review\cite{leittodarev} the progress made in classical phase space approach to non-statistical IVR and quantum state space corrections to RRKM dynamics. Here, the interest is in bringing out the close correspondence between the two. Note that in what follows, for the sake of uniformity and avoiding needless notational confusions, we follow the notation of Logan and Wolynes to a large extent.

\subsection{Introducing the IVR state space}
\label{sec:qns}

Consider a generic vibrational Hamiltonian of the form $H=H_{0}+V$ with
\begin{eqnarray}
H_{0} &=& \sum_{\alpha=1}^{f} \epsilon_{\alpha}(n_{\alpha}) \\
V &=& \frac{1}{3!}\sum_{\alpha \beta \gamma} \Phi_{\alpha \beta \gamma} (a_{\alpha}^{\dagger}+a_{\alpha})
(a_{\beta}^{\dagger}+a_{\beta}) (a_{\gamma}^{\dagger}+a_{\gamma}) + \ldots
\label{hres}
\end{eqnarray}
$H_{0}$ is  the zeroth-order part for a molecule with $f \equiv 3N-6$ vibrational degrees of freedom. Various perturbations
are contained in $V$ with strengths $\{\Phi_{\alpha \beta \gamma},\ldots\}$ arising from
the cubic and higher order mode couplings in the potential energy function. 
The mode energies $\epsilon_{\alpha}$ depend on the number of excitation quanta 
$n_{\alpha} = a_{\alpha}^{\dagger}a_{\alpha}$, expressed in terms of 
harmonic creation ($a_{\alpha}^{\dagger}$) and annihilation ($a_{\alpha}$) operators\cite{papaliev},
and include anharmonicities as well. Thus, one can associate $H_{0}$ 
with the usual Dunham expansion\cite{dunham,lefefield} 
\begin{equation}
H_{0} = \sum_{\alpha=1}^{f} \omega^{(0)}_{\alpha} n_{\alpha}
       +\sum_{\alpha,\beta=1}^{f} x_{\alpha \beta} n_{\alpha} n_{\beta} + \ldots
\end{equation}
with $\{\omega^{(0)}_{\alpha}\}$ and $\{x_{\alpha \beta}\}$ being the set of
harmonic frequencies and anharmonicities respectively. Note that the actual 
mode frequencies $\omega_{\alpha}$ depend on their occupancies and satisfy
\begin{eqnarray}
\hbar \omega_{\alpha} &\equiv& \frac{\partial \epsilon_{\alpha}}{\partial n_{\alpha}} \nonumber \\
                      &=& \omega^{(0)}_{\alpha} 
    + \sum_{\alpha,\beta=1}^{f} x_{\alpha \beta} (n_{\beta}+n_{\alpha} \delta_{\alpha \beta}) + \ldots
\end{eqnarray}
where the last relation follows from the Dunham expansion above. The oscillators are therefore nonlinear and one can define
\begin{equation}
\omega'_{\alpha} \equiv \hbar^{-1} \frac{\partial \omega_{\alpha}}{\partial n_{\alpha}}
\end{equation}
as a nonlinearity parameter (anharmonicity) associated with the $\alpha^{\rm th}$ mode.
For typical molecular systems, the anharmonicities are small when compared to
the frequencies and hence one assumes $\hbar |\omega'_{\alpha}| \ll \omega_{\alpha}$ for
all $\{n_{\alpha}; \alpha=1,2,\ldots,f\}$. 

Hamiltonians such as in Eq.~\ref{hres}, and its classical counterpart shown as Eq.~\ref{hrescl},  are examples of effective or spectroscopic Hamiltonians $H_{\rm eff}$ and are quite commonly used in the spectroscopic community\cite{lefefield}.  There are two routes for obtaining such Hamiltonians. One route is to obtain the effective $H$ via a fit to the extensive data coming from the experimental spectra. In this case there is no explicit connection between the parameters of the actual multidimensional ab initio potential surface and those of $H_{\rm eff}$. Nevertheless, one believes that the dynamics of $H_{\rm eff}$ is is qualitatively correct and accounts for the patterns observed in the spectral intensities and splittings. A second route to $H_{\rm eff}$ comes from performing canonical van-Vleck perturbation theory\cite{papaliev} (CVPT) on the ab initio potential surface. In this case the connection between the parameters of the two Hamiltonians is explicit. Moreover, it is possible to take CVPT to higher orders and estimate the errors involved in various observables. Details can be found in the reviews by Sibert\cite{sib_cvpt} and Joyeux and Sugny\cite{joy_cvpt}. There is a close connection between the quantum CVPT and the classical Lie transform\cite{liept} or equivalent perturbation theories. Essentially, both methods yield an effective Hamiltonian with the most relevant resonant perturbations at a given order. The close parallels between CVPT and the classical Birkhoff-Gustafson normal form\cite{bgnf} (BGNF) theory are clearly discussed by Fried and Ezra\cite{frezra_bgcv}. It is also worth mentioning that Ishikawa {\it et al.} have  compared\cite{hcp_cph} the ab initio $H$ and the corresponding $H_{\rm eff}$ for the $HCP \leftrightarrow CPH$ isomerization case in terms of both the dynamics and the eigenstates. The $H_{\rm eff}$ indeed yields correct qualitative insights even at fairly high levels of excitations. Note that Eq.~\ref{hrescl} is an example of a resonant normal form and, in general with more then one independent resonance present, indicates the nonintegrability of the Hamiltonian.

The use of a harmonic basis is not restrictive
for the following analysis
and one can write $H_{0}$ within, for example, a Morse oscillator basis. 
It is essentially a question of whether one wants to work with a normal mode or local mode
basis and at the typical high levels of excitation of interest the identification
of a mode as purely local or normal is not particularly useful. 
What is important here is that the eigenstates 
$|{\bm n}\rangle \equiv |n_{1},n_{2},\ldots,n_{f}\rangle$ of $H_{0}$ form a convenient,
and often experimentally relevant, basis to understand the concept of a molecular
state space. 
The molecular state space, also called as the quantum number space (QNS),
is nothing but the $f$-dimensional space with axis labeled by
the zeroth-order quantum numbers $(n_{1},n_{2},\ldots,n_{f})$. A point in 
the QNS represents a specific zeroth-order state which we imagine to be prepared by
the experimentalist and often referred to as the zeroth-order bright state 
(ZOBS)\footnote{We remark here that given the anharmonic zeroth-order Hamiltonian
one should more appropriately refer to them as ``feature" states\cite{ivr_rev8}. 
In fact, the bright spectral features observed in experiments directly
relate to the feature states and perhaps ZOFS
is a better acronym. However, we use the ZOBS terminology throughout.}.
The term ``bright", as opposed to ``dark", arises due to the accessibility of the state via the usual
electric-dipole transitions.
In order to establish the IVR $\leftrightarrow$ AL mapping it is useful to think of a specific
ZOBS as a site $|i \rangle \equiv |n_{1}^{(i)},n_{2}^{(i)},\ldots,n_{f}^{(i)} \rangle$
in the $f$-dimensional QNS with site 
energy $\epsilon_{i} \equiv \sum_{\alpha} \epsilon_{\alpha}(n_{\alpha}^{(i)})$.
Note that the number of possible sites depends on the total energy of interest.

\begin{figure}[t]
\begin{center}
\includegraphics[height=80mm,width=100mm]{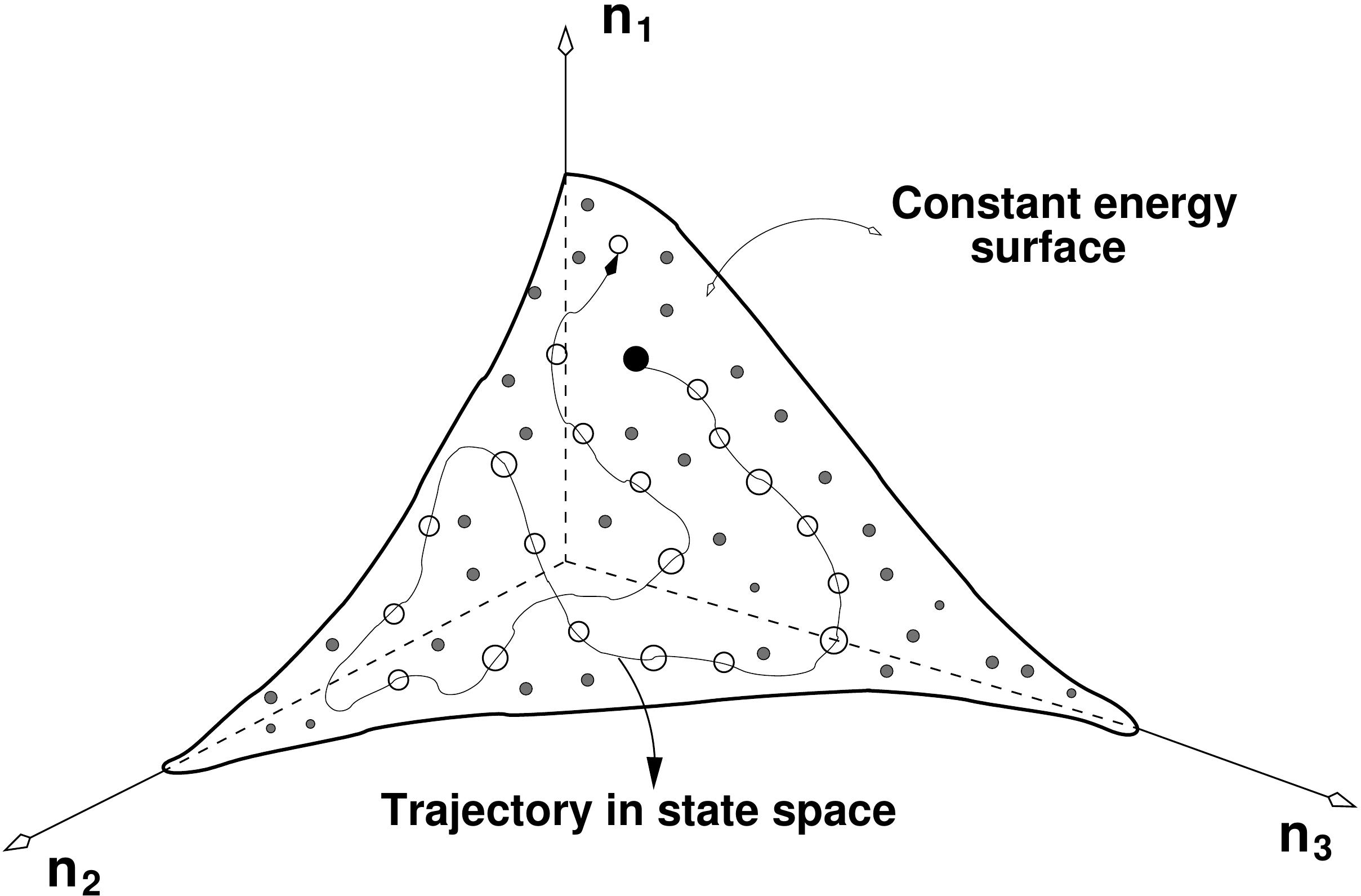}
\caption{Simple sketch of the discrete molecular state space or quantum number space
for a three mode $f=3$ case. An initial zeroth-order state (dark circle) diffuses
on the $f-1=2$ dimensional constant energy surface. The diffusion involves
some of the dark states (thick circles) whereas several other dark states (shaded
circles) are not involved. Reproduced from reference\cite{manithesis} with permission.}
\label{intro_qns}
\end{center}
\end{figure}

Clearly, under the influence of $H_{0}$ a point in the QNS
is stationary. However, 
under the influence of the perturbations $V$ in Eq.~\ref{hres}, a ZOBS $|i \rangle$
is no longer stationary and evolves in the QNS. It is convenient to think\cite{sw} of this motion
of $|i \rangle$ as a diffusion over the constant energy surface
$H \approx \epsilon_{i}$ and such a motion is 
nothing but the process of IVR (see Fig.~\ref{intro_qns}). This can
be understood very simply, since in going from $|i \rangle \rightarrow |j \rangle$
there is an exchange of quanta due to change in the mode occupancies $\{n_{\alpha}^{(i)}\}$
and $\{n_{\alpha}^{(j)}\}$, characterized by a ``distance'' 
$Q_{ij} \equiv \sum_{\alpha} |n_{\alpha}^{(i)}-n_{\alpha}^{(j)}|$ in the QNS.
Exchange of quanta occurs due to the finite coupling elements $V_{ij} \equiv \langle j|V|i \rangle$
and such terms in the AL context are referred to as the {\em hopping terms}. Thus, it is possible to recast\cite{loganwoly} the molecular Hamiltonian formally as a
tight-binding model
\begin{equation}
H = \sum_{i} \epsilon_{i} |i\rangle \langle i| + \sum_{i \neq j} V_{ij} |i \rangle \langle j|
\label{tightham}
\end{equation}
and hence bears close resemblance to the AL Hamiltonian.  It is useful to
briefly recall the interpretation of the various terms in the above Hamiltonian in case of AL.
The various sites in case of AL form a lattice in real space and are occupied by atoms. Every site 
is associated with a unperturbed electronic energy level $\epsilon_{j}$ corresponding
to an atomic orbital $|j \rangle$ centered on the site. The matrix elements $V_{ij}$
allow the electron to hop from site $|i \rangle$ to site $|j \rangle$. The analog of IVR
in this case has to do with the transport of the electrons through the medium.

At this stage it is appropriate to introduce a key parameter $K$ which is called as the
{\em connectivity} of the $f$-dimensional QNS. Specifically, $K+1$ is identified with
the number of distinct sites in the QNS to which a given site $|i \rangle$ of interest is connected
by non vanishing coupling elements. Later it will be seen that the parameter $K$ directly
enters into the criterion for quantum ergodicity threshold. For a given form of
the coupling terms, the determination of $K$ is a combinatorial problem. For instance,
taking into account only the cubic terms $\Phi_{\alpha \beta \gamma}$ and rewriting
the coupling as
\begin{eqnarray}
V &=& \frac{1}{2} \sum_{\alpha \neq \beta} \Phi_{\beta \beta \alpha} (2n_{\beta}+1)(a_{\alpha}^{\dagger}
                          + a_{\alpha}) \nonumber \\
&+& \frac{1}{2} \sum_{\alpha \neq \beta} \Phi_{\beta \beta \alpha}
                      (a_{\beta}a_{\beta}+a_{\beta}^{\dagger} a_{\beta}^{\dagger})
                      (a_{\alpha}^{\dagger}+ a_{\alpha}) \nonumber \\
&+& \sum_{\alpha<\beta<\gamma} \Phi_{\alpha \beta \gamma} 
                              (a_{\alpha}^{\dagger}+ a_{\alpha})(a_{\beta}^{\dagger}+ a_{\beta})
                               (a_{\gamma}^{\dagger}+ a_{\gamma})
\end{eqnarray}
one can show\cite{loganwoly} that
\begin{equation}
K + 1 \approx \frac{2f}{3}\left(2f^{2}+1\right)
\end{equation}

The flow of energy, starting with a given ZOBS $|j \rangle$, among the various vibrational modes
is determined from the evolution ($\hbar=1$)
\begin{eqnarray}
|\Psi(t) \rangle &=& e^{-iHt} |j \rangle = \sum_{l} \langle l|e^{-iHt}|j \rangle |l \rangle \\
                &\equiv& i\sum_{l} {\cal G}_{lj}(t) |l \rangle
\end{eqnarray}
with ${\cal G}_{ij}(t)$ being the state space time-dependent site Green function. The survival
probability {\it i.e.,} the probability that the system will be found at the starting site $|j \rangle$
at a later time is $P_{jj}(t) = |{\cal G}_{jj}(t)|^{2}$ and below we will focus on this quantity
in the context of a scaling theory. For now, note that the evolution of an intial ZOBS $|j \rangle$
for a given state space connectivity $K$ can be thought of as a kind of motion in the QNS. It is crucial to
observe that such a motion in the QNS cannot be naively associated with a unique trajectory,
as in classical dynamics, since the site
$|j \rangle$ is coupled to several other sites at any given time. However, for the sake
of visualization, if one only takes into account the site with maximum probability at a given time
then an approximate picture as shown in Fig.~\ref{intro_qns} emerges. Clearly, such an approximation
ignores the quantum interference effects and hence supposed to be taken with some 
caution\footnote{This might be being overly pessimistic. In reality, as several model
studies have shown, the state space dynamics is locally correlated and mediated by one or
more anharmonic resonances. Classically, even in the regions wherein two or more resonances overlap
leading to stochasticity, it is premature to think of the chaotic regions to be
structureless - as seen later in the review, stickinessm and partially broken tori can lead to long time correlated dynamics.
 Does one have enough evidence that such sort of effects are absent in systems with
high enough DoF? Logan-Wolynes are careful enough to claim that the IVR $\leftrightarrow$ AL
analogy is to be taken seriously only for systems with large DoF.}.
Nevertheless, it is still interesting to ask questions about the nature of this dynamics in
the QNS. For example, is it possible to interpret the motion as a diffusion in the QNS?
If so, then what is the nature of this diffusion? Is it typically Brownian or are there
reasons to expect anomalous diffusion? Does the dynamics of a ZOBS in the state space
have any relation to the nature of the transport in the underlying classical multidimensional
phase space? These dynamical questions will be addressed later
in the context of the scaling approach and discussions on the possible classical-quantum
correspondence aspects of the issue. 
One still has to recognize the fact that the motion
in the state space is inherently quantum and hence it is necessary to identify features
of the classical phase space transport which are robust under the quantization.

However, before stepping into the nature of the state space dynamics we outline
the arguments of the Logan-Wolynes paper which requires a time-independent viewpoint.

\subsection{Quantum ergodicity threshold}
\label{sec:qet}

In order to study the localized-extended eigenstates transition for the Hamiltonian in
Eq.~\ref{tightham} the time-independent (energy) Green function
\begin{eqnarray}
G_{lj}(E) &=& \lim_{\eta \rightarrow 0^{+}} \int_{0}^{\infty} e^{i(E+\eta)t} {\cal G}_{ij}(t) \\
          &=& \langle l|(E+i\eta-H)^{-1}|j \rangle
\end{eqnarray}
is the central object.
Since the survival probability $P_{jj}(t)$ is of particular interest, consider the
site-diagonal Green function $G_{jj}(E)$ which can be expressed as
\begin{eqnarray}
G_{jj}(E) &=& \left(E+i\eta-\epsilon_{j}-S_{j}(E) \right)^{-1} \\
          &\equiv& \left(g_{jj}^{-1}-S_{j}(E) \right)^{-1}
\end{eqnarray}
where we have defined the self-energy term $S_{j}(E)={\cal E}_{j}(E)-i\Delta_{j}(E)$
and the unperturbed site-diagonal Green function $g_{jj}(E)=(E+i\eta-\epsilon_{j})^{-1}$
in the above equation\footnote{One way of thinking about the self-energy is that it
{\em renormalizes} the unperturbed Hamiltonian $H_{0}$ in eq.~\ref{tightham} to
$\sum_{i} (\epsilon_{i}+S_{i}) |i \rangle \langle i|$, which is non self-adjoint
due to the fact that $S_{i}(E)$ is complex. As usual, the real part of $S_{i}(E)$
leads to a shift of the energy level whereas the imaginary part leads to
the broadening of the, otherwise, sharp lines.}. 
The motivation to write the site-diagonal Green function in the
above form has to do with the fact that the imaginary part of the self-energy $\Delta_{j}(E)$
is the crucial quantity which characterizes the transition. To appreciate the importance of
$\Delta_{j}(E)$, consider an exact eigenstate $|\Psi_{\lambda} \rangle$, with exact eigenvalue
$E_{\lambda}$, of the full Hamiltonian. The decomposition
\begin{equation}
|\Psi_{\lambda} \rangle = \sum_{l} \langle l|\Psi_{\lambda} \rangle |l \rangle
                       \equiv \sum_{l} C_{l \lambda} |l \rangle
\end{equation}
shows that the amplitudes $C_{l \lambda}$ determine the extent of (de)localization of
$|\Psi_{\lambda} \rangle$ in the QNS. Using the spectral representation of the
site-diagonal Green function
\begin{equation}
G_{jj}(E) = \sum_{\lambda} \frac{|C_{j \lambda}|^{2}}{E-E_{\lambda}+i\eta}
\end{equation}
and assuming a point spectrum for the full Hamiltonian it follows that
\begin{equation}
|C_{j \lambda}|^{2} = \left(1+\frac{\Delta_{j}(E_{\lambda})}{\eta}\right)^{-1}
\end{equation}
Clearly, the state space nature of the exact eigenstates are directly related to the imaginary part of
the site self-energy. Moreover, the long time limit of the time-averaged survival
probability $\bar{P}_{jj}(\infty) = \sum_{\lambda} |C_{j \lambda}|^{4}$, also
known as the dilution factor of the ZOBS $|j \rangle$, is determined by $\Delta_{j}(E_{\lambda})$
as well. 

We can now state the distinction between extended and localized eigenstates of Eq.~\ref{tightham}
in terms of $\Delta_{j}(E_{\lambda})$. For an eigenstate $|\lambda \rangle$  
\begin{equation}
\Delta_{j}(E_{\lambda}):
\begin{cases}
 \propto \eta \rightarrow 0^{+} & \text{localized}, C_{j \lambda} \neq 0\\
 \text{finite} & \text{extended}, C_{j \lambda} \rightarrow 0
\end{cases}
\label{locexttrans}
\end{equation}
The above conclusions are strictly true only in the ``thermodynamic limit" {\it i.e.,} as
the number of sites tends to infinity. Indeed the above condition arises by investigating the
convergence properties of the so called {\em renormalized Feenberg perturbation series}
(see below) which is an explicit series for the site self-energy\cite{feenberg}.
However, in practice, the state space is finite
and hence there is a finite mean energy level spacing which scales inversely with
the total density of states $\rho(E)$. Therefore, the finiteness of the QNS 
leads to a timescale $\tau_{H} \sim 2\pi\hbar \rho(E)$ ({\em Heisenberg time})
beyond which the survival probability $P_{jj}(t)$ can exhibit recurrences. The existence of a finite
$\tau_{H}$ interferes with the clean distinction between localized and extended eigenstates
as given above. Nevertheless, for sufficiently large densities this timescale is much larger
then the experimental timescales of interest ($\sim$ a picosecond) and one can safely 
assume the thermodynamic limit for large molecules. This is one of the reasons that
Logan and Wolynes expect the IVR $\leftrightarrow$ AL analogy to be useful for
describing the IVR process in high dimensional quantum systems. 

\subsubsection{Ensemble of Hamiltonians: probabilistic approach to the transition}
\label{sec:qet_prob}

From the discussions above it is clear that in order to make progress one needs to 
compute the various site self-energies $\{S_{j}(E)\}$ and invoke Eq.~\ref{locexttrans}
to determine the nature of the eigenstates. Suppose for a given ZOBS $|j \rangle$
one finds that every eigenstate is extended then one anticipates the IVR dynamics
out of $|j \rangle$ to ``ergodically" sample the QNS. The problem with such an approach
is that one is faced with a fairly challenging numerical (as well as analytical) problem. Indeed, using the 
Feenberg renormalized perturbation approach\cite{feenberg} it is possible to show that
\begin{eqnarray}
S_{j}(z) &=& \sum_{k \neq j} V_{jk} \frac{1}{z-\epsilon_{j}-S_{k}^{(j)}(z)} V_{kj} \nonumber \\
  &+& \sum_{k \neq j} \sum_{l \neq j,k} V_{jl} \frac{1}{z-\epsilon_{l}-S_{l}^{(jk)}(z)} V_{lk}
   \frac{1}{z-\epsilon_{k}-S_{k}^{(j)}(z)} V_{kj} + \ldots
\end{eqnarray} 
with $z \equiv E+i\eta$ and the notation $S_{k}^{(j)}$ stands for the self-energy of site $|k \rangle$
with the site $|j \rangle$ removed {\it i.e.,}
\begin{eqnarray}
S_{k}^{(j)}(z) &=& \sum_{l \neq j,k} V_{kl} \frac{1}{z-\epsilon_{l}-S_{l}^{(jk)}(z)} V_{lk} \nonumber \\
&+& \sum_{l \neq j,k} \sum_{m \neq j,k,l} V_{jm} \ldots
\end{eqnarray}
An iteration procedure leads to an infinite continued fraction
expression\footnote{Notice that a geometric representation of the perturbation series
involves {\em only non intersecting} paths in the QNS.} 
for $S_{j}(z)$ whose analyticity is intimately tied to the criteria for
the existence of localized or extended states. Furthermore, apart from the connectivity,
it is also necessary to
know the ``topology" (coupling network) of the QNS as determined by the various coupling matrix elements.

Anderson, in his seminal paper\cite{andloc}, made a detailed analysis of the perturbation series for $\{S_{j}(E)\}$
and found a way around the daunting problem by providing a 
probabilistic criterion for
the localized-extended states transition. Several aspects of the original Anderson paper
have been critically reexamined by a number of people including Ziman\cite{andloc_ziman}, Thouless\cite{andloc_thouless} and Economou and Cohen\cite{andloc_econcohen}.
Currently there is no doubt that the central point of Anderson's work is correct, as shown
by several numerical and experimental studies\cite{andloc_rev}. A detailed account of the progress and technical
aspects of AL are beyond the scope of the current review. However, in the context of the
IVR $\leftrightarrow$ AL analogy, a brief mention of the approach taken by Logan-Wolynes is appropriate.
Specifically, Logan-Wolynes adopt\cite{loganwoly} an ingenious approach to AL which was suggested nearly four decades ago by 
Abou-Chacra, Anderson and Thouless\cite{andloc_abouchac}. The key ideas behind the self-consistent theory of localization are as follows:

\begin{itemize}
\item The first step is to consider only the first term of the renormalized perturbation series
\begin{eqnarray}
S_{j}(z) &=& \sum_{k \neq j} V_{jk} \frac{1}{z-\epsilon_{j}-S_{k}^{(j)}(z)} V_{kj} \label{aat11}\\
S_{k}^{(j)}(z) &=& \sum_{l \neq j,k} V_{kl} \frac{1}{z-\epsilon_{l}-S_{l}^{(jk)}(z)} V_{lk}
\label{aat1}
\end{eqnarray}
This is in contrast to the original approach wherein the importance of successive order terms
had to be ascertained.
\item In the next step, one ignores the fact that the self-energy on the right hand side
of the Eqs.~\ref{aat11} is different from the self-energy on the left hand side and writes
\begin{equation}
S_{j}(z) = \sum_{k} V_{jk} \frac{1}{z-\epsilon_{j}-S_{k}(z)} V_{kj} 
\label{aat2}
\end{equation}
The essential idea here is to demand that the probability distributions of the self-energies
on the two sides of the above equation be {\em self-consistent}. 
\item The crucial observation made by Abou-Chacra {\it et al.} is that the above approximations become exact
if the topology of the lattice is
assumed to be a {\em infinite regular Cayley tree} (Bethe lattice) with connectivity $K$
and a nonzero coupling $V_{jk}$ only if $|j \rangle$ and $|k \rangle$ are neighboring sites.
Consequently, Eq.~\ref{aat2} is exact and
the Cayley tree approximation to the true topology is expected to be good 
in the limit of large DoFs. Hence, the statement by Logan-Wolynes that the 
IVR $\leftrightarrow$ AL analogy is useful only for large molecules.
\end{itemize}

\begin{figure*}
\begin{center}
\includegraphics[height=60mm,width=120mm]{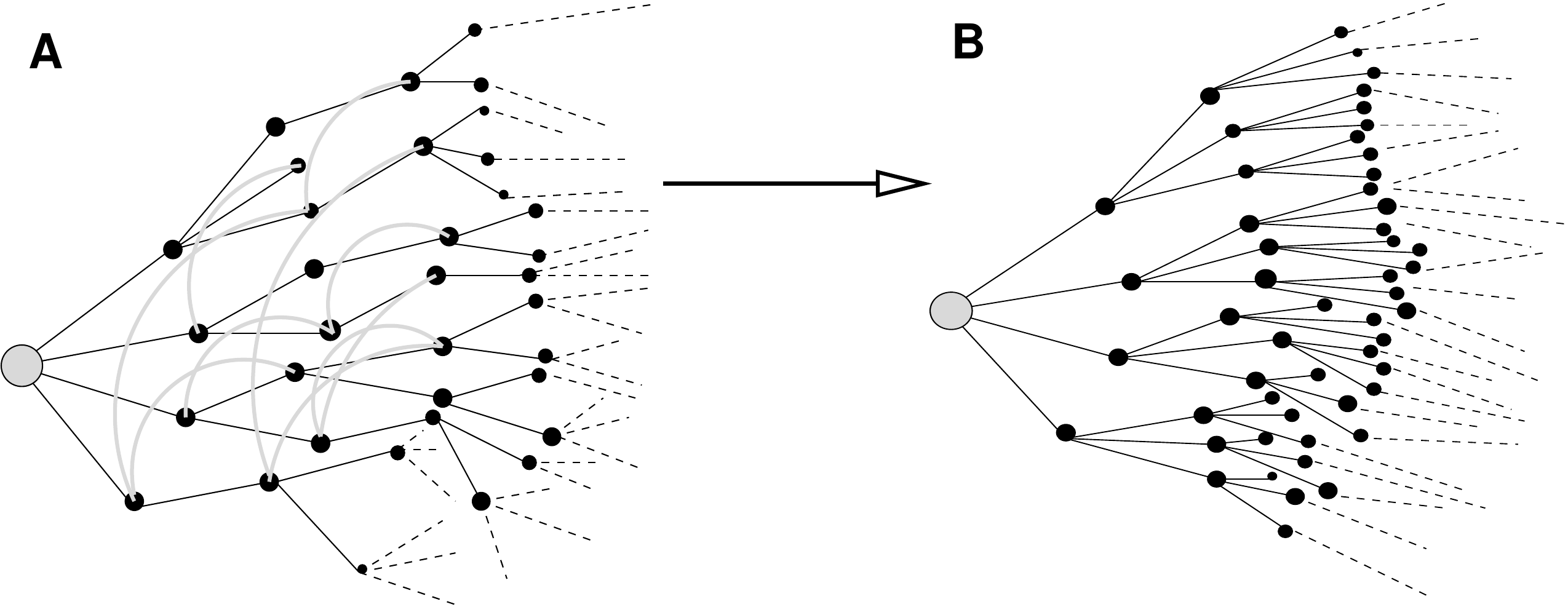}
\caption{(A) An example of a ``real" state space connectivity compared to its
Cayley tree approximation shown in (B). The large grey circle represents an
interior ZOBS in the state space. Note that (A) corresponds approximately to
the dominant direct resonant coupling structure for the effective Hamiltonian
of the SCCl$_{2}$ (thiophosgene) molecule. The thick grey lines connecting
some of the sites in (A) are not accounted for by the Cayley tree approximation.}
\label{qnsapproxcay}
\end{center}
\end{figure*}

Following the approach\cite{andloc_abouchac} of Abou-Chacra {\it et al.}, Logan-Wolynes consider an ensemble of
Hamiltonians and ask, with unit probability, whether an eigenstate of energy $E$ is
localized in the vicinity of site $|j \rangle$ ({\it i.e.,} $\Delta_{j}(E) \propto \eta 
\rightarrow 0^{+}$) or whether the eigenstate is extended ($\Delta_{j}(E)$ finite). 
However, there are essential differences between the IVR problem and the AL problem
which need to be taken into account. Firstly, in AL the $\{V_{ij}\}$ are taken to be
constant and finite only between nearest neighbor sites. In IVR, the $\{V_{ij}\}$ are
not all identical and depend on the location of site $|j \rangle$ in the QNS.
Secondly, in AL the site energies $\{\epsilon_{j}\}$ are treated as independent random
variables with identical distribution $P(\epsilon_{j})$ for every site. In contrast,
for the IVR problem the state space is not statistically homogeneous and the
site energies depends on the location in the QNS. Thirdly, a true thermodynamic
limit does exist for AL whereas for IVR the state space is finite and hence
the issue of {\em self-averaging} is very relevant. In other words,
are the ensemble averaged IVR characteristics typical for a 
specific realization of the Hamiltonian ensemble? Addressing the issue of self-averaging is
key for ensuring the applicability of the statistical model to a specific molecule
which, for instance, has an effective Hamiltonian fit coming from 
high resolution spectroscopy.

In Fig.~\ref{qnsapproxcay} we illustrate the key difference between the
Cayley tree approximation of the state space network topology and an actual
coupling network of the thiophosgene (SCCl$_{2}$) molecule. Note, that
in this case an interior-most state has been chosen within the
constraints of the effective Hamiltonian. It should, however,
be remembered that the comparison in Fig.~\ref{qnsapproxcay} is not quite fair
since the effective spectroscopic Hamiltonian\cite{heff_thio} for SCCl$_{2}$ is essentially a three
degrees of freedom ($f=3$) example.
Nonetheless, even in such an extreme case one observes a Cayley tree like
connectivity with $K=3$ (cf. Fig.~\ref{qnsapproxcay}A). For comparison
Fig.~\ref{qnsapproxcay}B shows\footnote{Remember, the connectivity of a Cayley tree
of order $K$ is given by $K+1$.} the Cayley tree structure for $K=3$.
A crucial difference is also seen and indicated by thick gray lines
in the actual coupling network. These ``rewirings" correspond to dominant and direct interactions
between sites on different levels of the tree. Clearly, such additional interactions lead
to ``loops" in the state space and would thus spoil the exactness
of Eq.~\ref{aat2}, leading to the violation of the assumption of a strictly
local energy landscape. Logan-Wolynes feel that this would make only quantitative
errors - something that has not been investigated/studied in detail\footnote{\em We
suspect that neglecting the loops is tantamount to certain dynamical approximations.
Would this spoil the nice scaling picture described in the next section? What one requires
is an estimate of how such "rewired" connections grow with the connectivity $K$\cite{refcor2}.}.

Logan-Wolynes argue that the above differences can be minimized by focusing attention
on ZOBS that lie in the interior of the QNS. Such states ({\em interior states})
have similar excitation levels (say of order $M$)
in every mode (modulo symmetry and selection constraints) and thus 
can be represented as $|j \rangle = |n_{1}^{(j)},\ldots,n_{f}^{(j)} \rangle = |M,M,\ldots,M \rangle$.
The class of states that are excluded are the so called {\em edge states} for whom
$|M,0,0,\ldots,0 \rangle$ is an extreme example. In the next section it will be seen that
the IVR dynamics from these two classes of states might also be expected to be sufficiently different.
Therefore, considering only the interior states, the following assumptions are made\cite{loganwoly}:
\begin{enumerate}

\item Although the site energy $\epsilon_{j}$ depends on the location in the QNS, the site energies
$\epsilon_{l}$ of the states $\{|l \rangle \}$ that are directly connected to $|j \rangle$ are
distributed symmetrically about $\epsilon_{j}$. In other words, writing
\begin{equation}
\epsilon_{l} = \epsilon_{j} + \zeta_{l}
\end{equation}
one assumes $\{\zeta_{l} \}$ to be symmetrically distributed with a width of the order of
$\hbar \bar{\Omega}(M)$. Here $\bar{\Omega}(M)$ is the mean value of the mode frequencies
with every mode occupancy being of the order $M$.

\item Instead of solving for the coupled stochastic equations for the site self-energies,
assume them to be independent random variables. The assumption is valid provided
\begin{equation}
\frac{\hbar \bar{\Omega}(M)}{K} < \hbar^{2} |\bar{\Omega}'(M)| \ll \hbar \bar{\Omega}(M)
\end{equation}
where $\bar{\Omega}'(M)$ is a measure of the mean anharmonicity of an interior state.
Note that the above condition is valid only for a large enough $K$ and hence, again, in
the limit of large molecules. Of course, in the limit of large enough vibrational DoF $f$
one expects the violations of self-averaging to be minimal.

\item Given the sequence of tiers
\begin{equation}
|j \rangle \rightarrow \{|l \rangle \} \rightarrow \{|p \rangle \} \rightarrow \ldots
\end{equation}
assume the distributions $P(\zeta_{l}) \approx P(\zeta_{p})$. Similarly, ignoring the
small changes in the coupling matrix elements as one progresses along the tree,
assume their distributions satisfy the condition $g(V_{jl}) \approx g(V_{lp})$. 

\end{enumerate}

Based on the above approximations Logan-Wolynes conclude that the distribution
for ${\cal E}_{l}$ and $\Delta_{l}$ {\it i.e.,} the real and imaginary parts
of the self-energy of site $|l \rangle$, denoted by $F({\cal E}_{l},\Delta_{l};\epsilon_{l})$,
is {\em universal}. Thus, one is interested in the conditional probability distribution
\begin{equation}
f(\Delta_{l};\epsilon_{l}) \equiv 
                 \int_{-\infty}^{\infty} F({\cal E}_{l},\Delta_{l};\epsilon_{l}) d{\cal E}_{l} 
\end{equation}
of $\Delta_{l}$ for given $E$ and $\epsilon_{l}$. In particular, the quantity of interest here
is the {\em most probable value} $\Delta_{\rm mp}(E;\epsilon_{l})$ of $\Delta_{l}$. 
Furthermore, a look at the expression (Eq.~\ref{aat2})
\begin{equation}
S_{j}(z) = \sum_{l}' \frac{|V_{jl}|^{2}}{z-\epsilon_{j}-\zeta_{l}-S_{l}(z)}
\end{equation}
suggests that the dominant perturbative couplings arise from sites $|l \rangle$
with $\zeta_{l} \approx 0$ and $E \approx \epsilon_{j}$. 
Consequently, in the equation for $S_{j}(z)$ above, one can approximate
$\Delta_{l}(E;\epsilon_{l}) \approx \Delta_{\rm mp}(E;\epsilon_{j})$, which
provides one approach to determine the conditional probability distribution.

Analyzing the asymptotic properties of the Fourier transform
\begin{equation}
\hat{f}(k;\epsilon_{j}) = \int_{-\infty}^{\infty} d\Delta_{j} e^{ik\Delta_{j}} f(\Delta_{j};\epsilon_{j})
\end{equation}
in the limit of small (localized) and large (delocalized)
values for the parameter $\mu \equiv \eta + \Delta_{\rm mp}(E;\epsilon_{j})$
one obtains the criteria for transition between localized and extended states as
\begin{equation}
\sqrt{\frac{2 \pi}{3}} K \langle |V| \rangle D_{L}(E;\epsilon_{j}) = 1
\label{locdeloccrit}
\end{equation}
In the above equation the quantity
\begin{equation}
\langle |V| \rangle = \int_{-\infty}^{\infty} dV |V| g(V)
\end{equation}
is the average of the magnitude of the coupling strengths charcterized by the distribution $g(V)$.
The term $D_{L}(E;\epsilon_{j})$ represents the average local density of states defined by
\begin{equation}
D_{L}(E;\epsilon_{j})  = \int d\zeta f_{o}(E-\epsilon_{j}-\zeta;\epsilon_{j}+\zeta)P(\zeta)
\end{equation}
with
\begin{equation}
f_{o}({\cal E}_{k};\epsilon_{k}) = \int d\Delta_{k} F({\cal E}_{k},\Delta_{k};\epsilon_{k})
\end{equation}
being the conditional distribution for the real part of the self-energy for a given site energy.

The key observation in the mean field approach, outlined above, is that it is the local quantities that are  
important for the transition criterion rather than the averaged total density of states. Further technical details on the asymptotic properties of the above quantities and derivations of related observables can be found in the original paper\cite{loganwoly}. 

It is interesting that Logan and Wolynes end their paper by stating that ``An understanding of the connection of our model with classical stochasticity criteria and Arnold diffusion will require a more complete understanding of the role of the correlations of state space site energies, which is connected with the (accumulation of) quantum states in classically resonant regions." As will be seen in Sec.~\ref{sec:cqcorrsec}, understanding the above remark requires a detailed classical-quantum correspondence insight into the AL $\leftrightarrow$ QNS analogy.

\section{Scaling theory of IVR}
\label{sec:scaling}

Subsequent to the Logan-Wolynes paper, 
Schofield and Wolynes\cite{sw}, in analogy to the scaling theory of AL\cite{andloc_scaling} given by Abrahams {\it et al.}, 
provided a single parameter scaling perspective for IVR. The argument is based on two important timescales. The first timescale is denoted by  $\tau_{IVR}(Q)$ (analogous to the Thouless timescale in AL) which corresponds to the timescale over which a ZOBS explores a distance $Q$ in the state space. The second timescale is the Heisenberg timescale $\tau_{H} \equiv 2 \pi \hbar \rho(E,Q)$ with $\rho(E,Q)$ being the local density of states at energy $E$ over a distance $Q$ in the state space. Associated with these timescales are the Thouless energy $E_{\rm th}(Q) \equiv 2 \pi \hbar/\tau_{IVR}$ and the mean level spacing $\langle \Delta E \rangle \equiv 2 \pi \hbar/\tau_{H}$. As in AL, the ratio of these energies yields the dimensionless quantity
\begin{equation}
g(Q) \equiv E_{\rm th}(Q) \rho(E,Q)
\label{dimlessconduct}
\end{equation}
known as the Thouless number or dimensionless conductance in context of AL. 

A useful interpretation of $E_{\rm th}(Q)$, originally due to Thouless\cite{andloc_thouless,thprl77}, is that it
roughly corresponds to the matrix element coupling energy levels in two adjacent
regions of average extent $Q$ in the state space.  From the
eigenstates perspective, if the coupling far exceeds the mean level spacings then
one anticipates significant mixing between the two regions leading to
eigenstates delocalized over a region of extent $2Q$ in the QNS. Existence of delocalized states means
facile IVR in the QNS and Eq.~\ref{dimlessconduct} 
can now be used to identify the transition from localized to delocalized states. Thus, For $g(Q) > 1$ there is
facile IVR and the initial ZOBS explores most of the state space. On the
other hand $g(Q) < 1$ signals localization and restricted IVR. The critical
value $g(Q) \equiv g_{c} = 1$ denotes the border between facile and
restricted IVR. From the eigenstates perspective, $ g > 1$ and $g < 1$ imply delocalized and localized eigenstates respectively. At the critical value $g = g_{c}$ one expects multifractal eigenstates\cite{multifraceigen}.

A critical step at this stage is to obtain the scaling of the conductance $g(Q)$ with $Q$ at a given energy $E$. Such a scaling leads to insights on how IVR explores the state space at an energy of interest. Following the AL analogy one introduces\cite{sw} the single parameter
\begin{equation}
\beta(g) = \frac{d \ln g(Q)}{d \ln Q}
\label{gscal}
\end{equation}
to determine the scaling of $g(Q)$ with $Q$ {\it i.e.,} $g(Q) \sim Q^{\beta}$.  Note that there are assumptions of statistical homogeneity and isotropic state space diffusion involved in using a one parameter scaling. While the latter assumption  can still be accommodated within the scaling approach, the former assumption requires careful study. The local density can be written as
\begin{equation}
\rho(E,Q) \sim \rho_{0}(E) Q^{d_{f}}
\label{locden}
\end{equation}
in terms of the effective dimension $d_{f}$ of the QNS and the 
density of states per unit volume $\rho_{0}(E)$.
Moreover, in the delocalized (facile IVR) limit, the average displacement in the QNS scales as 
\begin{equation}
Q(t) \sim (Dt)^{1/\alpha}
\label{rwdist}
\end{equation}
with $D$ being a constant (diffusion) and $\alpha$ being the degree of
the random walk. The random walk is called uncorrelated {\it i.e.,} ordinary Brownian motion
for $\alpha=2$ while a correlated random walk is implied by $\alpha > 2$. For
 $0 < \alpha \leq 2$ the random walk corresponds to the general class of 
L\'{e}vy diffusion\cite{zaslavbook}.
Combining Eq.~\ref{locden} with Eqs.~\ref{gscal} and \ref{rwdist} yields 
\begin{equation}
\beta = d_{f} - \alpha
\end{equation}
relating the scaling of the Thouless number with the effective dimensionality of the QNS and the nature of the
random walk. 

The above result on the scaling of the conductance can be used to infer the scaling of the 
survival probability $P_{j}(t) \equiv |\langle j|j(t) \rangle|^{2}$
of the ZOBS. More appropriately, given the assumption of statistical homogeneity, the appropriate object of interest here is the survival probability averaged over all isoenergetic states - a microcanonical average. 
Although the so obtained average $\langle P(t) \rangle$ is a quantum object, semiclassically one expects it to
be inversely proportional to the volume $V(t)$ explored by the random walk. Furthermore,
assuming the volume to be of dimensionality $d_{f}$ (possibly fractal) and the
random walk to be isotropic leads to the scaling
\begin{equation}
\langle P(t) \rangle \sim \frac{1}{V(t)} \sim \left(\frac{1}{Q(t)}\right)^{d_{f}}
\end{equation}
and therefore implies
\begin{equation}
\langle P(t) \rangle \sim \left(\frac{1}{Dt}\right)^{d_{f}/\alpha}
\end{equation}
for the average behavior of the survival probability.

The complete range of behavior for $\langle P(t) \rangle$ can be obtained by  
noting that at the threshold (critical) $g_{c}=1 \implies \beta=0$ and independent of $Q$
whereas below the threshold $g < 1 \implies \beta < 0$.
Assuming exponential timescales for the latter case one has $\beta \sim \ln g$.
Thus, one has the following predictions for the average survival probability
\begin{equation}
\langle P(t) \rangle =
\begin{cases}
 (\omega t)^{-1} & \text{threshold IVR} \\
 (D t)^{-d_{f}/\alpha} & \text{facile IVR}
\end{cases}
\label{powscaling}
\end{equation}
In the above $\omega$ is a microscopic frequency for local motion and
both $\omega, D$ depend on the couplings $\{\Phi_{ijk},\ldots\}$
present in the molecular Hamiltonian. An important prediction of the scaling theory is the critical scaling {\it i.e.,} the dimensionality independent behavior of the survival near the IVR threshold. Moreover the effective QNS dimensionality $d_{f}$ can have the maximum value of $(f-1)$ (accounting for the constant energy). Nevertheless, as seen below, several studies show that 
 $d_{f} \ll (f-1)$ {\it i.e.,} the ZOBS explores only a small part of
the full state space even in the delocalized regime.

The prediction in Eq.~\ref{powscaling} is surprising in that even in the regime of facile IVR one expects a power law behavior for the survival probability on intermediate timescales. By intermediate one means timescale greater then the $1/e$ falloff of $\langle P(t) \rangle$ and the long time limit due to the finite size of the QNS. This is in contrast to the standard golden rule/RRKM exponential decay and therefore has significant repercussions for the IVR dynamics. Specifically,
 the average survival probability is directly linked
to the non-RRKM corrections for the average reaction rate. We quote the
final result and refer to the original literature\cite{schowoly_quote,schowoly_jpc} for a detailed derivation. 
In essence, the RRKM rate expression
\begin{eqnarray}
k_{RRKM}(E) &=& \frac{N^{\ddagger}(E)}{2 \pi \hbar \rho(E)} \nonumber \\
            &=& \left(\frac{N(E)}{2 \pi \hbar \rho(E)}\right)
                    \left(\frac{N^{\ddagger}(E)}{N(E)}\right) \equiv \omega_{0} P_{r}^{eq}
\end{eqnarray}
is modified into
\begin{eqnarray}
k(E) &=& \frac{N_{\rm eff}(E)}{2 \pi \hbar \rho(E)} \\
     \frac{1}{N_{\rm eff}} & = & \frac{1}{N^{\ddagger}(E)} 
          + \omega_{0} \int_{0}^{\infty} dt \left(P_{r}(t)-P_{r}^{eq}\right)
\end{eqnarray}
where $N^{\ddagger}(E)$ is the number of states (``open channels") of the activated complex,
$\omega_{0}$ is the RRKM frequency, $\rho(E)$ is the total density of states,
and $P_{r}^{eq}$ is the equilibrium probability of
being in the transition region. The key quantity here is $P_{r}(t)$ which is a dynamical
object and represents the probability of returning to the
transition region after the QNS trajectory's sojourn into nonreactive regions.
Clearly the effective number of open channels $N_{\rm eff}$ is dependent on $P_{r}(t)$ which,
in turn, is determined by the nature of the average survival probability $\langle P(t) \rangle$.

A direct confirmation of the scaling predictions above was done in a study by Schofield, Wyatt and Wolynes by considering an explicit example of a six dimensional near-degenerate Morse oscillator system with cubic mode-mode 
couplings\cite{scprl95,scjcp96}. The relevant Hamiltonian is given by
\begin{equation}
H({\bf q},{\bf p}) = \sum_{k=1}^{6} \left[\frac{1}{2} p_{k}^{2} + D\left(1-e^{-a_{k}q_{k}}\right)^{2}\right] + \phi \sum_{i,j,k} q_{i} q_{j} q_{k}
\label{6dmorseham}
\end{equation}
 Note that the dissociation energy of all six Morse oscillator are the same ($32924$ cm$^{-1}$) and the $\{a_{k}\}$ are chosen by randomly choosing the harmonic frequencies $\omega_{k} \equiv a_{k} \sqrt{2D}$ in the range $(1000,1004)$ cm$^{-1}$ - hence the term near-degenerate. The coupling strength $\phi$ was then systematically varied to characterize the dynamics from several initial states. The choice of the parameters is made to mimic the notion of statistical homogeneity. The couplings in Eq.~\ref{6dmorseham} are restricted to nearest neighbors {\it i.e.,} couplings of the form $q_{1}q_{2}q_{3}$ are allowed whereas couplings like $q_{1}q_{2}q_{5}$ or $q_{3}q_{5}q_{2}$ are not allowed. Thus, the system is tailored to test the scaling predictions for the IVR dynamics. Quantum dynamical calculations indeed reveal the diffusional scaling $\langle P(t) \rangle \sim (Dt)^{-5/2}$ for $\phi$ above the threshold strength and the critical scaling $\langle P(t) \rangle \sim (\omega t)^{-1}$ at the threshold. Further details including in depth analysis of the individual state IVR dynamics and fluctuations can be found in the original references\cite{scjcp96}.

A few comments are in order at this stage. Firstly, in the above arguments and the example cited, the assumption of a strongly chaotic underlying classical phase space is implicit. Consequently, it is reasonable to assume normal diffusion, $\langle Q(t) \rangle^{2} \sim  D t$ in state space and the diffusion constant $D$ is essentially a classical quantity. Note, however, that the detailed classical dynamics of Eq.~\ref{6dmorseham} has not been studied until now. Secondly, the finiteness of $\hbar$ can lead to quantum corrections to the diffusion constant $D$, which would also strongly depend on the nature of the diffusion {\it i.e.,} the value of $\alpha$ in Eq.~\ref{powscaling}.  Finally, although Schofield and Wolynes have extended\cite{schowoly_jpc} the scaling arguments to cases with anisotropic diffusion, the possibility of some of the mode diffusions being anomalous with others being normal raises  the issue of the validity of a single parameter scaling model.

\subsection{State space predictions}
\label{sec:qnspred}

The state space picture of IVR as a diffusive process in the QNS has proved to be rather useful in understanding the IVR process in a number of systems. Thus, the isomerization of stilbene\cite{qns_stilbene}, IVR dynamics in a small molecule like thiophosgene\cite{qns_sccl2} (SCCl$_{2}$), and heat flow in large systems like proteins\cite{leit_ivrproteins} are a few examples. In both small and large systems the energy flow rates are insensitive to the global density of states and are primarily determined by the various anharmonic resonances and their disposition in the state space\cite{ivr_rev8,ivr_rev12}. 

Note that the local nature of IVR dynamics and the importance of anharmonic resonances has been recognized earlier\cite{ivr_rev5,ivr_rev6,ivr_rev9}, resulting in the well known tier model of IVR. What the state space formalism has achieved is to elucidate the ``tiering coordinate" and, combined with the analogy to AL, provided quantitative predictions for the IVR observables. 
Before proceeding further it is useful to summarize some of the key features of the state space approach to IVR.

\begin{figure}[tbp]
\begin{center}
\includegraphics[height=60mm,width=100mm]{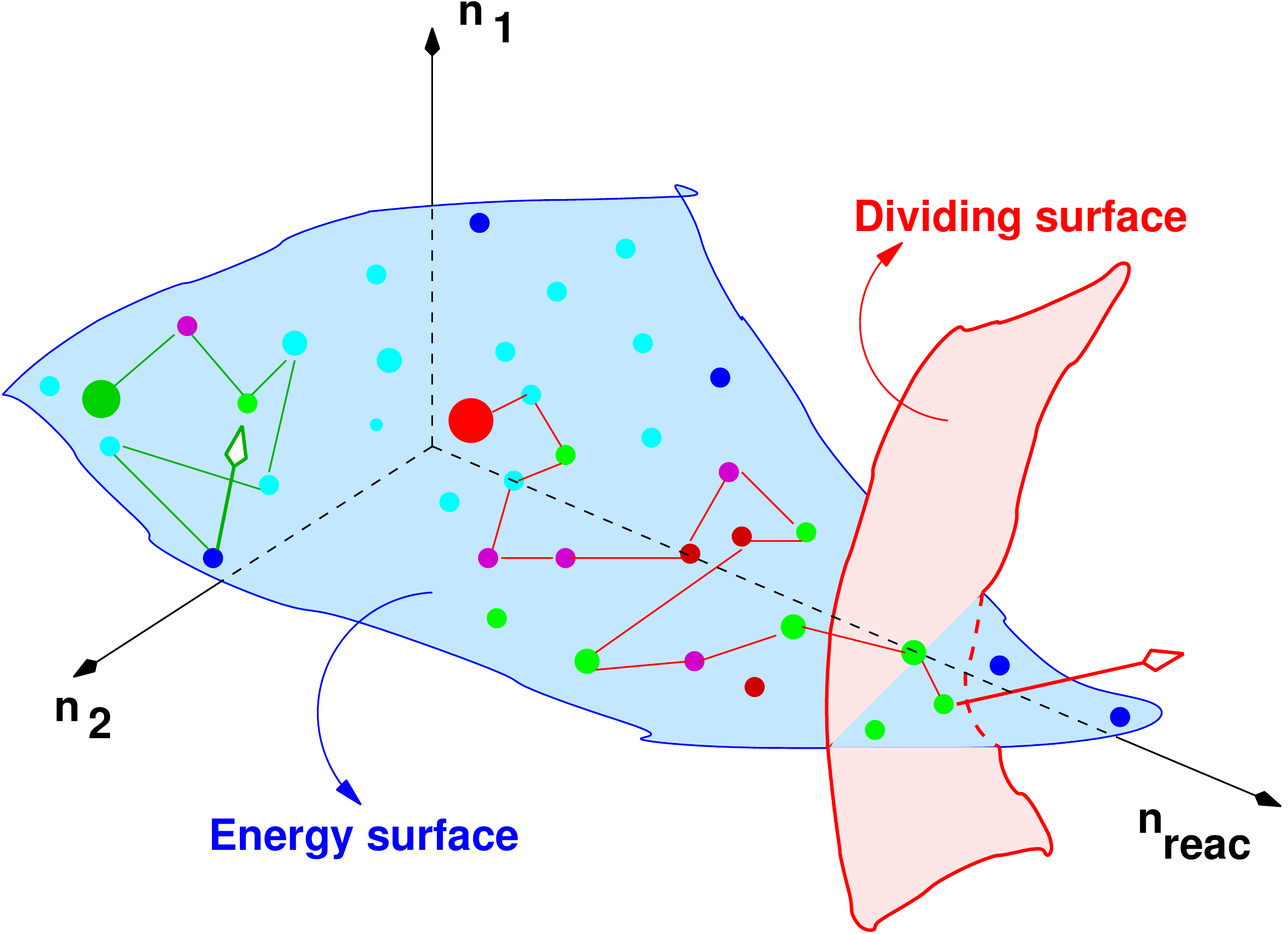}
\caption{A sketch of the state space illustrating the
different nature of IVR dynamics expected
from an edge state (large green circle) and an 
interior state (large red circle). Also shown is a dividing surface
which distinguishes reactive versus nonreactive regions of the
state space. Reproduced from reference\cite{manithesis} with permission.}
\label{intro_qns2}
\end{center}
\end{figure}

In the QNS one can distinguish between two classes of states, edge and interior, based on their location.  In the more familiar spectroscopic language, an edge state corresponds to
an overtone state with most of the excitation in one mode. On the other
hand an interior state corresponds to a combination state with many of the modes
having moderate to low excitations. 
A simple sketch is shown in Fig.~\ref{intro_qns2}.
For example, in the six-mode SCCl$_{2}$ a ZOBS is denoted as $|n_{1},n_{2},n_{3},n_{4},n_{5},n_{6}\rangle$ and the state $|7,0,0,0,0,0\rangle$ is an example of an edge state. In contrast, a state like $|3,2,3,2,3,3\rangle$ is a typical interior state.
 A measure for the ``edgeness"\cite{chjcp09} of a 
ZOBS $|b\rangle \equiv |n_{1}^{(b)},n_{2}^{(b)},\ldots,n_{f}^{(b)}\rangle$ can be provided as
\begin{equation}
e_{b} = \left[\frac{1}{f(f-1)} \sum_{i=1}^{f} \left(\frac{n_{i}^{(b)}}{\bar{n}^{(b)}}-1\right)^{2}
\right]^{1/2}
\end{equation}
with the two limits $e_{b}=1$ (edge) and $e_{b}=0$ (interior). The quantity
$\bar{n}^{(b)}$ is the mean occupation number of $|b\rangle$. Thus, states $|1,6,3,0,0,9\rangle$ has $e_{b} \approx 0.47$ whereas the state $|3,2,1,0,2,5\rangle$ has $e_{b} \approx 0.32$ and thus one may say that the latter is relatively more interior in the QNS as compared to the former state.

From a geometric point of view one expects that
the edge state has fewer dark states to interact with as compared to the interior
state. Hence, the rate of energy flow out of an interior state would be relatively
fast in comparison to that of an edge state. However, there is a balancing factor
in the sense that the anharmonicities felt by an interior state will be smaller
when compared to those felt by the edge states. Consequently, the influence of
an anharmonic resonance would be greater for edge states as opposed to interior
states. The actual IVR rate is an outcome of the subtle competition between the
two opposing factors.  Recently\cite{chprl08}, Chowdary and Gruebele have
argued that if the fraction of edge states in a $f$-dimensional
state space exceeds a certain threshold
\begin{equation}
x_{min} = \frac{\ln f}{f}
\label{fracedge}
\end{equation}
then one should expect localized eigenstates and hence nonstatistical IVR.
The edge states, although typically protected from IVR, can nevertheless undergo IVR via the phenomenon of dynamical tunneling\cite{dyntunrefs}. 

The survival probability of a ZOBS $|b\rangle \equiv |n_{1}^{(b)},n_{2}^{(b)},\ldots,n_{f}^{(b)} \rangle$
\begin{eqnarray}
P_{b}(t) &=& |\langle b|b(t) \rangle|^{2} = \sum_{\alpha,\beta} |\langle b|\alpha \rangle|^{2} |\langle b|\beta \rangle|^{2} e^{-i(E_{\alpha}-E_{\beta})t/\hbar}  \nonumber \\ 
&\sim &
\sigma_{b} +
(1-\sigma_{b})\left[1 + \frac{2t}{\tau D_{b}}\right]^{-D_{b}/2}
\label{powscal}
\end{eqnarray}
exhibits power law behavior\cite{mgpnas_powlaw} on intermediate time scales. Note that the above is the survival probability of a specific ZOBS and not the average $\langle P(t) \rangle$. In the above equation $\{|\alpha \rangle\}$ are the eigenstates of the full Hamiltonian with eigenvalues $\{E_{\alpha}\}$ and the dilution factor $\sigma_{b} = \sum_{\alpha} |\langle b|\alpha \rangle|^{4}$
has the intuitive interpretation of $\sigma_{b}^{-1}$ being the number of
eigenstates over which an initial feature is fragmented (or diluted)
in the frequency domain spectrum. Note that the power law exponent has now been
expressed as $\gamma = D_{b}/2$ which should be compared to the scaling prediction
$\gamma = d_{f}/\alpha$ with an appropriate definition of $D_{b}$.

The surprising fact that has emerged from several studies\cite{ivr_rev12,mgpnas_powlaw} is that
the effective dimension of the IVR ``manifold" $D_{b}$ is much less than $3N-6 \equiv f$ (dimensionality
of the state space for a molecule with $N$ atoms) even for fairly large organic molecules
at significant levels of excitation. For instance, Gruebele reanalyzed\cite{mgpnas_powlaw} the experimental results on IVR in several large organic molecules like fluorene, anthracene, and cyclohexylaniline  and found $D_{b} \approx 2-3$, suggesting a very small fraction of the QNS being involved in the IVR dynamics. Similarly, IVR in SCCl$_{2}$ from several ZOBS around $8000$ cm$^{-1}$ have an average $D_{b} \approx 2.8$ which is much smaller then the maximum value of $f-1=5$.  Further detailed analysis by Bigwood {\it et al.} revealed that the parameter
\begin{equation}
N_{loc}(|b\rangle) \equiv \sum_{i} L_{ib}^{2} =
\sum_{i}\left[ 1 +
\left( \frac{E^{0}_{i} - E^{0}_{b}}{\langle i|V|b\rangle}\right)^{2} \right]^{-1}
\end{equation}
is central to understanding several aspects of the IVR dynamics. The perturbative measure $N_{loc}$ is directly related to the local density of states and has close connections\cite{mgleitwoly_pnas} to the quantum ergodicity parameter $T(E)$ (see below). In particular, $N_{loc} \sim 1$ at the threshold for IVR and $N_{loc} \sim 10$ for unrestricted IVR. Essentially, for the later value of $N_{loc}$ the ZOBS in state space form an extensively connected network. A crucial point here is that around $N_{loc} \sim 1$ the nature and mechanism of the IVR dynamics can be very sensitive to specific anharmonic resonances.

Within the LRMT approach, the distribution $W(\sigma_{b})$ of the dilution factors
$\sigma_{b} \equiv \lim_{t \rightarrow \infty}P_{b}(t)$ at energy $E$  can be calculated\cite{leitwoly_dist} as
\begin{equation}
W(\sigma) = \frac{\gamma(T)}{\sqrt{\sigma(1-\sigma)^{3}}} 
\exp\left[-\frac{\pi \gamma^{2}(T)}{1-\sigma}\right]
\end{equation}
with $\gamma(T)$ depending on $E$ via $T(E)$, the IVR threshold parameter 
 \begin{equation}
T(E) = \frac{2\pi}{3} \left(\sum_{Q} K_{Q} \langle|V_{Q}|\rangle
D_{Q}(E) \right)^{2}
\label{lwqet}
\end{equation}
$T(E)$ is essentially the criteria in Eq.~\ref{locdeloccrit},  written in a slightly different notation, and represents the quantum ergodicity threshold.
More precisely, $\langle|V_{Q}|\rangle$
represents the average effective coupling of $| b \rangle$ to the states,
$K_{Q}$ of them with density $D_{Q}(E)$, a distance $Q$ away in state space.
The coupling $V_{Q}$ itself is determined by both the low order and high order
anharmonic resonances in the Hamiltonian.
Transition to facile IVR is signaled by $T(E) \rightarrow 1$ ($\gamma \rightarrow \infty$
and $W(\sigma) \sim \delta(\sigma)$), dubbed as the ``quantum
ergodicity threshold", and the very existence of such a threshold
implies that IVR is not a global statistical process.

The distribution $W(\sigma)$ turns out to be  bimodal and hence implies that  certain initial 
zeroth-order states undergo facile IVR whereas
others, at the same energy, can exhibit highly restricted IVR. Sibert and Gruebele have analyzed the IVR dynamics of SCCl$_{2}$ and confirmed\cite{sibgruebJCP06} the existence of such bimodal behavior. Nevertheless, the ``all or nothing" prediction is a little surprising. Indeed, as pointed out in a recent extensive study\cite{chjcp09} on SCCl$_{2}$, the above distribution assumes a common mean value for the coupling strengths of all the relevant anharmonic resonances and does not take the edge states into account. The edge state do substantially influence $W(\sigma)$ and, at least in case of SCCl$_{2}$, result in a stretched exponential distribution.  

\section{Important questions}
\label{sec:qnsquestions}

In the previous sections we have outlined the central idea, some of the key features, and the predictions of the state space approach to IVR. Nevertheless, certain aspects of the model are yet to be understood completely. These issues, which go beyond merely correcting the RRKM rate constants, are related to the fundamental assumptions made in the state space models and addressing them allows us to assess the applicability of the scaling perspective to a wider class of systems. At the same time, deviations from the state space predictions  in the context of studying the dynamics of individual molecules provide additional opportunities to gain insights into the mechanistic aspects of IVR. Afterall, it is the deviations from RRKM and TST that have considerably enriched our understanding of reaction dynamics. Furthermore, as discussed in the next section, the universal predictions made above have close ties with universal features of multidimensional classical phase space transport. Therefore, the following questions are also meant to motivate a detailed study of the classical-quantum correspondence between the quantum state space and the classical phase space perspectives on IVR.

\begin{enumerate}
\item Is the nature of the diffusion in state space normal or anomalous? 

Most of the studies to date have implicitly assumed normal diffusion. However, molecular Hamiltonians with their characteristic hierarchical coupling structure are prime candidates for observing anomalous diffusion. For instance,  it is quite possible for
an edge state to have $\alpha=2$ (normal diffusion) and an isoenergetic interior
state to have $0 < \alpha < 2$ (subdiffusive) or vice versa. The differing nature of the diffusion
in QNS and hence IVR from the two states could lead to results contrary to our expectations
regarding the effective IVR dimensions {\it i.e.,} $D_{b}(\text{interior}) < D_{b}(\text{edge})$. Note that anomalous diffusion is a topic of current interest and several physical systems have been found to exhibit such a behavior\cite{anomdiffrev}. In the context of the state space model, it is then of some interest to ascertain if dynamical and/or static traps can exist which can potentially lead to subdiffusion. On the other hand, recent study\cite{qhyperdiff} by Zhang {\it et al.} has also shown the existence of quantum hyperdiffusion (mean square displacement scaling as $t^{\gamma}$ with $\gamma > 2$) in tight-binding lattices. Molecular effective Hamiltonians are akin to generalized tight-binding Hamiltonians and one cannot rule out the possibility of quantum or classical hyperdiffusion. From the scaling perspective, such anomalous behavior can compromise the self-averaging assumption. 

\item Can the effective IVR dimension $D_{b}$ be related to the parameters 
of the underlying molecular Hamiltonian? The answer to this question is
significant from the perspective of understanding and predicting mode-specific effects in
molecules.

An important step towards answering this  question was taken
by Wong and Gruebele\cite{wojpca99} wherein they estimated the
effective IVR dimension as
\begin{equation}
D_{b} \approx D_{b}(Q) =
\frac{\Delta \ln \sum_{i} L_{ib}^{2}}{\Delta \ln Q}
\label{dimeq}
\end{equation}
where the symbol $\Delta$ indicates a finite difference
evaluation of the dimension due to the discrete nature of the
state space.
Estimates based on Eq.~\ref{dimeq} have compared well with the numerical computations
for the six-mode SCCl$_{2}$ molecule\cite{wojpca99}. Note, however, that dynamical tunneling\cite{keirpc07}
effects are not accounted for in Eq.~\ref{dimeq} and such quantum effects can
be important near the IVR threshold.

\item Is it possible for the survivals of certain initial states to exhibit multiple power law 
behavior over different timescales? Existence of such multiple power laws connecting
the short time exponential to the long time dilution limit can be relevant
in terms of the IVR mechanism and hence local control strategies.

Schofield and Wolynes\cite{schowoly_jpc} have shown
 that multiple power law behavior can manifest in the more generic
case of anisotropic diffusion in the QNS. In such cases the average survival probability
still scales as $\langle P(t) \rangle \sim (Dt)^{-d_{f}/\alpha}$ above the IVR threshold. However, 
$d_{f}$ can become time-dependent,
primarily due to the finiteness of the state space.
 There are not many examples of such a behavior which is somewhat surprising since 
anisotropic IVR is the rule rather than the exception. In any case, finiteness of
state space is one reason but perhaps not the only reason. 
Dynamical localization along select directions in QNS can also play an 
important role in this context. Gruebele has also discussed\cite{ivr_rev8} possible time-dependence of $D_{b}$ in Eq.~\ref{powscal} which, as seen later, relates directly to the nature of the classical Arnold web.

\item How does the power law behavior manifest itself in terms of the 
eigenstates (frequency domain spectra)? Is it possible to decode the dynamical
information from the analysis of the eigenstates? 

Although IVR is understood more naturally
from a time-dependent viewpoint {\it i.e.,}
the dynamics of a ZOBS and its mixing with
the various dark states belonging to different tiers\cite{ivr_rev3},
the time-independent perspective has its own merits. The apparent contradiction,
that a molecular eigenstate does not `move' and hence
there is no IVR is easily resolved
since the natural representation is now the energy or frequency
domain. Therefore the infinite time limit of the IVR dynamics is
imprinted onto the eigenstates in the form of subtle
spectral features and patterns, both strong and weak\cite{ivr_rev6}.
The time and frequency domain information are Fourier related
\begin{equation}
\int_{}^{} dt e^{i \omega t} \langle b(0)|b(t) \rangle
\longleftrightarrow \sum_{\alpha} |\langle b|\alpha \rangle|^{2} \delta(\omega-\omega_{\alpha})
\end{equation}
with $E_{\alpha}=\hbar \omega_{\alpha}$.
Hence, it is possible to infer the time-dependent
dynamics from the fractionation pattern of a bright state as revealed
by a frequency domain experiment and 
assignment of eigenstates is tantamount to insights 
into the long time IVR dynamics of the molecule.

Since the intermediate time power
law scaling of the survivals are related to the correlation between 
spectral intensities $|\langle b|\alpha \rangle|^{2}$ and the 
phases $\omega_{\alpha}$, one expects profound influence on the nature
of the eigenstates. The hierarchical tree analysis of Davis\cite{hierdavis}, power law scaling
of the survival\cite{sw,mgpnas_powlaw} and analogy\cite{loganwoly} to metal-insulator transitions in general suggest
that one possible signature could be the existence of multifractal eigenstates\cite{multifraceigen}. However, very little work has been done in this direction. 

\item Do the original scaling arguments and BSTR/LRMT model predictions, expected
to be valid for relatively large systems, hold for small polyatomic molecules? 

Strictly speaking the answer should be in the negative.  Nevertheless, several features of the state space model are seen in studies on small molecules with $3-6$ vibrational modes\cite{skh2oscaling,kscdbrclf,kscdbreigen,kssccl2}. In particular, the dynamics of such systems have been studied using the spectroscopic or molecular Hamiltonians that are specific to the molecule of interest. Such Hamiltonians have specific mode frequencies and anharmonicities and varying number and type of anharmonic resonances with very different strengths. Thus, the validity of state space models is a bit surprising and hint towards a universal mechanism underlying the intermediate-time power law behavior. However,  our understanding regarding the origin of such a universal mechanism is at present incomplete.
\end{enumerate}

\section{Classical-quantum correspondence and IVR}
\label{sec:cqcorrsec}

Answering the questions raised above involves obtaining insights into the subtle competition between classical and quantum mechanisms of IVR. Thus, a clear understanding of the various issues involved would require both classical and quantum studies.  Put another way, classical dynamical studies provide the proper baseline for us to grasp the subtle influence of quantum effects like tunneling, localization, and coherence. The utility of such a classical-quantum correspondence approach
is being increasingly appreciated in a wide range of disciplines. 
In fact, many of the examples mentioned in the introduction have utilized classical mechanics to extract useful insights into the system dynamics.

The utility of such a classical-quantum correspondence approach
is being increasingly appreciated in a wide range of disciplines, especially atomic physics. For example,  concepts like macroscopic quantum self-trapping, $\pi$-oscillations, and Rabi-Josephson transition in the trapped cold atom area have benefited immensely from the mapping between the Bose-Hubbard model and the classical pendulum Hamiltonian\cite{oberthrefs}.  Many experiments\cite{becexpts} with exquisite control over the Hamiltonian parameters are revealing an incredible classical-quantum correspondence in trapped cold atoms which, ironically, is an example system for quantum coherence. As another example, the emerging field of microcavity lasers\cite{microlaser} has brought together the fields of quantum optics and quantum chaos. Thus, directional emission from such devices needs chaos in the phase space and a recent work explains the mechanism using the unstable manifolds in the phase space of the system\cite{shinohara}. 
Note that even that epitome of quantumness, entanglement, is not immune to the classical phase space structures\cite{entangchaos}.

IVR itself has been the subject of several pioneering classical-quantum correspondence studies which have enriched our understanding of the process. Note that TST, RRKM, Slater's dynamical rate theory are all inherently classical in their conception and  best understood from a phase space perspective. In a classic paper\cite{thielewilson} written half a century ago, Thiele and Wilson sum up their study of two kinetically coupled Morse oscillators as follows

``In view of the predominant effects of anharmonicity exhibited by a pair of coupled Morse oscillators, we feel that the use of normal modes in approximately describing a dissociation process, along with an extended concept of a gradual flow of energy among these normal modes, is of very doubtful validity. Indeed, in the present model the flow of energy is generally so rapid that normal modes are unrecognizable."

Interestingly enough the paper concludes by stating that 
``The referee has pointed out, however (and we agree), the danger of generalizing from triatomic to complex polyatomic molecules." There are two crucial foresights here. First, is the hint that the notion of modes itself can become invalid at high levels of excitations. The second, is the warning by the referee -  generalizing classical dynamical insights from two to three or more DoFs is indeed dangerous. In fact, the technical and conceptual difficulties associated with such a generalization are mainly responsible for most researchers moving away from classical-quantum correspondence studies of IVR. However, over the last few decades, rapid advances in the field of nonlinear dynamics have yielded new tools and perspectives which are proving to be of immense value in understanding IVR dynamics in multidimensional systems. 

An  example has to do with the recent impressive advances in the area of TST, a topic related to IVR but not discussed here at all. A central problem in TST is to  identify recrossing-free dividing surfaces in the phase space of systems with $f \geq 3$, thus generalizing the beautiful work\cite{pejcp73} by Pollak and Pechukas on periodic orbit dividing surfaces for $f=2$.  Significant progress in this direction has come from identifying  a $(2f-3)$-dimensional manifold in phase space, the so-called normally hyperbolic invariant manifold\cite{nhim} (NHIM), which can be identified as the ``activated complex" for the reaction.  Consequently, it is now possible  to construct, perturbatively, locally recrossing-free dividing surfaces for systems with higher then one index saddles\cite{gt1indsaddle},  time-dependent systems\cite{tdeptst}, and even dissipative systems\cite{disstst}.
As pointers for further details we mention the recent reviews by 
Schubert, Waalkens, and Wiggins\cite{schubtst} and Bartsch {\it et al.}\cite{bartschtst} which provide a detailed account of how the Wigner perspective of phase space based TST has been advanced for systems with $f \geq 3$.
 
On a related note, in the context of RRKM approximation to unimolecular rates, several studies have shown non-exponential lifetime distribution of the reactant. Such studies clearly establish the importance of IVR in the reactant well and the 
pioneering work by Thiele on the so called gap time distribution problem\cite{thielegap}, originating from Slater's work, already anticipated the important role played by various phase space structures.
Recently, Ezra, Waalkens and Wiggins\cite{ezjcp09} 
have beautifully shown the power of Thiele's gap time
formulation for the $HCN \leftrightarrow HNC$ isomerization reaction in light
of the rapid advances in TST. 

The phase space perspective on IVR for $f=2$ systems has been instrumental in gaining crucial insights into the mechanism of IVR. Thus, essential concepts such as reaction bottlenecks, dynamical barriers, and local mode -normal mode transitions have originated from detailed classical-quantum correspondence studies. The latter concept is a special case of the more general situation wherein low energy modes can disappear or undergo  complete metamorphosis with increasing energy {\it i.e.,} the phenomenon of bifurcation. For example, the appearance of counter-rotating modes in acetylene\cite{c2h2bifur} and the isomerization modes\cite{saddlenode} in several systems are unambiguously identified as bifurcations in the phase space of the system.  It would not be an overstatement to say that not only the lingo of IVR but also the understanding of the quantum spectrum and eigenstates in terms of sequences and splittings is greatly aided by classical-quantum correspondence studies. We refer the reader to earlier reviews\cite{2dofreviews} for details on the $f=2$ system and, instead, focus here on the systems with $f \geq 3$ - the last frontier, perhaps.
Not surprisingly, the
resolution of some of the issues in this context have direct bearing on the questions
raised in sec.~\ref{sec:qnsquestions}.

\subsection{State space-phase space correspondence}
\label{sec:qnspscorr}

A direct connection between the QNS and classical phase space can be established by the correspondence
\begin{equation}
a_{\alpha} \leftrightarrow \sqrt{I_{\alpha}} e^{i \theta_{\alpha}} \,\,;\,\, a_{\alpha}^{\dagger} \leftrightarrow \sqrt{I_{\alpha}} e^{-i \theta_{\alpha}} \,\,;\,\, n_{\alpha} \leftrightarrow I_{\alpha}
\end{equation}
applied to the generic molecular Hamiltonian considered in sec.~\ref{sec:qns}. Here, the variables $({\bf I},{\bm \theta})$ are the action angle variables\cite{aavariables} corresponding to the integrable zeroth-order Hamiltonian $H_{0}$. The full classical limit Hamiltonian is 
\begin{eqnarray}
H_{\text cl}({\bm I},{\bm \theta})& = & H_{0}({\bm I}) +
2\sum_{ijk}^{f} \Phi_{ijk} \sqrt{I_{i}I_{j}I_{k}} f({\bm \theta}) + \ldots \nonumber \\
&\equiv&\sum_{\alpha=1}^{f} \omega_{\alpha}^{(0)} I_{\alpha} + \sum_{\alpha,\beta=1}^{f} x_{\alpha \beta} I_{\alpha} I_{\beta} +
\sum_{ijk}^{f} \bar{\Phi}_{ijk}({\bm I}) f({\bm \theta}) + \ldots
\label{hrescl}
\end{eqnarray}
where we have denoted
\begin{eqnarray}
f({\bm \theta}) &=& \cos(\theta_{i}+\theta_{j}+\theta_{k}) + \cos(\theta_{i}-\theta_{j}-\theta_{k}) \\
&+& \cos(\theta_{i}+\theta_{j}-\theta_{k}) + \cos(\theta_{i}-\theta_{j}+\theta_{k}) \nonumber
\end{eqnarray}
Similar terms
as in $f({\bm \theta})$ appear at higher orders and we retain only the third order terms 
in this discussion for the sake of clarity. The above Eq.~\ref{hrescl} is a classic example
of a nonlinear, multiresonant $f$-degree of freedom Hamiltonian. Such Hamiltonians are
ubiquitous in nature and appear in the description of celestial mechanics to atomic
and molecular dynamics. Indeed, according to Poincar\'{e} the study of the perturbations of
the conditionally periodic motions of Eq.~\ref{hrescl} is the fundamental problem of
dynamics. Amazingly enough, this century old statement of Poincar\'{e} translates in
the current molecular context to the study of how low-energy normal or local modes get 
perturbed (transformed) into entirely different types of modes/motion with increasing
energy. 

In the absence of perturbations $\tilde{\Phi}_{ijk}=0$ Hamilton's equations of motion 
imply that the actions are constants of the motion and the conjugate angles are
periodic functions of time. The classical dynamics is integrable and the phase space if filled with
 $f$-dimensional tori with radii $(I_{1},I_{2},\ldots,I_{f})$.
At this stage one can already see the direct connection between the QNS and the classical zeroth-order action space from the semiclassical correspondence
\begin{equation}
I_{k} \longleftrightarrow \left(n_{k}+\frac{\mu_{k}}{2}\right) \hbar
\end{equation}
with $\mu_{k}$ known as the Maslov index\cite{maslov} for the $k^{\text th}$ mode. Note that the Hamiltonians in eq.~\ref{hres} and eq.~\ref{hrescl} are written with the appropriate zero point energies scaled out. Thus, a point in the QNS $|n_{1},n_{2},\ldots,n_{f} \rangle$ corresponds to a point in the classical action space $(I_{1},I_{2},\ldots,I_{f})$. 

For, $\tilde{\Phi}_{ijk} \neq 0$ the actions are no longer constants of the motion and typically the classical dynamics is nonintegrable. Therefore, one has the possibility of chaotic motion coexisting with regular motion at a specific energy of interest. 
Nevertheless, the motion of an initial state space point $|{\bm n}\rangle$ in the QNS
and that of an initial phase space point $({\bm I},{\bm \theta})$ in the
phase space are closely related. The main difference is that whereas 
QNS is a $f$-dimensional space, the phase space has the dimensionality of $2f$
due to the additional conjugate angle variables. A closer correspondence can be achieved by averaging the classical observables over the $f$-dimensional angle space or by simply projecting the phase space onto the $f$-dimensional action space. 

In any case, both in the QNS and in the classical phase space, the flow of energy amongst various modes corresponds to the nontrivial evolution of $n_{k}$ and $I_{k}$ respectively. In QNS the crucial terms are the anharmonic resonances and the analogous terms, contained in $f({\bm \theta})$, in classical phase space are known as nonlinear resonances. Consequently, the nature and geometry of the nonlinear resonances are central to understanding the IVR dynamics. We now give a brief description of the geometry of the resonance network. The discussion is kept at a nontechnical level and, in particular, results will be stated and used without providing proofs. The rigorous results with appropriate conditions and proofs can be found in several excellent texts\cite{wigginsbook,arnoldbook,marsdenbook}.

\subsection{Geometry of the resonance network: Arnold web}
\label{sec:argeom}

Consider a general Hamiltonian of the form $H({\bf I},{\bm \theta}) = H_{0}({\bf I}) + \epsilon V({\bf I},{\bm \theta})$ with $\epsilon$ being a parameter that measures the relative importance of the perturbation. Specifically, as written, for $\epsilon=0$ the system is integrable, for $\epsilon \ll 1$ the system is near-integrable, and for $\epsilon \sim 1$ the system is nonintegrable. 
The key quantities for IVR are the nonlinear frequencies
\begin{equation}
{\bm \Omega}({\bm I},{\bm \theta}) = {\bm \nabla}_{\bm I}H({\bm I},{\bm \theta}) = {\bm \Omega}^{0}({\bf I}) + \epsilon  {\bm \nabla}_{\bm I}V({\bm I},{\bm \theta})
\end{equation}
which are functions of energy via the dependence on the action-angle variables. 
For $\epsilon=0$ the actions are fixed and hence the frequencies ${\bm \Omega}({\bm I},{\bm \theta}) = {\bm \Omega}^{0}({\bf I})$ are also fixed. The motion is completely regular and, for a specific set of initial actions and angles, constrained to a $f$-dimensional torus. For increasing $\epsilon$ the celebrated KAM theorem\cite{kam} provides a vivid picture of the metamorphosis of the phase space from completely regular to a completely chaotic situation. Note that here we will restrict the discussion entirely to the non-degenerate case {\it i.e.,} 
\begin{equation}
{\rm Det}\left(\frac{\partial H_{0}({\bf I})}{\partial I_{k} \partial I_{l}} \right) \equiv {\rm Det}\left(\frac{\partial \Omega^{0}_{k}}{\partial I_{l}} \right) \neq 0
\label{nondeg}
\end{equation}
The above condition is very important since it allows us to study the dynamics in the action space or in the frequency space - a feature that will prove particularly powerful in the context of correspondence perspective of this article. The condition in Eq.~\ref{nondeg} is not restrictive in the molecular context since anharmonicities are essential and typically present in the effective Hamiltonians. 

In order to appreciate the geometric structure associated with the resonances we start by writing down the condition for a resonance in the integrable case
\begin{equation}
m_{1} \Omega^{0}_{1} + m_{2} \Omega^{0}_{2} + \ldots + m_{f} \Omega^{0}_{f} 
 \equiv {\bm m} \cdot {\bm \Omega}^{0}({\bf I}) = 0
 \label{rescond}
\end{equation}
with integers ${\bm m}=(m_{1},m_{2},\ldots,m_{f})$. The frequency vector ${\bm \Omega}^{0}$ is said to be resonant and the order of the resonance is defined as $|m|= \sum_{k=1}^{f} |m_{k}|$. Specifically, Eq.~\ref{rescond} defines what is called as the multiplicity-$1$ resonance.
For $\epsilon \ll 1$ the breakup of an unperturbed  resonant torus, according to
the Poincar\'{e}-Birkhoff theorem\cite{arnoldbook}, will result in 
the creation of alternating elliptic and hyperbolic fixed points in the
phase space. This leads to the formation of ``resonant islands" in the phase
space with finite widths. The widths of the resonances decrease exponentially with the order.

\begin{figure}[t]
\begin{center}
\includegraphics[height=80mm,width=100mm]{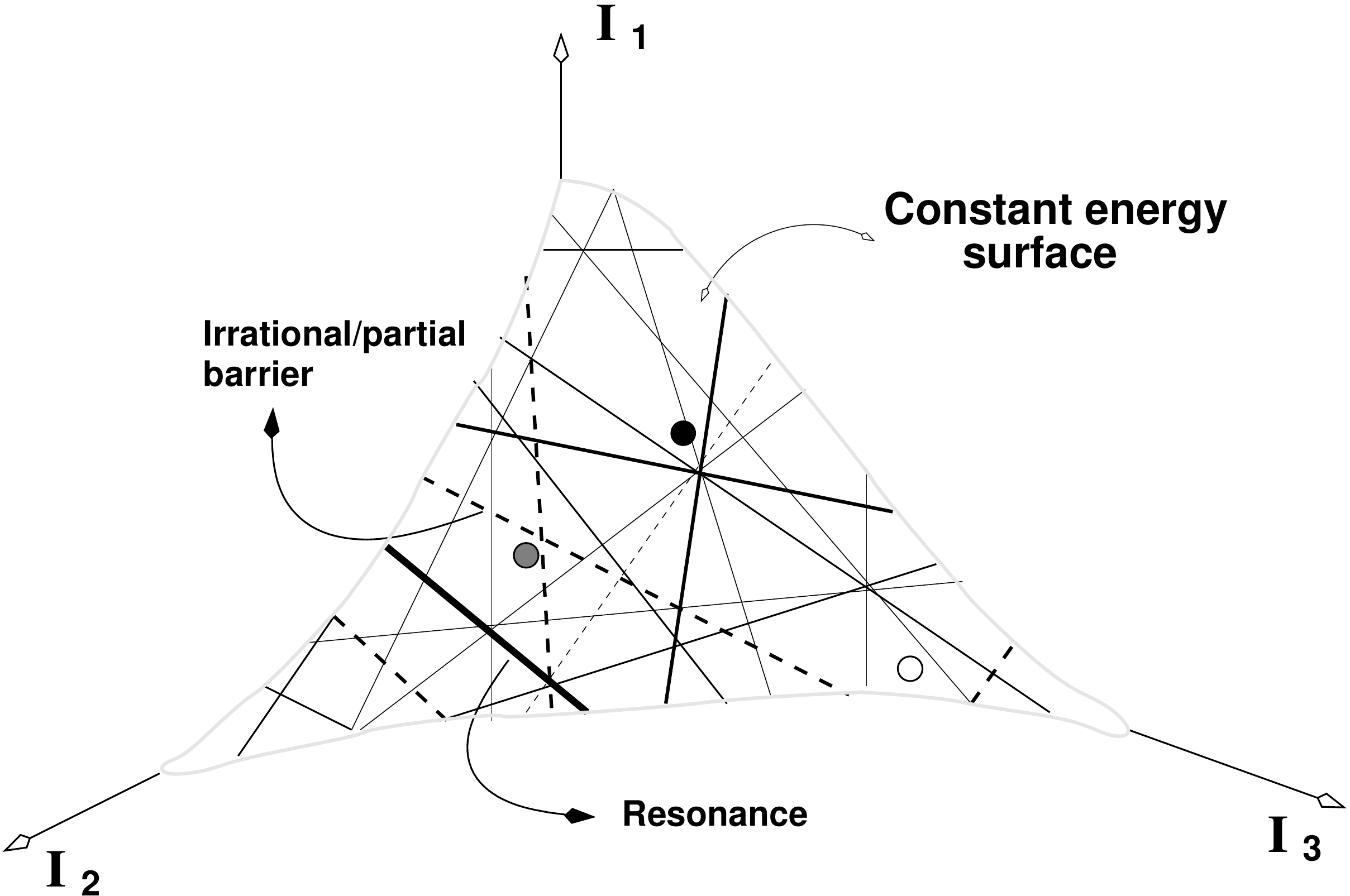}
\caption{Sketch of the resonance network {\it i.e.,} Arnold web for a three mode system.
The resonances, with varying thickness representing varying strengths, form an intricate
network over which dynamics of specific ZOBS (circles) occurs. Possible barriers to
the transport are indicated by dashed lines. Note the ``hubs" in the network corresponding
to the intersection of several low and high order resonances. Compare to the state space
picture shown in the earlier figures. Reproduced from reference\cite{manithesis} with permission.}
\label{intro_arweb}
\end{center}
\end{figure}

The condition in Eq.~\ref{rescond} defines a $(f-1)$-dimensional manifold in the action space. Focusing attention on the constant energy surface (CES) $H_{0}({\bf I}) = E$, we see that the resonance manifold restricted to the CES yields a $(f-2)$-dimensional object. Thus, for $f=2$ the resonance conditions are satisfied at points on the CES and hence energetically isolated. On the other hand, for $f=3$ example, the resonance manifolds intersect the CES along one-dimensional curves, implying that the resonances are not energetically isolated. Therefore, one can immediately see that the $f \geq 3$ cases are significantly different from the $f < 3$ cases. In particular,
for $f \geq 3$ it is possible for two or more independent resonance manifolds to intersect each other on the CES. In other words, the resonance frequency vector ${\bm \Omega}^{0}$ can satisfy ${\bf m}^{(r)} \cdot {\bm \Omega}^{0} = 0$ for independent vectors ${\bf m}^{(r)}$ with $r=1,2,\ldots,s$ and $s < f$. Such a vector ${\bm \Omega}^{0}$ is called as a multiplicity-$s$ resonant vector. For example, in the case of $f=3$ it is possible to have two independent resonances intersecting on the CES giving rise to a multiplicity-$2$ resonance.
Such intersection points are also called as ``hubs" or ``junctions" and play a critical role in determining the nature of the phase space transport in $f \geq 3$ systems. In principle, the vector ${\bf m}$ can run through all possible integer components and hence generate a dense set of resonances on the CES. For example, Fig.~\ref{intro_arweb} shows a typical situation in $f=3$ in the action space. An equivalent picture can be drawn in the frequency space since, thanks to the non-degeneracy condition, one can locally invert the map ${\bf I} \mapsto {\bm \Omega}^{0}({\bf I})$. Such a dense set of resonance lines is called as the Arnold web and this resonance network forms the basis for understanding the dynamics when $\epsilon \neq 0$. 

In the context of Fig.~\ref{intro_arweb}, it is useful to right away make contact with the QNS in Fig.~\ref{intro_qns}. A ZOBS is associated with the corresponding classical zeroth-order actions and hence a specific point in the action space. With respect to the Arnold web, a ZOBS can be near a resonance junction formed by low order resonances or ``away" from it. Since resonances are dense on the CES, by away we mean far from low order junctions. However, even though the ZOBS is away from a low order junction, it is quite possible that it is in the vicinity of some other junction formed by higher order resonance intersections on the CES. Dynamics is trivial for $\epsilon=0$, the ZOBS does not move. On the other hand, for any finite value of $\epsilon$ the ZOBS evolves with time and corresponds to the IVR dynamics in the system.
The state space picture of local, anisotropic IVR mediated by anharmonic resonances is clearly linked to the nontrivial evolution of the action variables on the Arnold web. The local nature is reflected in the dynamics of the ZOBS being influenced by a specific set of resonances and junctions. The anisotropy arises from the fact that the ZOBS is located in a region with resonances of different orders, corresponding to very different timescales for the evolution of the actions. Clearly, both the short and long time IVR dynamics is influenced by the topology of the Arnold web and the ensuing transport. 

In order to study the IVR dynamics we now need to understand the possible mechanisms of transport on the Arnold web for $\epsilon \neq 0$. This requires us to briefly discuss two regimes - Nekhoroshev regime ($\epsilon \ll 1$) and the Chirikov regime ($\epsilon \sim 1$). Note that considerable work\cite{nekhro,nekhrorefs1,nekhrorefs2} has been done in this context and only the salient features will be mentioned here. For details we refer the reader to some of the recent literature. Specifically, the article by Efthymiopoulos provides\cite{efthynekarnold} a connection between Nekhoroshev theorem and Arnold diffusion and the survey article\cite{celfrolega} by Celletti, Froeschl\'{e}, and Lega provides a readable introduction to Nekhoroshev's work.
According to KAM\footnote{More precisely, KAM proved that for small $\epsilon$ the tori that survive satisfy the Diophantine condition $|{\bf m} \cdot {\bm \Omega}^{0}| \geq \gamma |m|^{-f}$.}, for a finite but small $\epsilon$ the tori that get affected the most are
the  resonant tori. The destroyed tori are replaced by resonance layers of appropriate widths and stochastic regions are formed due to the destruction of the separatrices between KAM tori and the regular resonant tori. Thus, in this regime the stochastic layers form a connected network, the Arnold web, in the action space. Nekhoroshev proved\footnote{Apart from the requirement of nondegenerate Hamiltonians, Nekhoroshev also requires the Hamiltonian to be ``steep", a condition that we have not discussed here. Usually the so called steepness condition is replaced by the much stronger condition of convexity or quasi-convexity on $H_{0}$.} that the drift of the actions\cite{nekhro}
\begin{equation}
||{\bf I}(t)-{\bf I}(0)|| \leq c_{1} \epsilon^{\alpha} \,\,\,\,; \,\,\,\, |t| \leq c_{1} \exp(c_{2} \epsilon^{-\beta})
\label{nekhrosh}
\end{equation}
happens on exponentially long times. The Nekhoroshev exponents are estimated to be $\alpha=\beta=(2f)^{-1}$. In addition, if the initial actions are close to a resonant torus of multiplicity $r$, then the timescales on which the actions drift are given as above (with different constants) and exponents $\alpha=\beta=(2(f-r))^{-1}$. Remarkably, this predicts that  trajectories initiated close to higher multiplicity resonances can be trapped for very long times. For example, in a $f=3$ system, a ZOBS located near the resonance hub or junction is expected to be trapped for longer times near the junction as compared to another, isoenergetic, ZOBS located initially near an isolated resonance.

A novel phenomenon that occurs only in  $f \geq 3$ systems for $\epsilon \ll 1$ is that of Arnold diffusion\cite{ardif}. In essence, an initial condition in the stochastic layer in the action space can, due to the connected nature of the Arnold web, explore the entire action space by diffusing along resonances and changing directions at the resonance hubs. The timescale involved is exponentially large as suggested by Eq.~\ref{nekhrosh} and it is important to note that the term ``Arnold diffusion" is reserved\cite{lochak_ardif} for a very specific mechanism envisaged by Arnold in his seminal paper. Nevertheless, several studies have also indicated\cite{fastardif} the possibility of an Arnold-like diffusion which happens on a faster timescale then 
the original Arnold diffusion. 

Not surprisingly, given the subtleties associated with observing Arnold or Arnold-like diffusion, far fewer studies have worried about the effect of quantization on Arnold diffusion. Specific examples like the driven quartic oscillator\cite{maly_qardif} and the stochastic pump model\cite{leit_qardif} do suggest that Arnold diffusion might be localized due to quantum effects. In any case, 
the Nekhoroshev regime is expected to be very interesting and relevant in situations wherein the dominant IVR mechanism involves dynamical tunneling\cite{dyntunrefs,refcor3}. For instance, IVR from edge ZOBS is expected to have a strong contribution from dynamical tunneling. In fact, Stuchebrukhov and Marcus clearly showed\cite{stumar_dyntun} that the dynamical tunneling mechanism of IVR, originally conjectured by Davis and Heller\cite{davhel}, can be explained on the basis of a vibrational superexchange model. More recently, is has been established\cite{ksprerapid} that the quantum vibrational superexchange picture has a much cleaner interpretation in terms of the Arnold web. We note, however, that the phenomenon of dynamical tunneling is quite complex and can be broadly classified in terms of resonance-assisted\cite{rat} and chaos-assisted tunneling\cite{cat}. There is little doubt that these phenomena are intimately connected to the topology of and transport on the Arnold web.  At the present moment, however, not much is known about the competition between tunneling and novel classical transport mechanisms like Arnold diffusion. Details on dynamical tunneling and its consequences for IVR can be found in a recent review\cite{keirpc07}.
As a cautionary note, we have been, intentionally, ``mathematically cavalier" in describing this novel phenomenon in $f \geq 3$; this, admittedly brief and hand waving, description is provided to highlight the significant differences that arise in studying the dynamics of systems with $f \geq 3$ as compared to the dynamics of systems with lower DoFs. For a lucid and critical overview of  Arnold diffusion the article by Lochak\cite{lochak_ardif} is highly recommended.  

The more typical scenario in context of the scaling picture of IVR is moderate to large resonant couplings and hence beyond the Nekhoroshev regime, also called as the Chirikov regime.
The transition to the Chirikov regime occurs with increasing perturbation strengths for which the resonances overlap\cite{refcor4} leading to significant amounts of chaos. Due to the extensive overlap, the dynamics in the action space exhibits large scale diffusion over short timescales. The overall transport in the resonance network has both along and across resonance components. However,  several numerical studies\cite{honjokaneko,haller,laskar} have shown that a typical across resonance diffusion dominates over any along resonance diffusion. It is worthwhile noting that across resonance diffusion also happens in systems with $f < 3$ and hence not a new phenomenon. Nevertheless, presence of the two components of transport combined with the dynamics near various resonance junctions is already sufficient to give rise to nontrivial dynamics\cite{honjokaneko} specific to higher dimensional systems and hence, by correspondence,  nontrivial IVR dynamics of the system. An extensive study of the critical role of the hubs to transport on the resonance network in the Chirikov regime has been done by Honjo and Kaneko\cite{honjokaneko}. Using the coupled standard maps as a paradigm, it is argued that the global transport depends on the heterogeneity of the Arnold web induced by the various hub structures. 

A significant difference between the Nekhoroshev and Chirikov regimes arises due to the fact that in the latter regime the zeroth-order frequencies ${\bm \Omega}^{0}({\bf I})$ are, in general, no longer sufficient to describe the dynamics. Substantial deviations from ${\bm \Omega}^{0}({\bf I})$ come from the perturbation $V({\bf I},{\bm \theta})$ and the frequency associated with chaotic trajectories, in particular, exhibit  significant variations with time. Consequently, the frequency map has to be generalized as ${\bf I} \mapsto {\bm \Omega}(t)$ with the nonlinear frequencies ${\bm \Omega}(t)$ being determined numerically. From a correspondence point of view the implementation of the frequency map proceeds as follows:

\begin{itemize}
\item For a specific ZOBS, whose IVR dynamics of interest, with energy $E_{\bf n}^{0}$ the initial classical action-angle variables are chosen such that $H({\bf I}_{0}=\hbar {\bf n},{\bm \theta}_{0}) = E_{\bf n}^{0}$. Alternatively, if one is interested in the IVR dynamics at the energy of interest, then $H({\bf I}_{0},{\bm \theta}_{0}) = E_{\bf n}^{0}$ and note that the initial conditions constrained to a specific ZOBS are a subset of those generated for a specific energy.  

\item The classical equations of motion are numerically solved to obtain the trajectory $({\bf I}(t),{\bm \theta}(t))$ which is then subjected to a time-frequency analysis to obtain the time-dependent frequency vector ${\bm \Omega}(t)$. 

\item The ${\bm \Omega}(t)$ are monitored to check for various possible frequency lockings ${\bf m}^{(r)} \cdot {\bm \Omega}(t)=0$ and other useful information like residence time statistics (trapping) in or near specific resonances and nature of the diffusion. 
This information, in turn, is translated back to the zeroth-order action space and eventually compared to the quantum IVR dynamics in the QNS.

 \end{itemize}

\begin{figure}[t]
\begin{center}
\includegraphics[height=60mm,width=90mm]{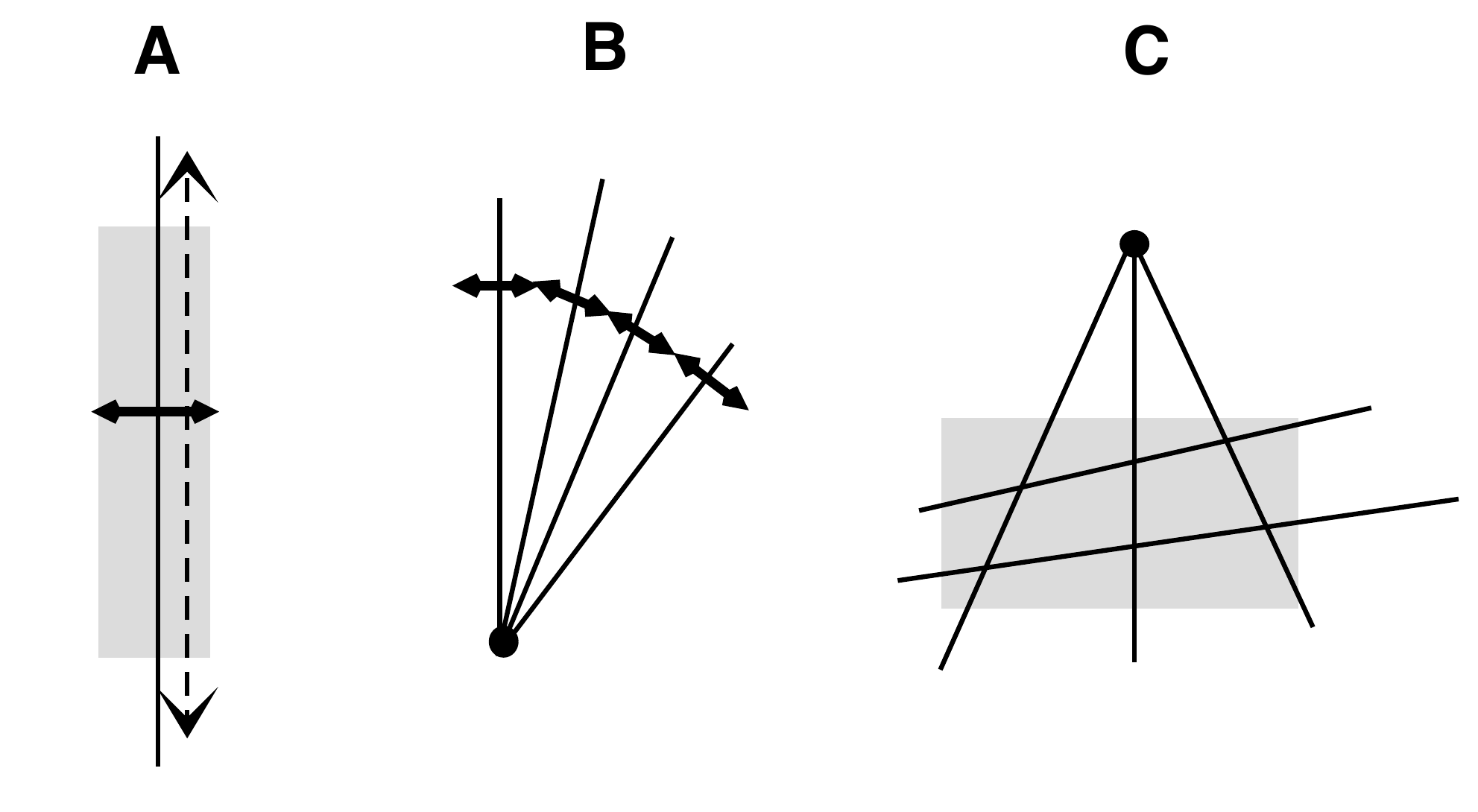}
\caption{Different scenario for diffusion in the frequency ratio space, which relates to the diffusion behavior in action space as well. (A) An isolated resonance (thin line) with the corresponding width (shown in gray). Possible fast, but bounded, across resonance diffusion (solid line with arrows) and slow along resonance diffusion (dotted line with arrows) are indicated. (B) Several resonances emanating from a hub. Overlap of resonances can induce extended fast across resonance diffusion. (C) Resonances (both high and low orders) in many different directions can overlap to generate diffusion over an extended region (gray area) of the frequency ratio space. In cases (B) and (C) high order resonances can play an important role and case (C) is fairly generic for molecular systems. Adapted from reference\cite{laskar}.}
\label{frsdiffcases}
\end{center}
\end{figure}

 The frequencies ${\bm \Omega}(t)$ obtained as above contain valuable information about the IVR dynamics of the system. However, since  primarily one is interested in the various frequency lockings at specific energy of interest, it is useful to construct  the so called frequency ratio space (FRS). This involves writing the resonance condition in Eq.~\ref{rescond} as 
 \begin{equation}
 m_{1}\left(\frac{\Omega_{1}^{0}}{\Omega_{f}^{0}}\right) + m_{2} \left(\frac{\Omega_{2}^{0}}{\Omega_{f}^{0}}\right) + \ldots + m_{f} = 0
 \label{frseq}
 \end{equation}
 where we have assumed $\Omega_{f}^{0} \neq 0$. More precisely, the construction of FRS assumes that the system is isoenergetically degenerate {\it i.e.,} the $(f-1)$ frequency ratios $\Omega_{k}^{0}/\Omega_{f}^{0}, k=1,2,\ldots,(f-1)$ are functionally independent on the CES. Another way of stating the condition is that the derivative of the frequency ratios with respect to actions constrained to the CES are not zero and hence the isoenergetically degenerate condition can be expressed as (see appendix $8D$ of\cite{arnoldbook})
 \begin{equation}
 {\rm Det}\left|\begin{array}{cc} \frac{\partial^{2} H_{0}}{\partial {\bf I}^{2}} & \frac{\partial H_{0}}{\partial {\bf I}} \\
                                                        & \\
                                  \frac{\partial H_{0}}{\partial {\bf I}} & {\bf 0} \end{array} \right| \neq 0                 
 \end{equation}
Note that the condition of nondegeneracy in Eq.~\ref{nondeg} is independent from the isoenergetically nondegenerate condition given above. 
In general, even for systems far from integrable, the phase space dynamics can be visualized in the FRS, only now the frequencies are the actual numerically computed ones rather than the zeroth-order approximations. Examples shown later will amply illustrate the usefulness of the FRS in terms of identifying key phase space structures that regulate the IVR dynamics. Based on the discussion above of the various transport regimes, Fig.~\ref{frsdiffcases} summarizes\cite{laskar} the typical scenario that one can encounter while studying the IVR dynamics in the frequency representation of the Arnold web.

\begin{figure}[t]
\begin{center}
\includegraphics[height=70mm,width=80mm]{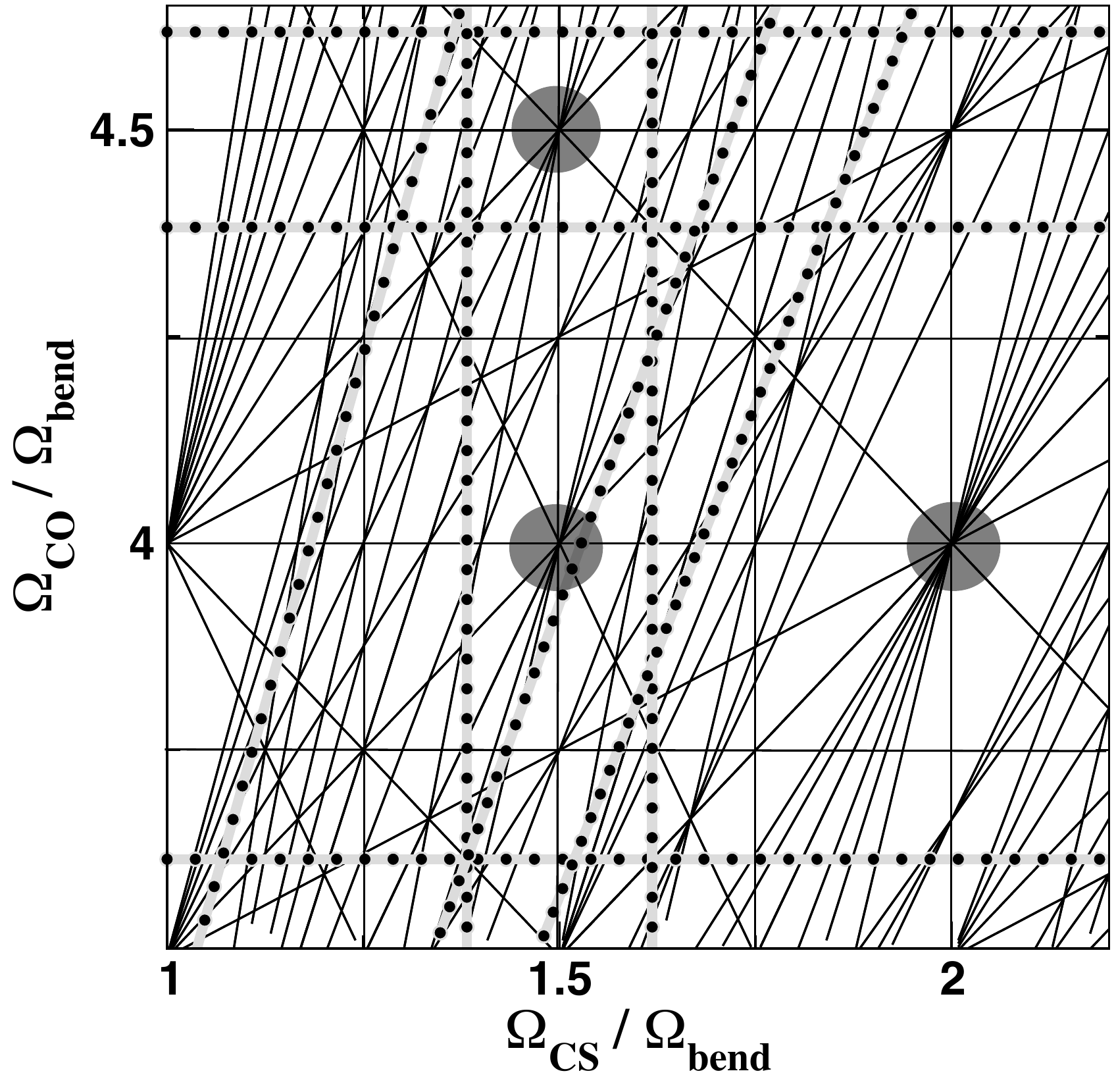}
\caption{Dynamically accessible frequency ratio space for OCS at $E \approx 20000$ cm$^{-1}$. Resonances (lines) of total order $|m| \leq 10$ are shown. Some of the resonance junctions are highlighted with shaded circles. A few of the pairwise noble frequency ratios are also indicated (gray lines with dots). What parts of this web are dynamically important/relevant to the IVR dynamics?}
\label{OCStune}
\end{center}
\end{figure}

In order to illustrate the importance of and the need for constructing the dynamical FRS we briefly recount the case of OCS - a molecule which has received considerable attention over the past few decades. Historically, planar OCS was the first system for which explicit efforts\cite{martezdav_ocs} to characterize the FRS were made. In Fig.~\ref{OCStune} the static FRS for $E \approx 20000$ cm$^{-1}$ is shown, highlighting several possible junctions. The key issue, however, is to identify the dynamically relevant regions of the FRS for the IVR dynamics. Martens, Davis, and Ezra have shown\cite{martezdav_ocs} that the $\Omega_{CS}/\Omega_{\rm bend} = 3/2$ resonance plays a particularly important role. They also identified several other important regions, but stopped short of constructing the full Arnold web or dynamical FRS for the system. Very recently, Paskauskas, Chandre, and Uzer have made critical progress in understanding the mechanism of IVR by identifying families of two-dimensional invariant tori  that act as dynamical traps, leading to significant dynamical correlations\cite{paska_ocs}. This understanding comes from time-frequency analysis of the classical dynamics, identifying the key structures in the FRS of Fig.~\ref{OCStune}, to which we now turn our attention. Note that the quantum manifestations of such lower dimensional tori is not known at this point of time.

\subsection{Computing the Arnold web}
\label{sec:arwebcomput}

The construction of the Arnold web or the resonance network is a challenging task. In particular, constructing and visualizing the Arnold web in the Nekhoroshev regimes requires accurate long time dynamics and novel tools for detecting the various stability zones. The numerical effort involved in exploring the dynamics near high multiplicity junctions is rather high due to the scaling of the width of the resonances with the order of the resonance and the long time stability near such junctions. However, a detailed understanding of phase space transport in higher degrees of freedom and characterizing the transition from the Nekhoroshev regime to the Chirikov regime is of paramount importance in several areas of study - from stability of celestial motion  to the energy flow dynamics in molecules. In the context of IVR it is interesting to observe that the long time stability of an initial ZOBS translates to significant deviations from RRKM. Therefore, in recent years, considerable efforts have been made towards mapping out the Arnold web for specific systems. A detailed account of the various tools and techniques is not attempted here. Instead, we only mention a few of the techniques without getting into their detailed description and focus on one specific approach in the next section. 

\subsubsection{Variational approaches}

The group of Froeschle and coworkers have studied  the Hamiltonian
\begin{equation}
H({\bf I},{\bm \phi}) = \frac{1}{2}(I_{1}^{2}+I_{2}^{2}) + I_{3} + \epsilon \left(\frac{1}{4 + \cos \phi_{1} + \cos \phi_{2} + \cos \phi_{3}}\right)
\label{fliham}
\end{equation}
in great detail for more then a decade\cite{froeschle_fli}. Specifically, using the so called fast Lyapunov indicator (FLI) as a tool they have mapped the Arnold web with increasing $\epsilon$. Note that the form of the perturbation implies that a large number of resonances already appear at $O(\epsilon)$, thereby facilitating the characterization of the transition between different stability regimes. More recently, Todorovic {\it et al.} have used\cite{todorovic_fli} the FLI technique on a four-dimensional map, related to the Hamiltonian above, and observed both fast and slow diffusion in the system. The FLI technique requires some care in the choice of initial conditions and final time of integration. Barrio has proposed\cite{barrio_ofli} a modification of the FLI technique, a second order variation based orthogonal FLI, and applied to several systems including escape from open systems.
The FLI method has not been explored much in the context of intramolecular dynamics. A recent example\cite{schekinova_fli} can be found in  Schekinova {\it et al.} wherein they have used the FLI to investigate the IVR dynamics in the $f=3$ planar OCS molecule. 

Another variational method, the mean exponential growth of nearby orbits (MEGNO)  proposed by Cincotta and Simo, has been extensively studied\cite{cincsim_megno} on various model systems. In particular, a detailed study of the Arnold web of the $f=3$ coupled system
\begin{equation}
H({\bf p},{\bf q}) = \frac{1}{2}(p_{x}^{2}+p_{y}^{2}+p_{z}^{2}) + \frac{1}{4}(x^{4} + y^{4} + z^{4}) + \epsilon x^{2}(y+z)
\label{megnoham}
\end{equation}
showed the existence of stable regions at the resonance junctions. However, this approach has not yet been utilized to study the IVR dynamics in molecular systems. Nevertheless, the above Hamiltonian being a coupled anharmonic oscillator model, some of the observations made in this context could prove to be very relevant to IVR studies. 

A detailed comparison of the various methods, including the possible appearance of spurious structures, is provided in a recent study\cite{barriocomp} which also illustrates the subtlety in distinguishing between chaotic, regular and resonant motions over a finite time scale. Another recent paper \cite{maffione}by Maffione {\it et al.} also provides a critical comparison of the above methods including ones that we have not mentioned - the small alignment index (SALI) method. Note that the techniques mentioned above are all variational in nature and do not directly attempt to obtain the nonlinear frequencies of the system. From the IVR dynamics perspective it is advantageous to work with the spectral methods which obtain the various  ${\bm \Omega}(t)$ numerically. We turn our attention now to some of the spectral techniques with special emphasis on wavelet based time-frequency analysis. 
 
 \subsubsection{Time-frequency analysis}
 \label{sec:frswavelet}

\begin{figure*}
\begin{center}
\includegraphics[height=60mm,width=120mm]{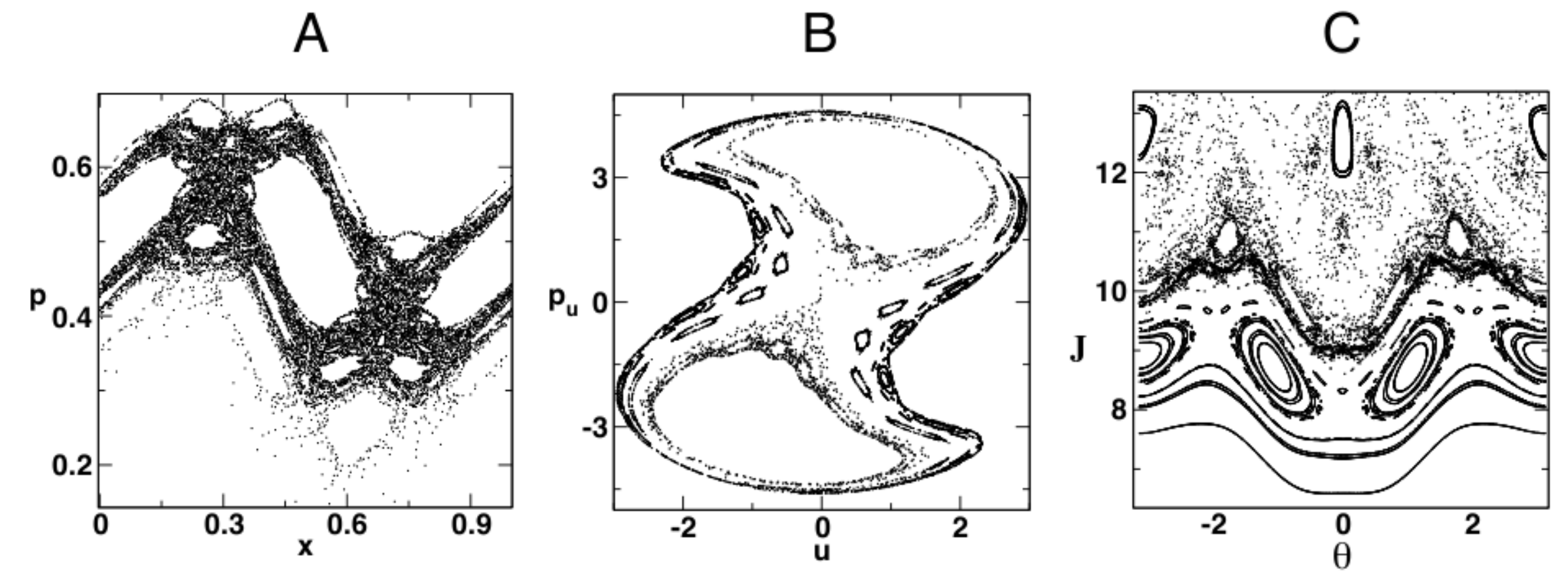}
\caption{ Example of three different systems with mixed phase spaces exhibiting
 ``stickiness",  a hallmark of Hamiltonian systems.
(A) A single irregular but sticky trajectory for the standard map. (B) The Davis-Heller Hamiltonian $(u,p_{u})$ surface of section shows sticky trajectory near the large regular islands. (C) Stroboscopic section of a monochromatically driven Morse oscillator in action ($J$) and angle ($\theta$) coordinates. Considerable stickiness can be seen near the regular-chaos border.}
\label{stickyexamps}
\end{center}
\end{figure*}

The problem of extracting the time-dependent frequencies ${\bm \Omega}(t)$ in a general nonintegrable system is particularly challenging due to the fact that dynamics in mixed regular-chaotic phase spaces is incredibly rich. In the context of this review, mixed phase spaces are generic to molecular Hamiltonians even at very high energies. Typically, in mixed phase space regimes one can broadly classify trajectories into three classes. The first are the regular KAM trajectories for which the frequencies do not change with time and the motion is quasiperiodic. The second class of trajectories are the strongly chaotic ones for which the frequencies vary significantly with time due to the large scale exploration of different regions of the phase space. The third class, somewhat intermediate between the regular and chaotic ones, is considerably more difficult to characterize. These are the so called ``sticky" trajectories\cite{zaslavbook} and, as the name suggests, such trajectories  during their sojourn in the multidimensional phase space frequently get trapped for extended periods of time in the vicinity of a regular structure. The sticky trajectories are chaotic over a long timescale. However, in most physically relevant systems the trapping times can be of the same order as that of a dynamical event of interest. Thus, the ${\bm \Omega}(t)$ of a sticky trajectory can have episodes of near constancy, reflecting the frequencies of the regular structures to which it sticks,  followed by periods of substantial time dependence. Three different examples exhibiting significant stickiness are shown in Fig.~\ref{stickyexamps}, illustrating the generic nature of stickiness for Hamiltonian systems.

From the examples shown in Fig.~\ref{stickyexamps} it should be clear that in mixed regular-chaotic phase spaces all three classes of trajectories coexist. It is crucial to observe that the examples in Fig.~\ref{stickyexamps} are all $f \leq 2$ systems and hence one has the luxury of visualizing the global phase space structure by computing the appropriate Poincar\'{e}  sections. For systems with $f \geq 3$ such as the Hamiltonians in Eqs.~\ref{fliham}, and \ref{megnoham} or most molecular Hamiltonians the increased dimensionality of the phase space and additional new routes to transport, discussed in the previous section, call for an entirely different approach for gaining insights into the dynamics. Time-frequency analysis is precisely one such technique since studying the dynamics in FRS by keeping track of ${\bm \Omega}(t)$  yields the same information as in Fig.~\ref{stickyexamps} in terms of the various frequency ratios and their time-dependence. 
 
 The arguments above point to the need for joint time-frequency information for the trajectories in order to understand the nature of phase space transport in multidimensional systems\footnote{It is worth mentioning here that the various advances in classical dynamics from Newton to Nekhoroshev have always involved frequencies as the central objects.}. The central aim, therefore, is to extract local frequency information from a trajectory, naturally represented as a time series $z(t)$. Henceforth, we will refer to such analysis as the local frequency analysis (LFA). Note that the Fourier transform of $z(t)$
 \begin{equation}
 \hat{z}(\Omega) = \int_{-\infty}^{\infty} z(t) e^{-i\Omega t} dt
 \end{equation}
 yields only information about the frequencies but not time, and hence unsuitable for studying transport. Fourier transforms of dynamical variables can be used to obtain the power spectra for the system and several studies use this information to distinguish between regular and chaotic trajectories. However, such long time distinctions are of little use in the context of IVR. Even finite time power spectra need to be analyzed with care as discussed in detail in an early work by Dumont and Brumer\cite{dumbrumer}. Progress can be made by by using windowed Fourier transforms like the Gabor transform\cite{gabor}
 \begin{eqnarray}
 {\cal G}z(\Omega,t) &=& \int_{-\infty}^{\infty} g(t'-t) z(t') e^{-i\Omega t'} dt' \\
 &=& \frac{1}{\tau \sqrt{2\pi}} \int_{-\infty}^{\infty} e^{-(t'-t)^{2}/2\tau^{2}} z(t') e^{-i\Omega t'} dt'
 \end{eqnarray}
 with $g(t)$ being a ``window function", here chosen to be a gaussian. The joint time-frequency information is encoded in the quantity $|{\cal G}z(\Omega,t)|$, called as the spectrogram. In other words, the spectrogram yields the local frequencies as a function of time. There are other choices for the window function, notably the Blackman-Harris window, but the key idea is to extract time-dependent local frequencies via Fourier transforms of short segments of a long time trajectory. 
 
 A drawback of  windowed transforms has to do with differing accuracies for high and low frequencies arising due to the fixed width of the window function. Therefore, in practice the length of the trajectory segments needs to be chosen appropriately - long enough for obtaining accurate frequencies but short enough so that any interesting dynamical correlations arising from trappings near different phase space structures is not lost. Ideally, one would like the window function to dynamically adapt to the local frequencies characteristic of that particular trajectory segment. Such a method has been around for a number of decades and goes by the general name of wavelet transforms\cite{waveletgenrefs}. 
 
However, before giving a brief discussion of the wavelet method, we note that a very accurate windowed Fourier transform approach to LFA was already provided by Martens and Ezra\cite{martezra_windowft} in the context of semiclassical quantization of multidimensional systems. In fact, Martens, Davis, and Ezra utilized this approach in their pioneering work\cite{martezdav_ocs} on the IVR dynamics of the $f=3$ planar carbonyl sulfide (OCS) molecule, including a rather  detailed study of the dynamics in the FRS and possible manifestations of traps and bottlenecks to energy flow. Subsequent to the OCS study, Laskar proposed\cite{laskar} a similar approach, again very accurate but utilizing the Hanning instead of the Blackman-Harris window, in the context of celestial mechanics. The LFA analysis by Laskar has been instrumental in gaining insights into the long term dynamical evolution and stability of  terrestrial planets\cite{refcor5}. Interestingly, although the central issues involved in IVR and long time celestial dynamics have a certain mathematical similarity, the LFA approach has been utilized much more in the context of celestial mechanics when compared to molecular systems.  Borondo and coworkers have utilized this approach to gain insights into intramolecular dynamics of several systems\cite{tino1,tino2}, especially the LiCN $\leftrightarrow$ LiNC isomerization reaction. Without offering any reasons for this imbalance, and rather than getting into a discussion on the relative merits and demerits of the various approaches, we turn our attention to the wavelet-based LFA.

The history of wavelets starts as far back as the that of IVR itself and, understandably, the literature is too vast to be summarized here. Several\cite{waveletgenrefs} excellent texts and reviews can be consulted for the theory and practice of wavelet transforms - both discrete and continuous.  The description that follows was proposed\cite{arevwiggins} by Vela-Arevalo and Wiggins and used to study the IVR dynamics in a local mode effective Hamiltonian for water and the planar OCS molecule. Subsequently, the method has been applied to various systems like IVR in DCO radical\cite{kspccp}, non-RRKM dissociation dynamics of ethyl radical\cite{chenc2h5}, fractional kinetics in model conformational reaction Hamiltonians\cite{shojitoda},  IVR dynamics of CDBrClF\cite{kscdbrclf}, dynamical assignments of highly excited vibrational eigenstates\cite{kscdbreigen}, IVR in systems with methyl rotors\cite{manirot}, spectral diffusion in hydrogen-bonded systems\cite{amalenhbond}, transport in circularly restricted three body problem\cite{arefox}, and driven coupled Morse oscillators\cite{ksdrivenmorse}.  Some of the examples will be highlighted in the next section. In this rather brief discussion, we will focus on continuous wavelet transform and that too with a specific choice of the, so called, mother wavelet. 

The wavelet transform of $z(t)$ is defined as
\begin{eqnarray}
{\cal W}_{g} z(a,t) &=& \int_{-\infty}^{\infty} z(t') g^{*}_{a,t}(t') dt'  \nonumber \\
&\equiv& \frac{1}{\sqrt{a}} \int_{-\infty}^{\infty} z(t') g^{*}\left(\frac{t'-t}{a}\right) dt' 
\label{wavelett}
\end{eqnarray}
with the parameter $a$ being the scale and $g_{a,t}(t')$ is called as the mother wavelet. Assuming $g_{a,t}$ to be some localized function in time, Eq.~\ref{wavelett} yields the local frequency (inversely proportional to the scale $a$) over a small time interval around $t$ within the constraints of the time-frequency uncertainty principle. Choosing the Morlet-Grossman form for the mother wavelet
\begin{equation}
g(t) = \frac{1}{\tau \sqrt{2\pi}} e^{-t^{2}/2\tau^{2}} e^{2\pi i \lambda t}
\label{mgmother}
\end{equation}
one can see that $g_{a,t}(t')$ has a narrow width for small values of the scale $a$ and vice versa. Thus, although there is a similarity to the Gabor transform, the essential difference comes from the fact that the window function now is capable of adapting to the changing frequency. In essence, the wavelet transform involves translating the window function to the time of interest and dilating the window function depending on the local frequency content of the trajectory $z(t)$. The parameters $\tau$ and $\lambda$ can be tuned to improve the resolution. The modulus $|{\cal W}_{g} z(a,t)|$ contains all the relevant frequencies in a time window around $t$ and one can thereby generate the so called scalogram associated with a specific trajectory. However, in practice, one is interested in the dominant frequency component which is determined by the scale $a_{m}$ that maximizes the modulus of the wavelet transform. Specifically, the local frequency at time $t$ is given by
\begin{equation}
\Omega(t) = \frac{1}{2a_{m}}\left[\lambda + \left(\lambda^{2} + \frac{1}{2 \pi^{2} \tau^{2}} \right)^{1/2}\right]
\end{equation}
In order to use the wavelet based LFA approach to characterize the Arnold web and IVR dynamics in $f \geq 3$ systems it is useful to benchmark the method using $f \leq 2$ systems, where phase space can be readily visualized. Below we show two such examples before proceeding to systems with higher degrees of freedom.

\begin{figure*}[t]
\begin{center}
\includegraphics[height=60mm,width=120mm]{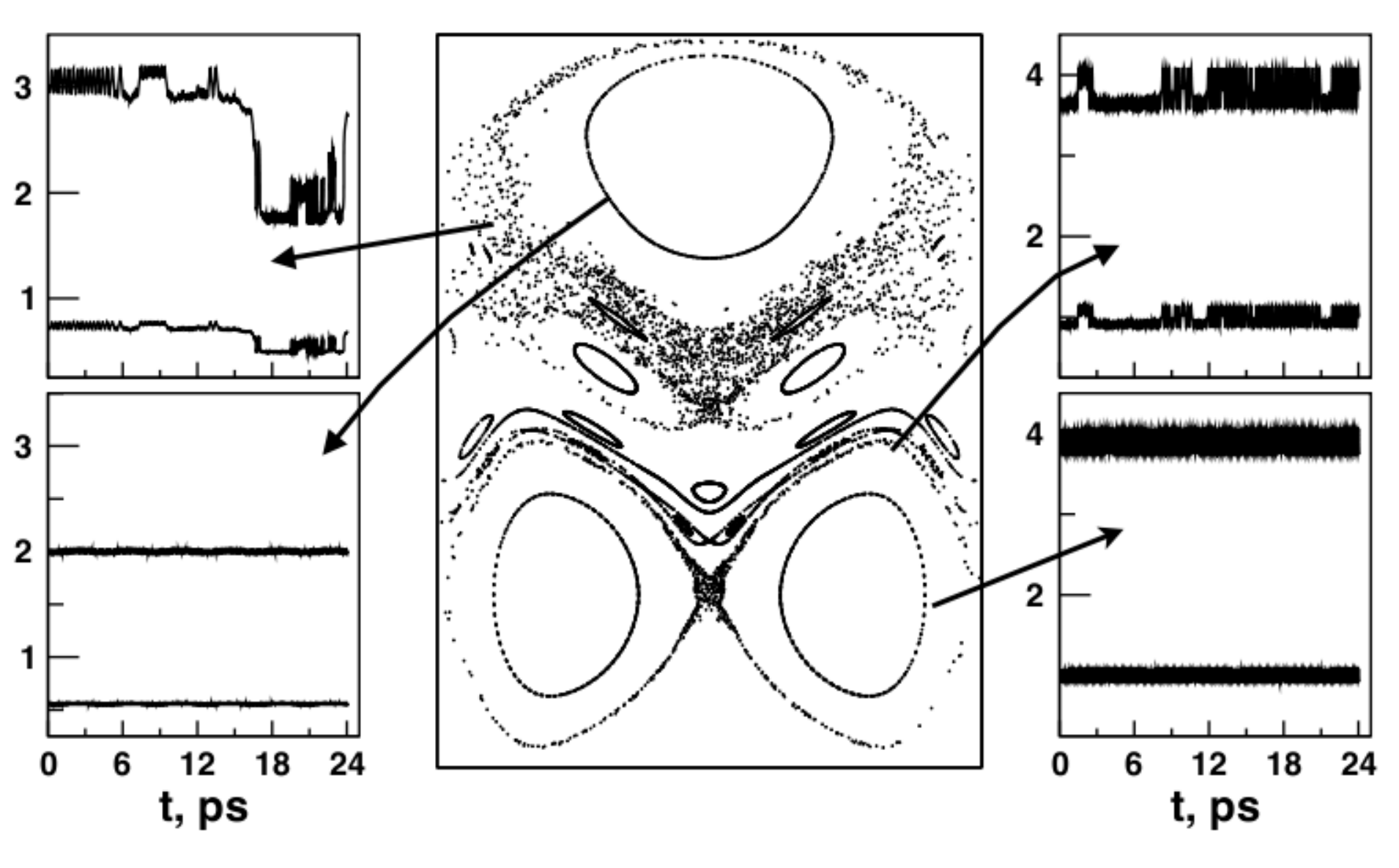}
\caption{Example of a mixed phase space (center) surface of section generated for the Baggott Hamiltonian for H$_{2}$O. Two independent frequency ratios for each trajectory (indicated by arrows) are computed using wavelet transform with $\tau=2$ and $\lambda=1$. Lower left and right panels show a $2$:$1$ and a $1$:$1$ resonant trajectories respectively. The upper left and right panels show chaotic trajectories exhibiting varying degree of stickiness near the regular regions.}
\label{waveexamps1}
\end{center}
\end{figure*}

As an example, in Fig.~\ref{waveexamps1} shows the result of the wavelet analysis for a typical mixed phase space situation. The phase space itself is generated for the three-mode Baggott spectroscopic Hamiltonian\cite{baggott,ksgseh2o} for water and shown as a surface of section in appropriate action-angle coordinates\cite{ksgseh2o}. Two independent frequency ratios $\Omega_{1}/\Omega_{b}$ (stretch-bend ratio) and $\Omega_{1}/\Omega_{2}$ (stretch-stretch ratio) are shown as a function of time. The frequency ratios are essentially constant over the entire $25$ ps duration for the large regular resonances. Note that the modulation observed in these cases are expected\cite{fmicordani}. The upper right panel of Fig.~\ref{waveexamps1} corresponds to a broken normal mode separatrix with a thin layer of chaos. The frequency ratios in this case exhibit  nontrivial time dependence and stickiness due to the $1$:$1$ resonance island can be clearly seen. An example of a chaotic trajectory with significant time-dependence is shown in the upper left panel. Again, sticking of the trajectory near the large $2$:$1$ resonance island and the higher order resonance island is captured very well by the wavelet analysis. It is important to note that despite retaining only the dominant frequency for each mode, the method differentiates between trajectories with varying degree of stickiness and chaos.
 
\begin{figure*}[t]
\begin{center}
\includegraphics[height=60mm,width=120mm]{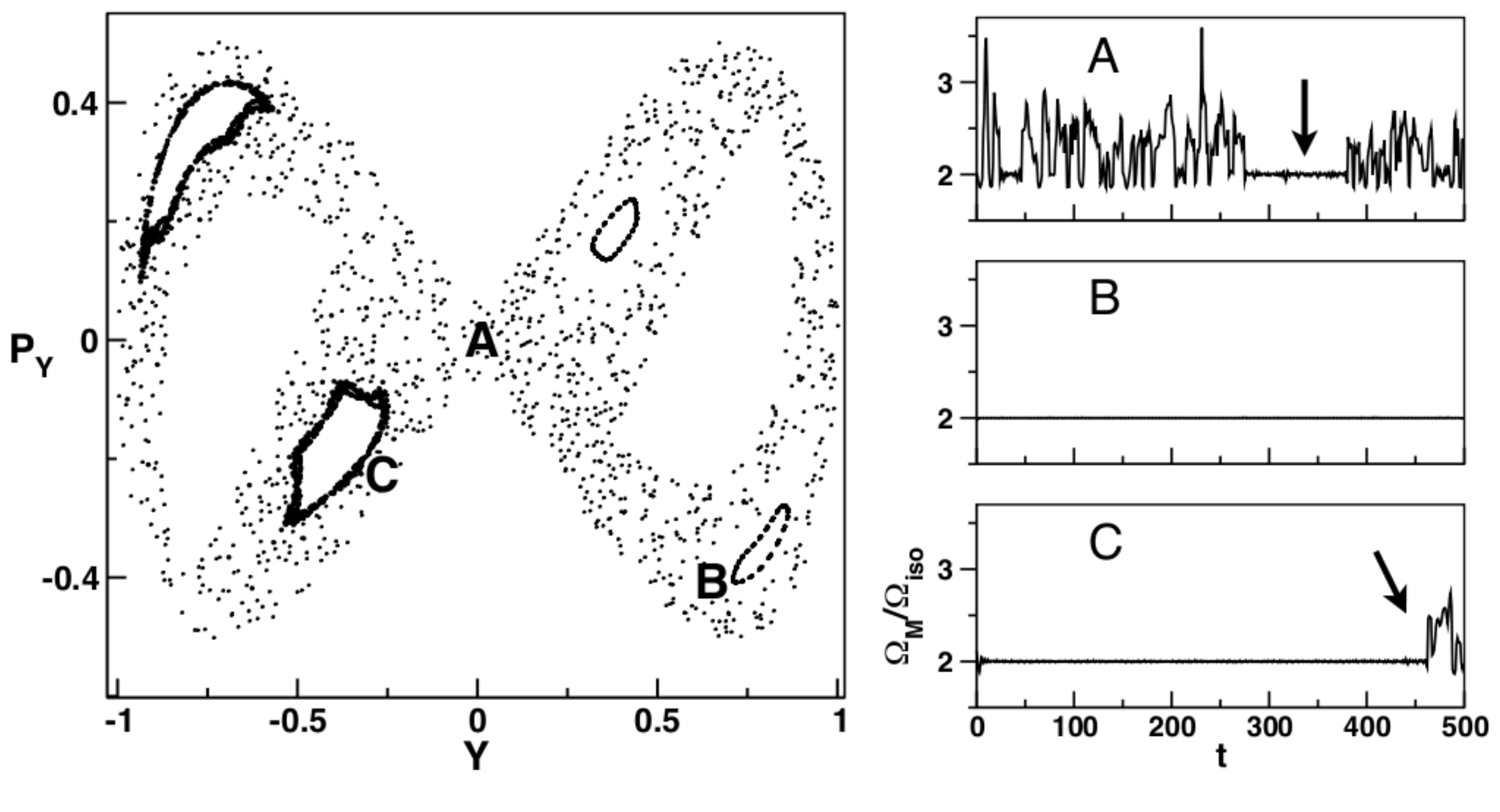}
\caption{Surface of section at an energy slightly above the isomerization barrier and $(\lambda,z)=(1.5,1.0)$ for the De Leon-Berne Hamiltonian.  Time dependent ratio of the Morse to isomerization nonlinear frequencies obtained via the wavelet based approach for (A) a chaotic trajectory with a long episode of stickiness (arrow), (B) a regular resonant island, and (C) a trajectory sticking around the left resonant island for long times and eventually escaping into the chaotic sea. Adapted from computations done by P. Manikandan.}
\label{waveexamps2}
\end{center}
\end{figure*}

Another illustrative example comes from an important work\cite{deleonberne} of De Leon and Berne wherein a model $f=2$ isomerization Hamiltonian
\begin{equation}
H(P_{X},P_{Y},X,Y) = H_{M}(P_{X},X) + H_{\rm iso}(P_{Y},Y) + V(X,Y)
\label{deleonberneham} 
\end{equation}
with the zeroth-order parts
\begin{eqnarray}
H_{M}(P_{X},X) &=& 4 P_{X}^{2} + \lambda_{3} \left[1 - e^{-\lambda X}\right]^{2} \nonumber \\
H_{\rm iso}(P_{Y},Y) &=& 4 P_{Y}^{2} + 4 Y^{2}(Y^{2}-1) + 1  
\end{eqnarray}
and the coupling term
\begin{equation}
V(X,Y) = 4 Y^{2}(Y^{2}-1) \left[e^{-z \lambda X} - 1 \right] 
\end{equation}
describing coupling of a Morse oscillator ($H_{M}$) to a double well potential ($H_{\rm iso}$) is studied in great detail. In particular, the authors observed irregular trajectories exhibiting a very high degree of correlation and executing coherent librational motion in one of the wells. In fact, these observations correlated well with deviations of the numerically computed rates from the RRKM predictions.  In Fig.~\ref{waveexamps2} we show a specific regime from the original work wherein the microcanonical reactive flux exhibits nontrivial behavior despite the system being at an energy slightly above the barrier to isomerization. In this case there is only one frequency ratio $\Omega_{M}/\Omega_{\rm iso}$ and Fig.~\ref{waveexamps2} clearly indicates the existence of extensive stickiness in the system.  For instance, the chaotic trajectory gets trapped for $t \approx 100$ near the regular island (marked C in the figure) whereas an initial condition near the same region sticks for $t \approx 450$ before escaping out to the stochastic region. Again, the ratio of the dominant frequencies is able to account for the rich intramolecular dynamics of the system.

Before leaving this section we briefly mention some of the systems for which the Arnold web has been constructed, in addition to the earlier mentioned examples using the FLI and the MEGNO approaches. Laskar has constructed\cite{laskar} and provided  a detailed analysis of the coupled standard maps using his LFA approach.  The same system has also been analyzed  by Martens, Davis, and Ezra\cite{martezdav_ocs_long} as well as Honjo and Kaneko\cite{honjokaneko}. The latter authors have performed very detailed studies, particularly on the role of the various resonance junctions to the transport. The Arnold web for the hydrogen atom in crossed electric and magnetic fields system has been constructed\cite{milczuzer} by Milczewski, Diercksen and Uzer. Cordani has proposed\cite{fmicordani} a frequency modulation indicator (FMI) method, implementable within the wavelet based approach, and constructed the Arnold web for the quadratic Stark-Zeeman and other perturbed Keplerian systems. Very recently Seibert, Denisov, Ponomarev, and H\"{a}nggi have determined the time evolution of the Arnold web for a model near-integrable three DoF using a novel computational method\cite{gpuarweb}.

\subsubsection{``Coarse-grained" frequency ratio space}
\label{sec:coarsefrs}

\begin{figure}
\begin{center}
\includegraphics[height=70mm,width=90mm]{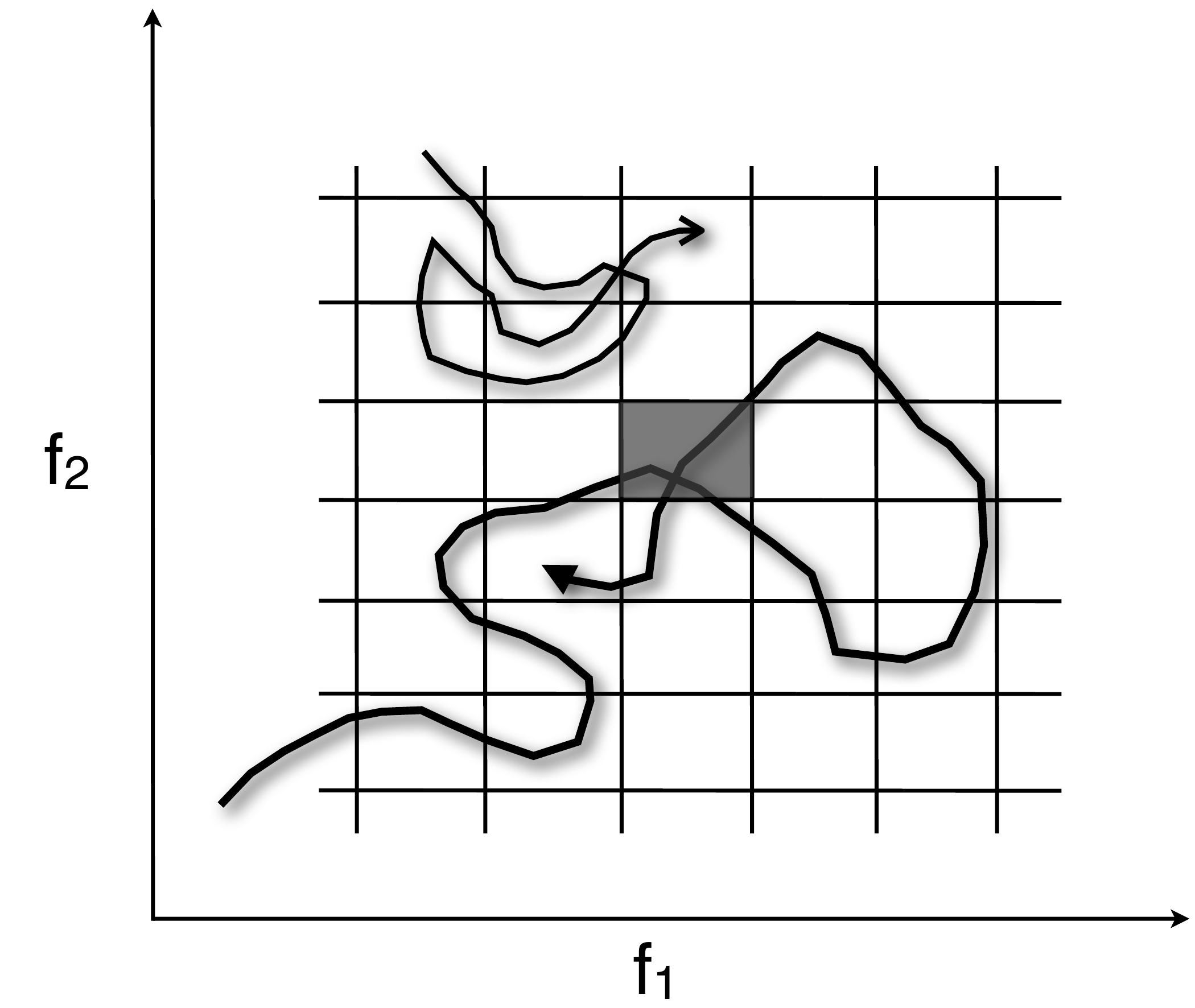}
\caption{Coarse graining the frequency ratio space. The FRS $(f_{1},f_{2})$ is divided into cells (grey) and dominant frequency ``trajectories" are followed, recording their visits to each cell. The result is a density plot of the FRS at specific times of interest. See text for details.}
\label{cfrs}
\end{center}
\end{figure}

Given the computational efforts involved in computing the Arnold web as a function of time, there is a need to construct simpler representations of the FRS which still retain some of the essential dynamical information. One such approach involves ``coarse-graining" the FRS, as indicated schematically in Fig.~\ref{cfrs}. Basically, the FRS is divided into cells and the total number of visitations of each trajectory in a particular cell is recorded. Repeating the procedure for a sufficiently large number of trajectories yields a density plot of the FRS. Such a density plot is capable of revealing the important regions in the FRS including specific resonances, resonance junctions, and the presence of sticky regions. 

The nature of the FRS so constructed can be determined in two extreme limits. In the first limit, if every trajectory on the CES explores the FRS uniformly then the resulting density plot will show no regions with enhanced density. The other opposite limit corresponds to the case when  the trajectories have very little or no excursion in the FRS {\it i.e.,} regimes typical for near-integrable systems. In this limit the density plot will look similar to the Arnold web provided sufficient number of trajectories are computed for long times. Note that typically one tracks only the dominant frequency through the FRS and hence the ``trajectory" shown in Fig.~\ref{cfrs} is usually not as smooth. It is possible to keep track of the subdominant frequencies as well and construct a more refined density plot. Examples of such FRS density plots are shown in the next few sections.

\subsection{Quantum state space $\leftrightarrow$ classical phase space}
\label{sec:qnsfrscor}

In this section we illustrate the correspondence between the state space and phase space IVR dynamics by revisiting the questions asked in Sec.~\ref{sec:qnsquestions} in context of the state space approach to IVR.  En route, apart from outlining some of the progress made to date, the discussions naturally lead to some of the open questions that remain in this field. 

\begin{enumerate}
\item {\em The nature of the state space diffusion}. Given the correspondence between the QNS shown in Fig.~\ref{intro_qns} and the Arnold web shown in Fig.~\ref{intro_arweb}, it is natural to expect that the nature of the classical transport in the FRS may significantly impact the nature of transport in the QNS. Nature of the classical transport depends on the extent of stickiness, existence of various barriers or bottlenecks, and the topology of the Arnold web. In fact stickiness is one possible reason for observing classical anomalous transport\cite{zaslavbook}. For instance, the stickiness of the standard map orbit shown in Fig.~\ref{stickyexamps}A happens due to a very special phase space structure, a noble cantorus, formed due to the breakdown of a KAM torus with highly irrational frequency ratio. Stickiness is generic\cite{perrywiggins} to Hamiltonian systems and has been observed in a wide variety of systems. Indeed, every single example shown in Fig.~\ref{stickyexamps} can be traced back to the effect of specific phase space structures. We refer the reader to the excellent review by Zaslavsky\cite{zaslavbook,zaslavreview} wherein the phenomenon of stickiness is put in the more general framework of dynamical traps and fractional kinetics. 

The relevance to IVR is all the more expected because a 
measure that has been central to all studies on trapping of chaotic
trajectories is the distribution of Poincar\'{e} recurrences\cite{zaslavbook} ${\cal P}(t)$ which is the probability to return to a given region in phase space with a
recurrence time larger than $t$. In fact there is an intimate link between
the decay of correlations and ${\cal P}(t)$. The fact that ${\cal P}(t)$
exhibits a power law behavior
\begin{equation}
{\cal P}(t) \sim t^{-\gamma},
\label{recprob}
\end{equation}
due to trapping of chaotic trajectories in the hierarchically structured
regular-chaotic border in the phase space implies that decay of
correlations is non-exponential. Again, an immediate link to RRKM can be made
which was realized in an important paper by De Leon and Berne nearly three decades ago\cite{deleonberne}. 
There is still considerable debate over the value of the exponent $\gamma$
and as to its universal nature. In particular, Weiss {\it et al.}
argue\cite{wepre03} that a single scaling is not sufficient to capture the
long time trapping dynamics accurately. Note that this is reminiscent
of similar arguments in the context of anisotropic diffusion
in the QNS. However, in a recent work Venegeroles\cite{veprl09}
has argued that the exponent $\gamma$ and the exponent $\beta$ 
characterizing the anomalous diffusion
\begin{equation}
\langle \left(\Delta_{t}x\right)^{2} \rangle \sim t^{\beta},
\end{equation}
for a dynamical variable $x(t)$
obey a universal relation $\gamma+\beta=3$ for unbounded phase spaces.
In case of bounded phase space one anticipates $3/2 \leq \gamma \leq 3$ and 
there seems to be considerable support from numerical studies on two-dimensional
systems. Interestingly, and this points to a strong connection to the state space
scaling arguments, the power law behavior in Eq.~\ref{recprob} has been linked
to conductance fluctuations and eigenfunctions of mesoscopic systems. 

\begin{figure*}
\begin{center}
\includegraphics[height=60mm,width=130mm]{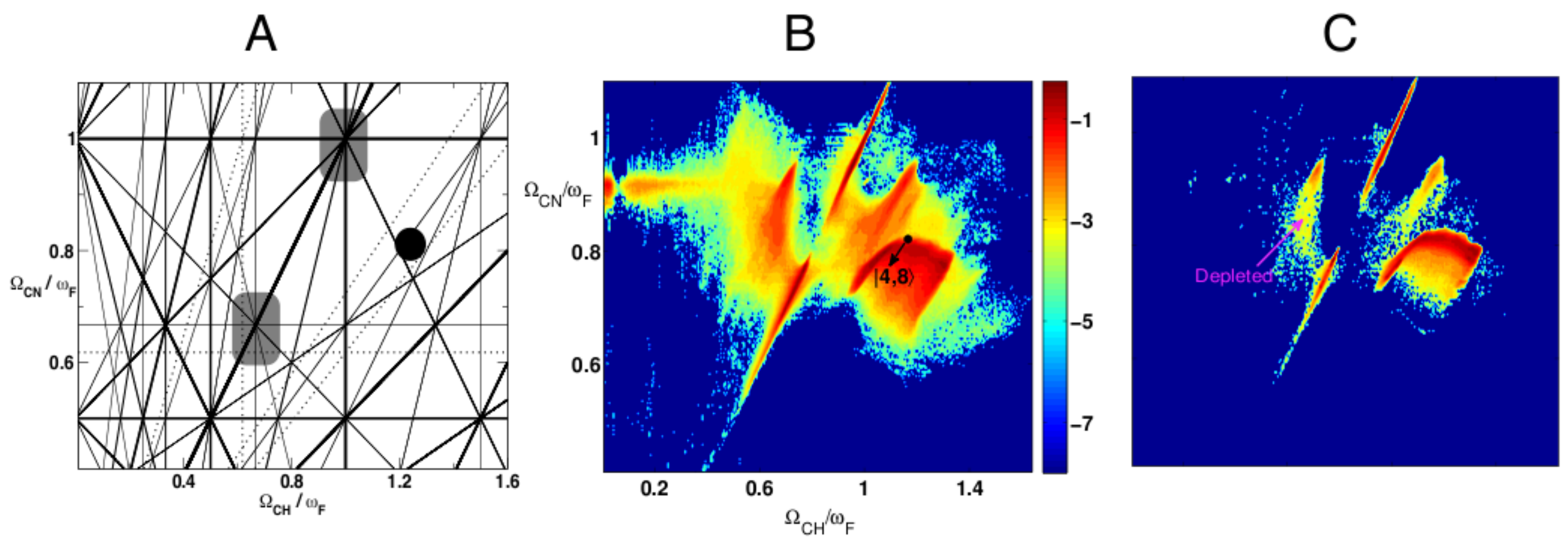}
\caption{Example of stickiness in a $f=2.5$ DoF system. (A) Arnold web $(\Omega_{\rm CH}/\omega_{F},\Omega_{\rm CN}/\omega_{F})$ for the driven coupled Morse system showing all resonances of order $|m| \leq 5$. Dotted lines indicate some of the pairwise noble frequency ratios. (B) FRS density plot for the zeroth-order state $|n_{\rm CH},n_{\rm CN}\rangle = |4,8\rangle$ generated by propagating $5000$ trajectories for $5000$ field periods. (C) FRS as in (B), but constructed using trajectories that do not dissociate until the final time.}
\label{2dmorse_frs}
\end{center}
\end{figure*}

A substantial portion of the work alluded to above is for systems with $f=2$ and extensions to higher DoFs is a problem of considerable current interest. As evident from the recent article of Bunimovich\cite{buni_open}, establishing and characterizing stickiness in $f \geq 3$ is still an open problem\cite{kurths}. There are several reasons for this and we mention a few of them. Firstly, the notion of cantori does not readily generalize. Is it still possible for lower dimensional objects to influence transport in $f \geq 3$? Secondly, the nature of transport near resonance junctions and its consequence for global phase space transport is not yet studied in detail. Hints for resonance junctions to act as dynamical traps\footnote{Strictly speaking, one cannot have absolute dynamical traps in Hamiltonian systems. Nevertheless, trajectories can be trapped for times much longer then any physically relevant  timescale in the system.} have come from numerical studies. However, do such dynamical traps entail anomalous dynamics? Thirdly, there is the issue of whether stickiness survives the ``thermodynamic limit" {\it i.e.,} $f \rightarrow \infty$. A recent work\cite{altmankantz} by Altmann and Kantz on coupled area preserving maps suggests reduced stickiness with increasing DoFs and hence bodes well for the Logan-Wolynes conjecture on the applicability of the AL $\leftrightarrow$ QNS for sufficiently large molecules. 

An example for sticky behavior and importance of pairwise noble barriers in higher DoFs comes from a recent study\cite{ksdrivenmorse} of driven, coupled Morse oscillators. The Hamiltonian is given by
\begin{eqnarray}
H({\bf p},{\bf q}) &=& \sum_{j=x,y}\left[\frac{1}{2M_{j}} p_{j}^{2} + D_{j}(1-e^{-a_{j} j})^{2}\right] \nonumber \\
&-& K p_{x} p_{y} -\lambda_{F} \mu(x) \cos(\omega_{F} t)
\label{2dmorseham}
\end{eqnarray}
with Morse parameters corresponding to the CH-stretching mode ($x$) and CN-stretching mode ($y$) of the HCN molecule. The modes are kinetically coupled and the entire system is driven by a monochromatic field with strength $\lambda_{F}$ and frequency $\omega_{F}$. This is a minimal system which has $f > 2$ and active IVR amongst the two modes. 

In Fig.~\ref{2dmorse_frs}A we show the Arnold web, including resonances upto order $|m| \leq 5$, for the dynamically accessible range corresponding to an initial state $|n_{\rm CH},n_{\rm CN}\rangle = |4,8\rangle$. The dominant nonlinear frequencies $(\Omega_{x}(t),\Omega_{y}(t))$ were determined by subjecting the dynamical variables $z_{j} = x_{j} + i p_{j}$ to the continuous wavelet transform. The resulting FRS density plot, shown in Fig.~\ref{2dmorse_frs}B, clearly identifies the dynamically relevant regions of the Arnold web. Interestingly, significant density is found along the $(1,1)$ line, and around the $\Omega_{x}/\omega_{F} \sim \gamma - 1$ region. In addition the maximal density is seen over a broad region centered around $(1,2,0.7)$ which, on comparing to the Arnold web, corresponds to a region comprised of several high order mode-mode resonances and some of the pairwise noble ratios. Analysis of residence time distributions in the various high density regions indicated considerable stickiness\cite{ksdrivenmorse}. Indeed, further confirmation comes from Fig.~\ref{2dmorse_frs}C wherein the FRS is constructed using only trajectories that do not undergo dissociation up to the final time. 

The FRS in Fig.~\ref{2dmorse_frs} clearly shows the important role played by resonance junctions and lower dimensional barriers in the dissociation dynamics. Preliminary work shows\cite{astha_thesis} that the $\Omega_{x}/\omega_{F} \sim \gamma - 1$ region acts as a significant barrier to the quantum dynamics of the state - the quantum dissociation probability is reduced by a factor of two in comparison to the classical dissociation probability. It is worth noting at this juncture that the state does not dissociate in the absence of the driving field, suggesting an intricate interplay between the IVR and the field induced dynamics. It remains to be seen, however, as to whether the dynamical information provided by the FRS can be utilized for controlling the dissociation dynamics.

\item {\em Estimating the effective IVR dimension}. If indeed the supposed state space - phase space correspondence holds, then the effective IVR dimensionality must have its origins in the dynamically relevant part of the resonance network. Such an expectation is based on assuming\cite{ivr_rev8} $D_{b}$ to be correlated with the number of effective anharmonic resonances which participate in the IVR dynamics of the ZOBS of interest. Recently, it was suggested\cite{kscdbrclf} that the fractal dimension of the FRS might correspond to the fractal $D_{b}$ of the QNS. The study involved investigating the femtosecond IVR dynamics from near-degenerate highly vibrationally excited initial states corresponding to the CDBrClF molecule. The Hamiltonian utilized in the study came from detailed experiments by Quack and coworkers\cite{quackexpt} on IVR amongst the high frequency modes of the molecule. Specifically, the Hamiltonian 
\begin{eqnarray}
H &=& \sum_{j=s,f,a,b} \omega_{j} a_{j}^{\dagger} a_{j} + \sum_{i \leq j} x_{ij} a_{i}^{\dagger} a_{i}a_{j}^{\dagger} a_{j} \nonumber \\
&+& \sum_{i,j }^{a,b,f} K_{sij} a_{s}a_{i}^{\dagger}a_{j}^{\dagger} + \sum_{i \neq j}^{a,b,f} D_{ij} (a_{i}^{\dagger})^{2}(a_{j})^{2} + h.c.
\label{cdbrclfham}
\end{eqnarray}
includes the CD-stretch ($n_{s}$), CF-stretch ($n_{f}$), and the two CD-bending ($n_{a},n_{b}$) modes.  The Hamiltonian in Eq.~\ref{cdbrclfham} has several strong anharmonic Fermi, multimode, and Darling-Dennison resonances which couple the various modes. Due to the existence of a polyad (conserved quantity) $N = n_{s}+(n_{f}+n_{a}+n_{b})/2$ the above Hamiltonian corresponds to a $f=3$ system. In Fig.~\ref{cdbrclf_frs}A we show the temporal autocorrelation function\cite{ketzgeis,huckstein}
\begin{equation}
C_{b}(t) = \frac{1}{t} \int_{0}^{t'} P_{b}(t') dt'
\label{coft}
\end{equation}
for two  ZOBS in polyad $N=5$. The edge state $|n_{s},n_{f},n_{a},n_{b}\rangle = |5,0,0,0\rangle \equiv |E\rangle$ ($E_{b}^{0} \approx 10571$ cm$^{-1}$) is near-degenerate with the interior state $|3,3,0,1\rangle \equiv |I\rangle$ ($E_{b}^{0} \approx 10567$ cm$^{-1}$). Both states show the expected intermediate time power law behavior with the interior state decaying faster then the edge state at short times. However, at longer times it is clear that the IVR dynamics from the interior state is more complex as compared to that of the edge state. Indeed, despite $N_{\rm loc}(|E\rangle) \sim 1.8$ and $N_{\rm loc}(|I\rangle) \sim 3.0$, the effective QNS dimension $D_{b} \approx 1.8$ and $1.3$ for  $|E\rangle$ and $|I\rangle$ respectively.

\begin{figure*}
\begin{center}
\includegraphics[height=60mm,width=120mm]{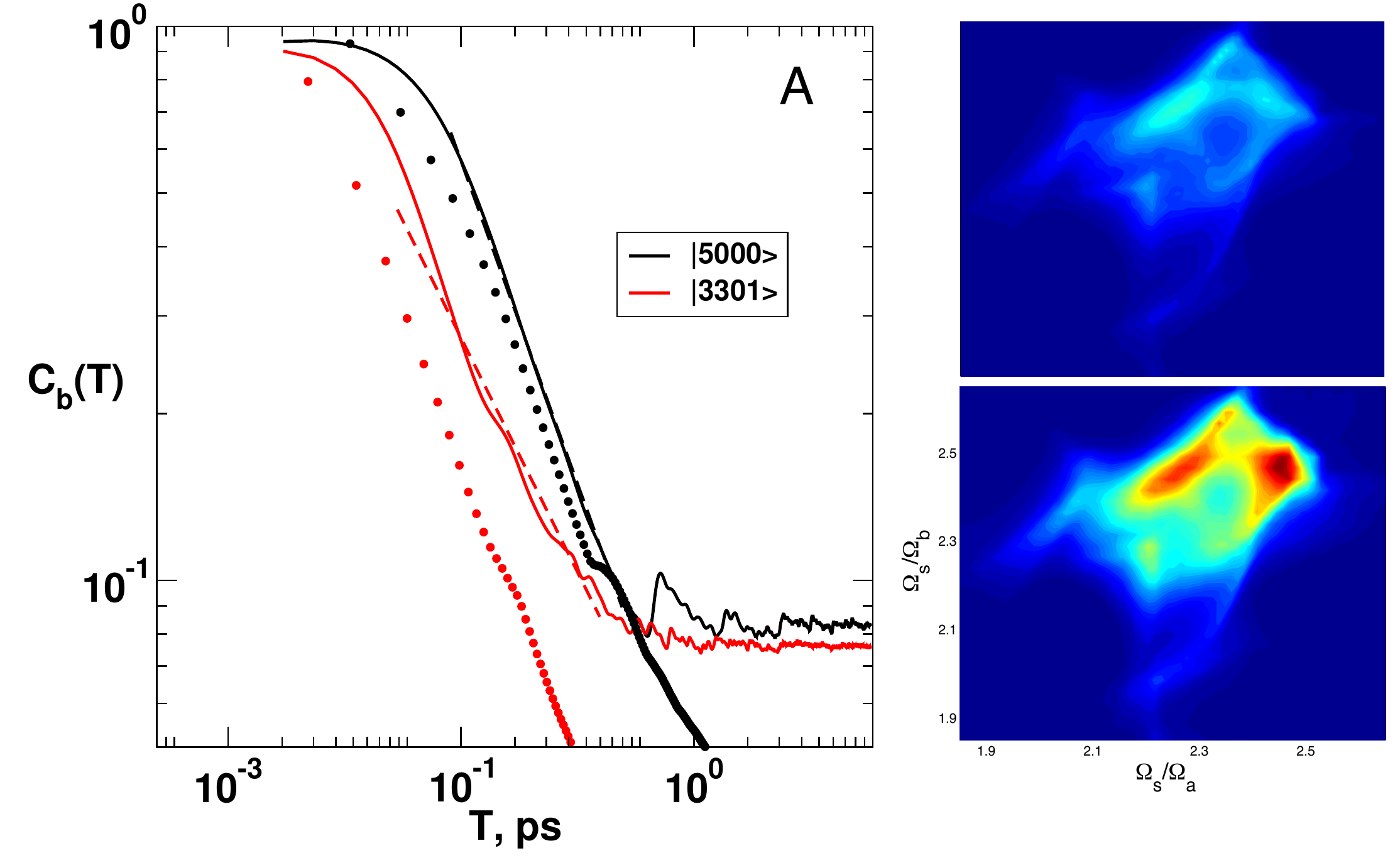}
\caption{(A) Quantum temporal autocorrelation Eq.~\ref{coft} for nearly degenerate edge $|E\rangle$ and interior $|I\rangle$ states of CDBrClF. Clear power law behavior is observed with effective IVR dimension $D_{b}(|E\rangle) > D_{b}(|I\rangle)$. The classical analogs are shown by dots. Right panel shows the classical $(\Omega_{s}/\Omega_{a},\Omega_{s}/\Omega_{b})$ FRS density plot for $|E\rangle$ (top) and $|I\rangle$ (bottom) computed using $5000$ action-constrained trajectories for a total duration of $T=10$ ps. See text for details.}
\label{cdbrclf_frs}
\end{center}
\end{figure*}

Insights into the differing IVR dynamics of the two near-degenerate states comes from looking at the classical IVR dynamics in the FRS. In Fig.~\ref{cdbrclf_frs} the $(\Omega_{s}/\Omega_{a},\Omega_{s}/\Omega_{b})$ FRS density plots for the two states are shown. A clear distinction emerges from comparing the FRS of the two states - FRS of $|I\rangle$ is more heterogeneous and exhibits a particularly strong $\Omega_{a} = \Omega_{b}$ resonance locking involving the bend modes. This strong trapping, absent in $|E\rangle$, is the reason for the more complicated behavior of the $C_{b}(t)$ seen in the figure. Looking at the extent of nonuniformity of the two FRS  it was conjectured that there must be a correspondence between $D_{b}(\rm FRS)$ and $D_{b}(\rm QNS)$. 

Another example pertaining to the molecule CF$_{3}$CHFI is shown in Fig.~\ref{tfie_cortune} wherein the IVR dynamics of three near-degenerate states for $N=5$ is analyzed. The Hamiltonian for this system\cite{tfieham} has the same form as Eq.~\ref{cdbrclfham} with $n_{j}$ for $j=s,a,b$ now denoting the CH stretch and bend modes. The various resonance strengths are quite different from the CDBrClF case. The interior state $|n_{s},n_{f},n_{a},n_{b}\rangle = |3,1,3,0\rangle$ shows a fairly complicated $C_{b}(t)$, particularly around $t \sim 0.5$ ps, and much before the long time dilution limit. The location of the states in the dynamically accessible FRS is also shown in Fig.~\ref{tfie_cortune} and it is clear that $|3,1,3,0\rangle$ is very close to a multiplicity two resonance junction. However, given the strong resonant couplings, it is necessary to compute the dynamically relevant part of the Arnold web using the wavelet based LFA. It would be interesting to see if there are any significant differences in the IVR dynamics of the three states. The results of such a computation are shown in Fig.~\ref{tfie_frs} and it is clear from the figure that the observed enhanced density in different regions for $|4,2,0,0\rangle$ and $|3,1,3,0\rangle$ correlates well with the $C_{b}(t)$ of the corresponding states. Note that the FRS also suggests different IVR mechanisms for the three states, indicating significant mode-dependent IVR even at such high levels of excitation.

Although the examples given above indicate that $D_{b}({\rm QNS})$ is correlated with $D_{b}({\rm FRS})$, an exact relation is not known since the power law scaling in Eq.~\ref{powscal} occurs at intermediate times and a careful study of the time evolution of the Arnold web is required to compute the appropriate fractal dimensions.  Moreover, at best, subtle quantum effects involving competition between localization and tunneling might render $D_{b}({\rm FRS})$ as an upper bound to $D_{b}({\rm QNS})$.
It is safe to say that very little is known about any possible link between $D_{b}({\rm FRS})$ and $D_{b}({\rm QNS})$ at this point of time. Nevertheless, the classical-quantum correspondence seen in Figs.~\ref{cdbrclf_frs} and \ref{tfie_frs} is pleasantly surprising.

\begin{figure*}
\begin{center}
\includegraphics[height=60mm,width=120mm]{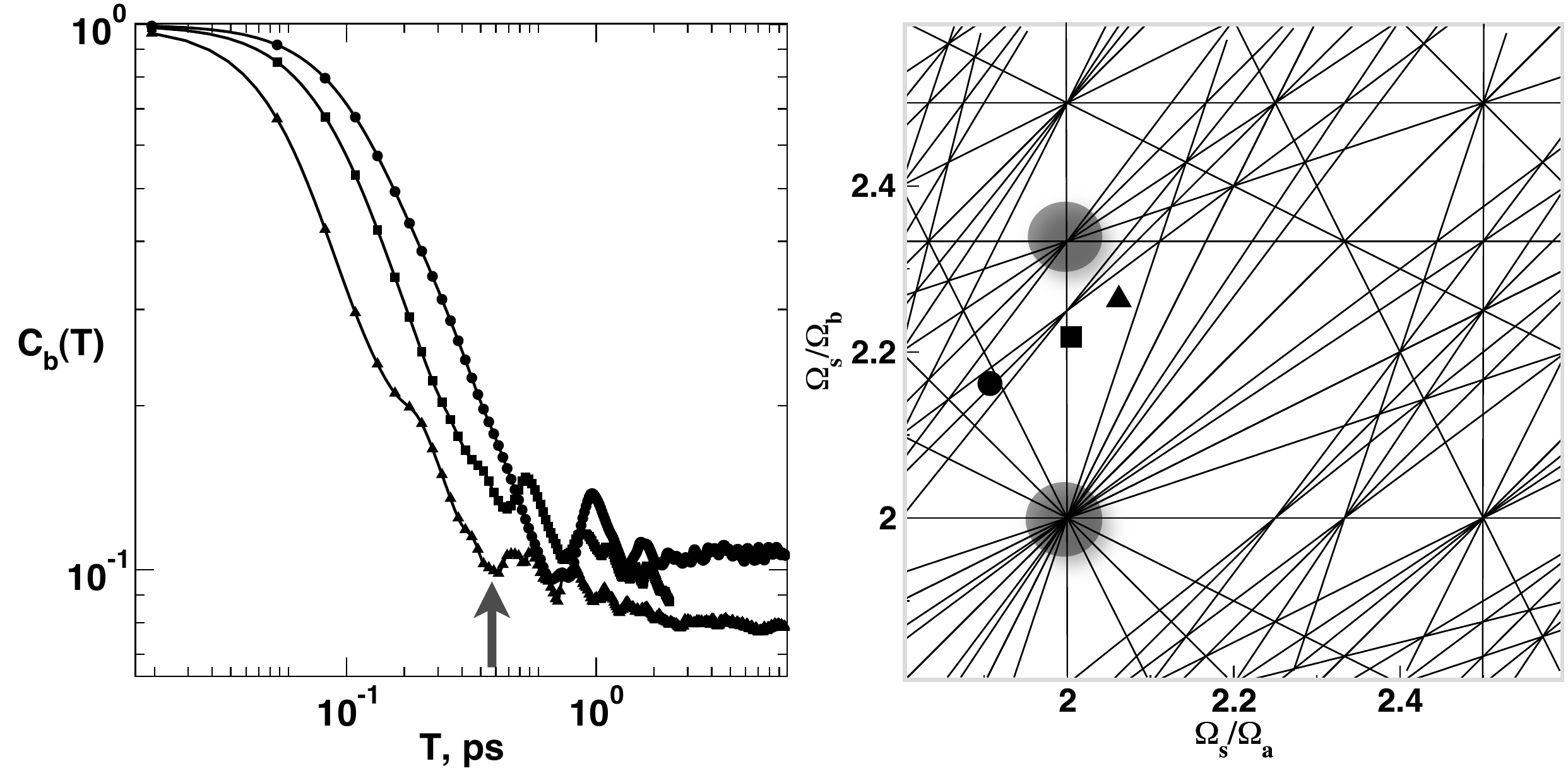}
\caption{Temporal autocorrelations (left panel) for three near-degenerate initial states of CF$_{3}$CHFI. The states $|n_{s},n_{f},n_{a},n_{b} \rangle$ $(E_{b}^{0})$ are $|5,0,0,0\rangle$ ($13793$ cm$^{-1}$, circle), $|4,2,0,0\rangle$ ($13797$ cm$^{-1}$, square), and $|3,1,3,0\rangle$ ($13795$ cm$^{-1}$, triangle). Right panel shows the dynamically accessible tune space $(\Omega_{s}/\Omega_{a},\Omega_{s}/\Omega_{b})$ with resonances of order $|m| \leq 10$ indicated. The location of the three states in the FRS is also shown.}
\label{tfie_cortune}
\end{center}
\end{figure*}

\item {\em Multiple power law behavior}. There is little doubt that anisotropy in the FRS (examples can be seen in Figs.~\ref{cdbrclf_frs} and \ref{tfie_frs}) is going to be reflected in the QNS as well. In addition, finite extent in the state space also translates to a finite extent of the dynamically accessible region in the FRS. Thus, time dependence $D_{b}(t)$ is possible if the system enters a very different region of the FRS, involving a different set of resonances, at a later time. Specifically, here we are referring to multiple power law scaling before the long time dilution limit due to the finiteness of the state space. An example for $D_{b}(t)$ can already be seen in Fig.~\ref{tfie_cortune} for the $|3,1,3,0\rangle$ state. Note that another example for multiple timescales due to the inhomogeneity of the Arnold web comes from a detailed study by Shojiguchi, Li, Komatsuzaki, and Toda for the dynamics of a $f=3$ Hamiltonian model for isomerization\cite{shojitoda}. The classical probability to survive in a reactant well was found to exhibit power law scaling for short residence times and an exponential scaling for longer residence times. The coexisting timescales can be traced back to two different classes of trajectories - those (power law) that mainly localize away from the primary resonance junctions and others (exponential) which did explore the primary resonance junctions. In particular, subdiffusive behavior for the actions were found\cite{shojitoda} in the power law scaling case. 

Based on several studies, another important source for observing multiple power law scaling has to do with the  presence of  major hubs in the FRS. These hubs, combined with anomalous diffusion, can significantly alter the effective IVR dimension. For instance, Martens, Davis, and Ezra observed\cite{martezdav_ocs} trapping near resonance junctions in case of planar OCS, resulting in strong dynamical correlations. Additional examples are starting to emerge in a recent study\cite{kssccl2} on the IVR dynamics in thiophosgene (SCCl${2}$) which lend further support to  the important role of hubs in the FRS. For instance, in Fig.~\ref{thio_powlaw} we show $C_{b}(t)$ for two zeroth order states using the highly accurate effective Hamiltonian for SCCl$_{2}$. One of them is a pure edge state ($|7,0,0,0,0,0\rangle$ with $N_{\rm loc} \sim 1$) and the other one is an interior state ($|3,2,1,0,2,5\rangle$ with $N_{\rm loc} \sim 5$), where the six quantum number labels correspond to the six normal modes of the molecule. As expected the edge state undergoes slow IVR when compared to the interior state. However, starting around $t \sim 0.5$ ps, the interior state slows down considerably and the initial $t^{-0.9}$ scaling becomes a $t^{-0.2}$ scaling. As seen from the inset in Fig.~\ref{thio_powlaw}, the slow down can be related to rapid and partial recurrences of the survival probability $P_{b}(t)$. An important point to note here is that the recurrences happen much before the long time dilution limit ($\sim 2$ ps). A careful study of the dynamical FRS establishes the critical role of specific resonance junctions. 

Since higher then one multiplicity resonances can only occur for systems with $f \geq 3$, we suspect that manifestation of multiple power law behavior in the QNS may be a genuine $\geq 3$ DoF effect. There can, however, be other sources for multiple power law behavior and it is quite possible that some of them might be genuine quantum effects with no easily detectable classical analog. Nevertheless, there are strong indications that inhomogeneous Arnold webs do significantly impact the effective IVR dimensionality.  Preliminary investigations indicate that the second timescale in Fig.~\ref{thio_powlaw} is related to the ``rewirings" shown in Fig.~\ref{qnsapproxcay}A, further evidence for the close QNS-FRS correspondence.

\begin{figure*}
\begin{center}
\includegraphics[height=50mm,width=120mm]{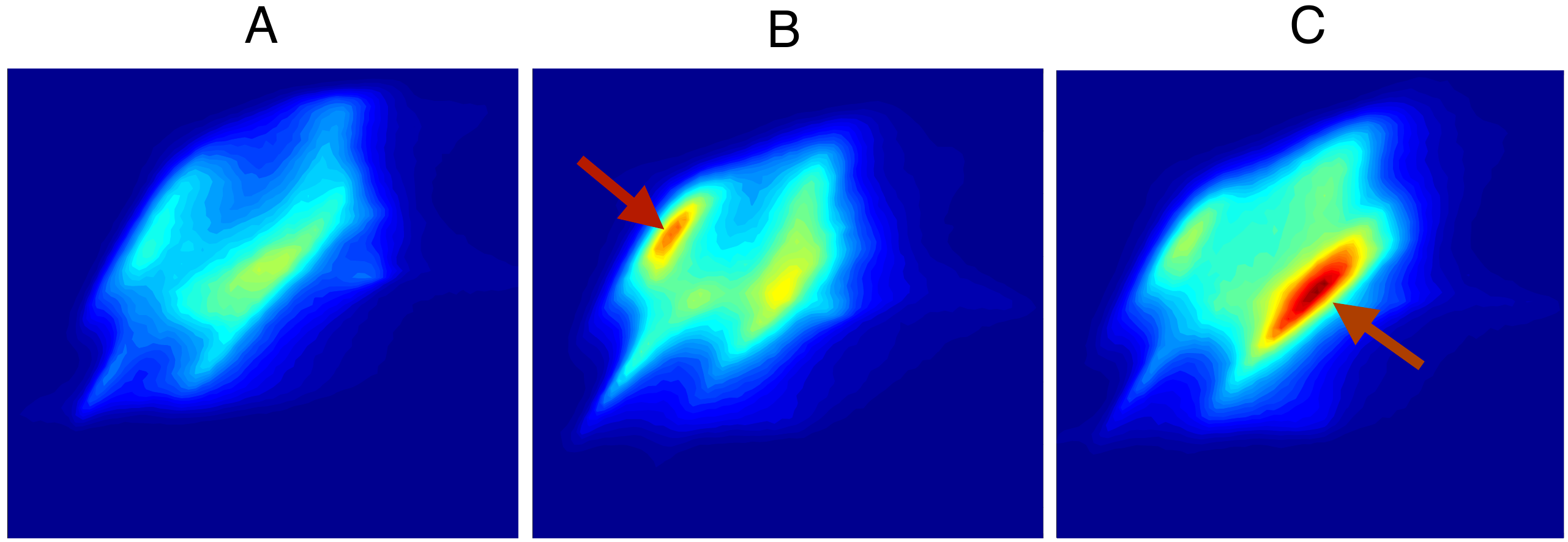}
\caption{Frequency ratio space for the three near-degenerate initial states of CF$_{3}$CHFI (A) $|5,0,0,0\rangle$ (B) $|4,2,0,0\rangle$  (C) $|3,1,3,0\rangle$ computed using $5000$ trajectories for $T=10$ ps. The axes and range are identical to the FRS shown in Fig.~\ref{tfie_cortune}, to which the current plot should be compared. Note the significantly enhanced density (indicated by arrow) for the latter two states. Based on the unpublished work of A. Semparithi.}
\label{tfie_frs}
\end{center}
\end{figure*}

\item {\em Manifestation of power law scaling in the eigenstates}. Based on the AL $\leftrightarrow$ IVR analogy it is natural to expect the vibrational eigenstates to undergo a localization-delocalization transition with multifractal eigenstates around the IVR threshold. At the same time, there exits a large body of work which has established detailed correspondence between the nature of the eigenstates and the underlying phase space structures and their bifurcations. The recent review\cite{faguojoyreview} by Farantos, Schinke, Guo, and Joyeux gives an up to date account of the recent developments.  For example, dynamical assignments of highly excited states eigenstates in several systems clearly indicate the existence of sequences of eigenstates sharing common localization characteristics\cite{jungtay,kseigpaps}. Note that states in such sequences are neither regular  nor ergodic, and certainly not measure zero. Such systems do exhibit intermediate time power law scaling of the survival probability with exponents larger than that for the threshold behavior but less then the dimension of the QNS (cf. Fig.~\ref{cdbrclf_frs}). 

However, it is nontrivial to make an explicit connection between the different classes of eigenstates at specific energies and the effective IVR dimension $D_{b}$. This, in turn, is directly related to the difficulty associated with understanding the nature of highly excited eigenstates in $f \geq 3$ - required for clearly distinguishing the critical scaling from the diffusive scaling regimes (cf. Eq.~\ref{powscaling}). As one example, amongst many others and relating to the first question in this list, we mention the existence of a class of states called as hierarchical states\cite{hierstates}.  In a detailed study, Ketzmerick, Hufnagel, Steinbach, and Weiss associate the hierarchical states in the standard map model with the phase space stickiness and argue that the fraction of such states scales as $\hbar^{\alpha}$ with the exponent being related to the one in Eq.~\ref{recprob} as $\alpha = 1-1/\gamma$. Existence and characterization of such states, let alone their influence on the IVR dynamics, in the $f\geq 3$ case is largely an unexplored area. Possible interesting connections, not elaborated in this review, between the quantum ergodicity threshold of Eq.~\ref{locdeloccrit}, the weak ergodicity theorem of Shnirelman-Colin de Verdi\'{e}re-Zelditch\cite{scdvz}, and the eigenstate thermalization hypothesis\cite{eth} of Deutsch-Srednicki may lead to a better understanding of the nature of the eigenstates, $D_{b}$, and the effective dimensionality associated with the scaling of the average survival probability. A start has been made in a recent work\cite{kscdbreigen} wherein quantum eigenstates have been ``lifted" onto the FRS, in analogy to the lifting of states onto Poincar\'{e} surface of sections using Husimi or Wigner functions.

\item {\em Validity of scaling for small molecules}. At the outset the original arguments of Logan and Wolynes, as elaborated in Sec.~\ref{sec:qet}, are expected to be valid only for sufficiently large molecules. Nevertheless, three decades ago Fishman, Grempel, and Prange in their pioneering work\cite{fishpran} on dynamical localization in the kicked rotor model established a connection between AL and quantum chaos. Recently, Chab\'{e} {\it et al.} have experimentally realized the Anderson metal-insulator transition in a quantum chaotic system\cite{chabeALexpt}. Furthermore, Garcia-Garcia and Wang have proposed a one-parameter scaling theory\cite{garciawang}, analogous to the presentation in Sec.~\ref{sec:scaling}, for understanding the various universality classes of quantum chaotic systems.  Clearly, the interesting parallel between the AL $\leftrightarrow$ IVR analogy and the quantum chaos $\leftrightarrow$ AL analogy can be ``closed" by completing the FRS $\leftrightarrow$ QNS correspondence. If indeed such an argument is tenable then we expect the Logan-Wolynes mapping along with the Schofield-Wolynes scaling approach to be applicable to small molecules as well. Note, however, that the scaling theory for quantum chaos assumes, as do Logan and Wolynes in their work, a completely chaotic phase space - a rarity in molecular context. Indeed,  a recent study\cite{iominzas} by Iomin, Fishman, and Zaslavsky clearly shows the subtle nature of the localization-delocalization transition in the mixed phase space regimes. Understanding such subtleties in $f \geq 3$ requires a comprehensive study of IVR from the perspective of transport in the FRS.

\end{enumerate}

\begin{figure}
\begin{center}
\includegraphics[height=80mm,width=80mm]{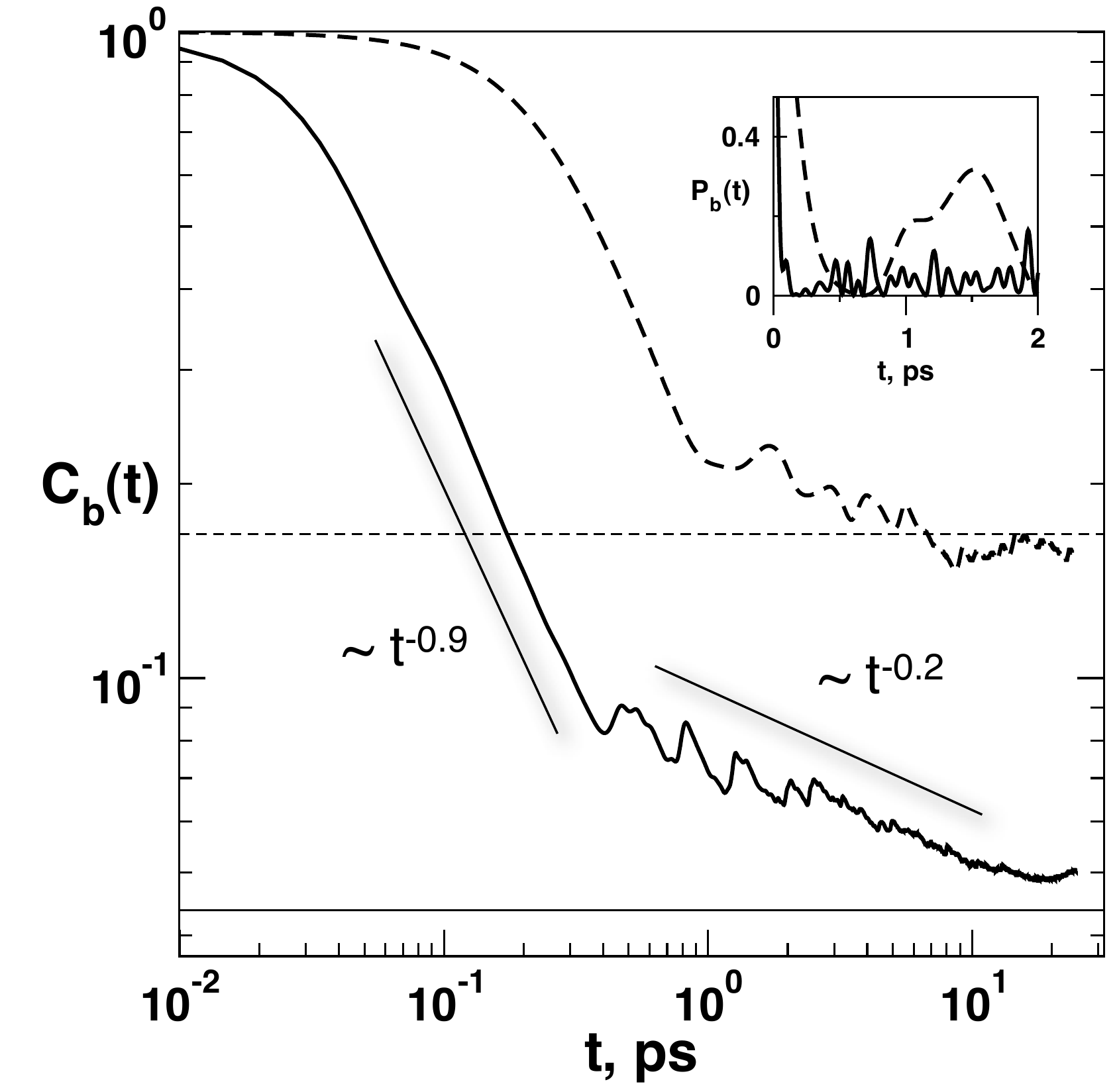}
\caption{Temporal autocorrelation function for two zeroth-order states $|7,0,0,0,0,0\rangle$ (dashed line) and $|3,2,1,0,2,5\rangle$ (solid line) of SCCl$_{2}$. The latter state clearly shows multiple power law timescales. The horizontal lines indicate the respective dilution limits and the inset shows the survival probabilities. See text for discussion.}
\label{thio_powlaw}
\end{center}
\end{figure}

\section{Concluding remarks}
\label{sec:conclusions}

The main aim of this review is to convey the message that the recent advances in our understanding of classical dynamics of systems with three or more degrees of freedom can be utilized to gain important insights into the quantum IVR dynamics in polyatomic molecules. More specifically, the scaling approach to IVR and the state space model of energy flow have a direct correspondence to the classical phase space transport characterized in terms of the Arnold web and local frequency analysis. Combined with the recent striking advances that have happened in the field of TST, one may hope to gain rather deep mechanistic insights into a reaction - all the way from the reactant well, through the transition state, to products. Controlling a reaction involves controlling IVR, directly or indirectly, and progress in answering the questions posed in this review will lead to a clearer understanding of the role of quantum coherence in our efforts to manipulate the dynamics. 

In this review we have focused on gas phase IVR and ignored the topic of condensed phase IVR entirely. Currently, several exciting studies are emerging in the context of condensed phase IVR\cite{orrewing} and it is of some interest to see whether the state space approach can prove useful in this context. An early attempt in this direction has been made by Assmann, Kling, and Abel\cite{akingabel}, which also highlights issues that require further theoretical studies. 

There are still open questions, some outlined in the previous section, which need to be answered in order to make progress. The finiteness of  $\hbar$ raises the interesting and age old issue of how much of the correspondence will actually survive quantization. In the context of $f \geq 3$ systems, this topic is being addressed in earnest only recently. The issues involved here are subtle, to say the least. For instance, partial barriers in phase space can become highly restrictive upon quantization. Nevertheless, quantum dynamics finds a way by tunneling through the barriers. At the same time, one may argue that the effective $\hbar$ of the system is too large for phase space structures with area less than $\hbar$ to influence the quantum dynamics. The naivety of such an argument is evident from recent examples involving coherent control of modified kicked rotors\cite{tuncoh_control}, ionization of atoms in pulsed microwaves\cite{perotti}, and dynamical tunneling\cite{dyntunrefs,keirpc07,entangchaos2}.  In this context, we also note that the argument\cite{leit_qardif} of Leitner and Wolynes on the quantum localization of Arnold diffusion in the stochastic pump model needs to be revisited. The analogy to disordered wires, whose eigenstates are localized, suggest quantum localization of Arnold diffusion. However, for instance, in $f=4$ Arnold diffusion happens on a two dimensional manifold and the analogy to disordered wires might not be very useful. Note that Basko has recently suggested\cite{basko} the possibility of destruction of Anderson localization by Arnold diffusion in the nonlinear Schr\"{o}dinger equation model. 

The remarks made above clearly point to the importance and utility of studying carefully constructed model Hamiltonians. In particular, there is a need here to construct and study the classical-quantum correspondence dynamics of model Hamiltonians which can be tuned from the KAM to the Chirikov, though the Nekhroshev, regimes in a controlled fashion.
Our ability to gain insights into novel mechanisms, uncover possible universal features in a class of systems, and hopefully come up with intelligent control methods is critically dependent on such studies. 

\section*{Acknowledgments}
\addcontentsline{toc}{section}{Acknowledgments}

It is a pleasure to acknowledge my students, Aravindan Semparithi,  Paranjothy Manikandan, and Astha Sethi who contributed significantly  to the results discussed in Sec.~\ref{sec:qnsfrscor}. I am also grateful to Greg Ezra, Martin Gruebele, David Leitner, and Steve Wiggins for several discussions which helped in clarifying, hopefully, several sticky issues in both state space and phase space.

{\em Notes added with this version} - Thanks to Shmuel Fishman and David Leitner for pointing out the references $78$ and $25$ respectively.

\end{document}